\documentclass{article}

\usepackage[english]{babel}
\usepackage[utf8]{inputenc}
\usepackage[T1]{fontenc}
\usepackage{amsmath,amsxtra,amssymb,amsfonts,amsthm,amscd,array}
\usepackage{dsfont}
\usepackage{graphicx,latexsym,wrapfig}
\usepackage{bm}

\DeclareMathOperator*{\argmin}{arg\,min}
\usepackage{color}
\usepackage{natbib}
\usepackage{mathtools}
\usepackage{float}
\usepackage{booktabs}
\usepackage{fancyhdr}
\usepackage{lscape}
\usepackage[colorlinks=true, allcolors=blue]{hyperref}
\usepackage{appendix}
\usepackage{tikz}
\usepackage{environ}
\usepackage{rotating}

\makeatletter
\newsavebox{\measure@tikzpicture}
\NewEnviron{scaletikzpicturetowidth}[1]{%
  \def\tikz@width{#1}%
  \def\tikzscale{1}\begin{lrbox}{\measure@tikzpicture}%
  \BODY
  \end{lrbox}%
  \pgfmathparse{#1/\wd\measure@tikzpicture}%
  \edef\tikzscale{\pgfmathresult}%
  \BODY
}
\makeatother

\usepackage{subcaption}
\usepackage{mwe}

\usepackage[official]{eurosym}
\usetikzlibrary{matrix,chains,positioning,decorations.pathreplacing,arrows}

\newcommand{\Ben}{\begin{enumerate}}
\newcommand{\Een}{\end{enumerate}}
\newcommand{\Bit}{\begin{itemize}}
\newcommand{\Eit}{\end{itemize}}
\newcommand{\Beq}{\begin{equation}}
\newcommand{\Eeq}{\end{equation}}
\newcommand{\Ba}{\begin{align*}}
\newcommand{\Ea}{\end{align*}}

\newcommand{\Mbb}{\mathbb}

\newcommand{\be}{\begin{equation}}
\newcommand{\ee}{\end{equation}}
\newcommand{\bea}{\begin{eqnarray}}
\newcommand{\eea}{\end{eqnarray}}
\newcommand{\beas}{\begin{eqnarray*}}
\newcommand{\eeas}{\end{eqnarray*}}

\usepackage{xparse}

\usepackage[top=3cm,bottom=2cm,left=3cm,right=3cm,marginparwidth=1.75cm]{geometry}

\theoremstyle{definition}

\newcommand{\R}{\mathbb{R}}

\newcommand{\PP}{\mathbb{P}} 

\newcommand{\QQ}{\mathbb{Q}}
\newcommand{\E}{\mathbb{E}}

\title{Design and pricing of a transparent parametric-modeled loss CAT bond: application to German windstorm}
\author{John Ery\footnote{RiskLab, Department of Mathematics, ETH Zurich, john.ery15@gmail.com} \and Erwan Koch\footnote{Expertise Center for Climate Extremes (ECCE), Faculty of Business and Economics (HEC) - Faculty of Geosciences and Environment, University of Lausanne, CH-1015 Lausanne. Corresponding author: erwan.koch@unil.ch}}
\date{\today}

\begin{document}
\maketitle

\begin{abstract}
Catastrophe (cat) bonds overcome some lack of reinsurance by sourcing capacity from the wider capital markets.
We present a new type of cat bond addressing the known trade-off between moral hazard and basis risk.
As our main contributions we propose a trigger mechanism which is entirely transparent and simpler to evaluate compared to indemnity modeling techniques, as well as a methodology to price this cat bond.
This is relevant for insurers and public authorities in a world where natural disasters are occurring with increasing frequency and severity due to climate change, but also for players willing to enter the cat bond market for whom the lack of transparency of this asset class has been a significant obstacle. 
Our trigger is derived from a cost random field which separates the physical hazard, a vulnerability function and the exposure. 
This allows the trigger to take a flexible form between parametric and modeled loss, in case exposure is taken into account.
We present a case study based on historical windstorm events impacting Germany. 
Using wind speed data from historical storms, we fit a max-stable random field on a resolution which is standard in the reinsurance industry. The availability of industry loss and exposure data allows us to calibrate the vulnerability component to historical observations. 
Besides measuring the basis risk associated with our trigger, we perform a full model assessment and discuss numerical results.
\end{abstract}

\section{Introduction}\label{sec:intro}
Catastrophe (cat) bonds transfer peak insurance risks in a securitized form to capital markets. The underlying risk most often being the occurrence of natural catastrophes, cat bonds are lightly correlated to other asset classes such as equities or traditional bonds, and generally offer attractive yields compared to other products with similar levels of risk; see, e.g., \cite{cummins}. They are typically fully collateralized and hence do not include any (or very low) counterparty default risk as the proceeds are invested in government bonds with very low credit risk. 

Cat bonds can be found in the market under various triggering schemes: parametric, modeled loss, industry loss index, indemnity or a combination thereof.
An essential aspect to consider when designing the trigger is to strike an appropriate balance between basis risk and moral hazard, both from the perspective of the sponsor and the investors. Basis risk captures the possibility of a mismatch between the losses suffered by the sponsor and the actual payout from the bond. Moral hazard corresponds to the risk that the parties involved may influence the payment outcome. Decreasing the potential for moral hazard leads to higher transparency.
Cat bond prices are obtained by running vendor models (Moody's RMS, Verisk and CoreLogic being the main ones) which are like black boxes as their modeling assumptions are not public. This means that the risk modeling and ultimately the pricing is not performed in a transparent way and cannot be computed independently by the investor. 

\begin{figure}[ht!]
\begin{center}
\includegraphics[
width=0.75\textwidth]{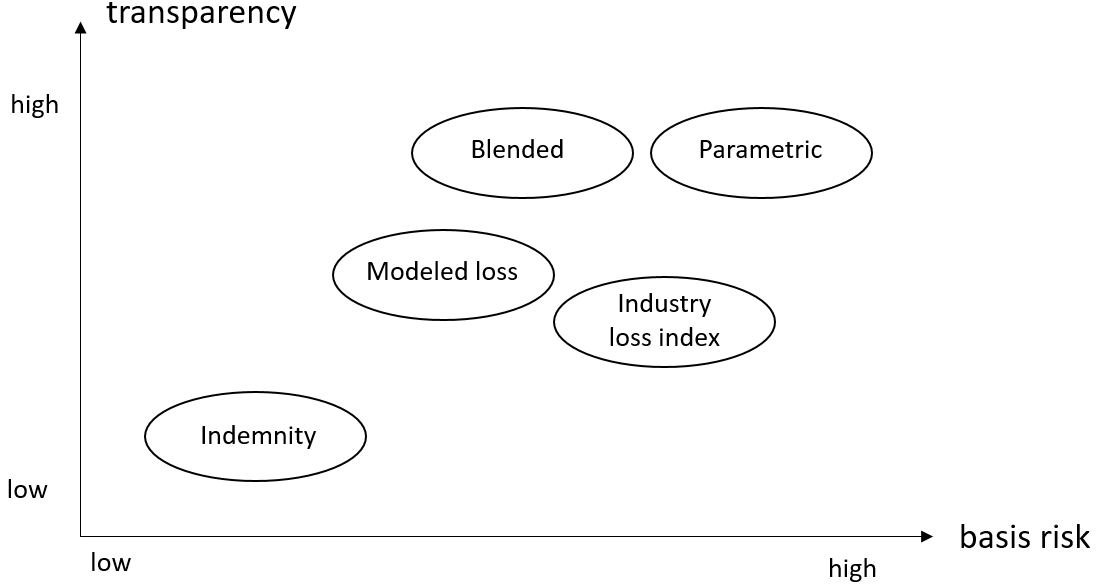}
\caption{Transparency-basis risk trade-off for common trigger types and for our blended trigger.}\label{fig:triggers}
\end{center}
\end{figure}

In this work, we propose a new blended trigger which is more complex than parametric products currently available in the market. Such triggers are typically based on physical characteristics of the event underlying the risk such as magnitude and depth for earthquakes, storm surge height for coastal windstorms, or minimal central pressure for hurricanes. Our approach allows for a more flexible modeling using a cost field which incorporates exposure and vulnerability, see Section \ref{sec:contrib}, while retaining the high degree of transparency typical of parametric triggers. It additionally offers the possibility to account for industry exposure, making the trigger closer to modeled loss triggers in this case, albeit without resorting to complex physical cat models from modeling agencies. 
We illustrate the positioning of our proposal for a blended trigger relative to common trigger types in terms of transparency and basis risk in Figure \ref{fig:triggers}. We discuss the various trigger types found in the cat bond market in Section \ref{sec:overview}, below.

Our approach builds on hazard data compiled by trusted and reputed institutions such as governmental weather agencies and which are available to both the investors and the sponsor. The availability of industry loss data further allows us to calibrate the trigger to observed losses. We believe this methodology is beneficial for combinations of regions and perils where cat models are either unavailable or cannot be trusted, and where insurers do not have any historical loss data at their disposal; otherwise more standard techniques such as frequency-severity loss models or vendor cat models can be used. For regions and perils where cat models are available, our model provides an additional view of risk based on a series of assumptions that are disclosed to both parties to the transaction, which is not the case for vendor models that need to be licensed.
Even though our trigger is also derived from a frequency-severity-type model, we stress that the severity component is based on the theory of spatial extremes and not on the more common Pareto distribution used in actuarial modeling. 

Underlying our loss model, we define a cost field which takes into account the hazard, vulnerability and exposure, as is typically done in cat modeling.
The cost field is itself derived from a max-stable random field, which is commonly used in the modeling of spatial extremes but has so far not been applied in the context of cat bonds.
Furthermore, we propose a fully transparent pricing formula in a contingent claims setting, allowing both parties in the transaction to compute the price from the loss model. 

We refer the reader to \cite{cummins} and \cite{ils} for a general overview of the cat bond market. 
Parametric trigger design with a discussion of basis risk is explored in \cite{goda}, \cite{pucciano} and \cite{goda19} with applications to Turkey earthquake bonds, and in \cite{franco} and \cite{calvet} for Costa Rica earthquake bonds. However, in each case, events are generated from complex cat models and not from models fitted to publicly available data. In more recent works, \cite{tanzhang} design weather index insurance products for US corn production using penalized splines, whereas \cite{bagnarosa} consider a spatio-temporal framework in the context of crop insurance risk modeling. A method using univariate extreme value theory (EVT) is discussed in \cite{zimbidis} and applied to the pricing of cat bonds on Greek earthquakes, however, spatial dependence between locations is not taken into account. In the context of hurricane frequency modeling, \cite{chang2020} propose a regime-switching Poisson process based on climate indices. An option pricing approach is presented in \cite{chang2022} for hurricane bonds with two severity formulations depending on physical characteristics of the hazard, however with no consideration of vulnerability and exposure. Extending the scope of covered risks under cat bond transactions, \cite{xu2021} present a modeling and pricing approach for (non-stationary) data breach risks. 

One of the first attempts to price cat bonds in a contingent claim framework goes back to \cite{cox}, where an equilibrium pricing model is introduced. It assumes that financial and insurance technical (catastrophe) variables are independent. We will use this setting as the basis for pricing our product. The approach proposed in \cite{lee02} relies on a contingent claim model to price default-risky cat bonds (which is less of an issue nowadays due to the SPV investing the proceeds in highly-rated securities) in the presence of moral hazard and basis risk, using a generic compound Poisson loss model with log-normal severities. These two methods consider cat bonds as zero-beta securities, i.e., no natural catastrophe risk premium is assumed, see also \cite{nowak13}. Since CAT bond payouts cannot be replicated by securities available in financial markets, pricing has to performed under an incomplete market framework. Indifference pricing techniques via expected utility are discussed in \cite{egami08} and \cite{zhu11}. The valuation of so-called Wincat coupons from the Winterthur Insurance convertible bonds on an indemnity basis is investigated in \cite{schmock99} with a discussion of model risk, whereas \cite{shao15} consider multiple financial and insurance variables with an application to cat bonds covering California earthquake risk. 

An alternative way to price securities in an incomplete market is by using the Wang transform, see  \cite{wang96}, \cite{wang00}, and \cite{wang02}. In the context of cat bond pricing, this approach has been investigated for mortality bonds in \cite{lin05}, \cite{lin08} and \cite{chen09}, for longevity bonds in \cite{denuit07} and \cite{chen10}, and for earthquake bonds in \cite{tang_yuan_2019}. A comparison of various premium calculation methods is undertaken in \cite{galeotti} and it is shown that the Wang transform performs best in terms of prediction accuracy. An arbitrage approach built on risk indices that obey jump-diffusion processes is presented in \cite{vaugirard03a}. The issue of non-tradeability of catastrophe risk in the framework of no-arbitrage pricing theory is tackled in \cite{gatzert19} using index-linked products.

Another line of research focuses on econometric models for pricing cat bonds, see, e.g., \cite{lane_2000}, \cite{bodoff}, \cite{papachristou} and \cite{gurtler}. Such models regress historical cat bond spreads on observable variables such as the expected loss, see \cite{braun16} where the whole universe of bonds issued between 1997 and 2012 is considered.
An approach based on random forests to predict spreads in the primary cat bond market has been introduced in \cite{makariou} and builds on a comparative analysis between various machine learning methods in \cite{gotze}. In our setup, we consider the spread as exogenously given and playing the role of an input variable to derive the cat bond price.

In the cat bond market, several transactions have been structured using a parametric trigger built on transformations of a variable characterizing the underlying risk, for example, wind speeds in the case of windstorm risk. However, these constructs are confidential and only very limited information is available in the public domain. The trigger we propose accounts for both vulnerability and exposure and as such, it aims at achieving lower basis risk while keeping the transparency inherent to parametric triggers. We are not aware of any academic publication that examines such a trigger within a detailed concrete application. Moreover, to the best of our knowledge, this is the first attempt to price cat bonds using max-stable models. This allows us to consider several locations and the spatial dependency between them, as opposed to univariate methods such as \cite{zimbidis}.
One can them embed the loss model into a pricing framework inspired by the approach developed in \cite{cox}, by assuming independence between financial and insurance technical variables. We price our cat bond under a risk-neutral measure implied by the Wang transform, see \cite{galeotti} and \cite{tang_yuan_2019}. 
Using this framework, investors are able to challenge each modeling assumption and to independently compute the bond price. Moreover, in case of an event, the trigger can be evaluated using publicly available data.
We investigate basis risk under several trigger specifications and perform a full model assessment by validating the hazard model using wind speed data, the vulnerability model using industry insured loss data per event as well as exposure data. Moreover, we specify a frequency distribution describing the number of events per year.

The remainder of the paper is structured as follows. In Section \ref{sec:general}, we provide a general description of the cat bond asset class, present a pricing formula, and lay out the necessary mathematical building blocks for our framework. Section \ref{sec:contrib} is dedicated to our main contributions: the novel trigger and the pricing methodology based on our loss model. Finally, we present a real case study based on German windstorm events to support our findings, accompanied with numerical results in Section \ref{sec:case study}. 

\section{Cat bonds and max-stable random fields}\label{sec:general}
In this section, we first provide a general overview of cat bonds and discuss how the pricing is performed, both in the literature and in practice. This opens the door to our main contributions presented in Section \ref{sec:contrib}. Then, we briefly review max-stable random fields, as we will use such fields to model the weather variable underlying the cat bond in our case study in Section \ref{sec:case study}. 

\subsection{Overview of cat bonds}\label{sec:overview}
Cat bonds play a major role in the alternative risk transfer (ART) market and, in particular, within the insurance-linked securities (ILS) asset class, see \cite{cummins}. Investors' appetite has concentrated in this segment of the market as a higher level of transparency as well as liquidity is present in the cat bond market compared to more private and bespoke ILS structures. As such, they complement the reinsurance market by providing additional risk-bearing capacity. We refer to \cite{ils} for a more complete introduction to the ART market and other standard financing tools such as industry loss warranties (ILW), reinsurance sidecars or collateralized reinsurance.

\begin{figure}[ht]
{\centering
\begin{scaletikzpicturetowidth}{\textwidth}
\begin{tikzpicture}[scale=\tikzscale]
\tikzstyle{arrow} = [thick,->,>=stealth]
\draw (0,0) rectangle (1.6,0.7);
\draw (3.2,0) rectangle (4.8,0.7);
\draw (6.4,0) rectangle (8,0.7);
\draw (3,-1.5) rectangle (5,-1);
\draw [arrow] (1.6,0.35) -- (3.2,0.35);
\draw [arrow] (6.4,0.35) -- (4.8,0.35);
\draw [arrow] (4,0) -- (4,-1);
\draw [arrow] (6.5,-0.3) -- (7,-0.3);
\node at (0.8,0.35) {Sponsor};
\node at (4,0.45) {Special purpose};
\node at (4,0.25) {vehicle (SPV)};
\node at (7.2,0.35) {Investors};
\node at (4,-1.25) {Collateral account};
\node at (2.4,0.5) {Premium};
\node at (5.6,0.5) {Investment};
\node at (5.6,0.2) {(Principal)};
\node at (5.4, -0.55) {Investment of funds received};
\node at (7.6, -0.3) {= cash flow};
\end{tikzpicture}
\end{scaletikzpicturetowidth}
\caption{Structure of an ILS transaction.}\label{fig:ILScreation} }
\end{figure}
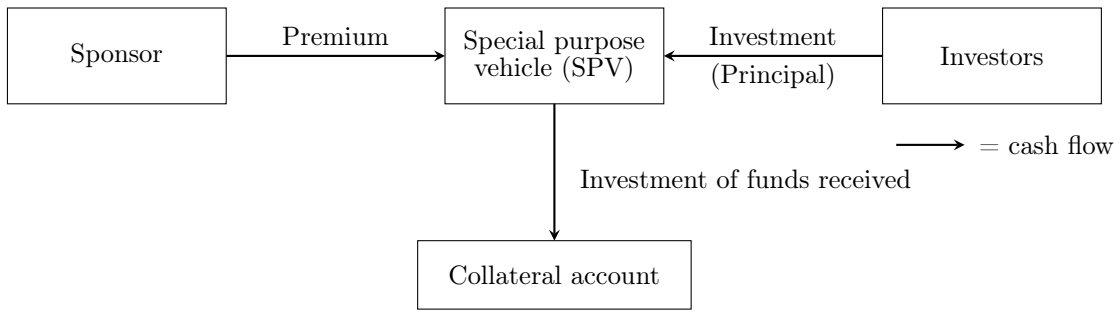

We illustrate the structure of a cat bond in Figure \ref{fig:ILScreation}.
The so-called special purpose vehicle (SPV) issues the cat bond notes to investors and invests the funds received in highly-rated securities, e.g., U.S. money market funds, making the risk transfer fully collateralized and highly reducing the counterparty default risk. In exchange for bearing the catastrophe risk, investors receive on top of the (floating) risk-free rate a fixed premium or risk spread from the sponsor, which is usually an insurer, a reinsurer or a state agency such as the California Earthquake Authority (CEA). Investors in cat bonds typically range among dedicated ILS funds, hedge funds, pension funds, and (re)insurers.

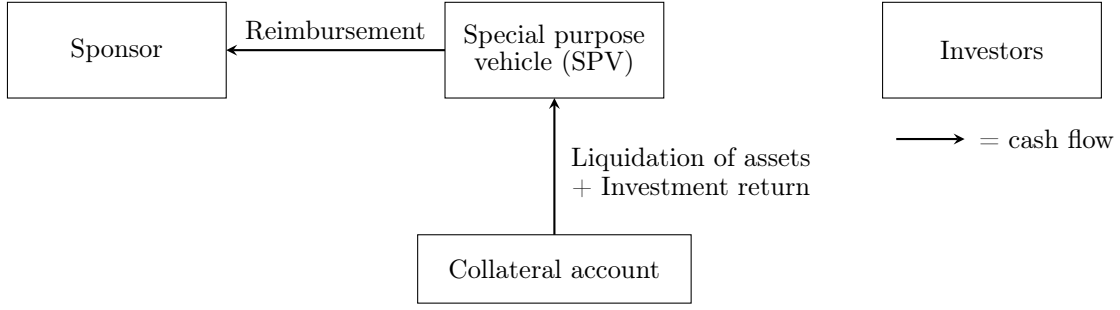
\begin{figure}[ht!]
{\centering
\begin{scaletikzpicturetowidth}{\textwidth}
\begin{tikzpicture}[scale=\tikzscale]
\tikzstyle{arrow} = [thick,->,>=stealth]
\draw (0,0) rectangle (1.6,0.7);
\draw (3.2,0) rectangle (4.8,0.7);
\draw (6.4,0) rectangle (8,0.7);
\draw (3,-1.5) rectangle (5,-1);
\draw [arrow] (3.2,0.35) -- (1.6,0.35);
\draw [arrow] (4,-1) -- (4,-0);
\draw [arrow] (6.5,-0.3) -- (7,-0.3);
\node at (0.8,0.35) {Sponsor};
\node at (4,0.45) {Special purpose};
\node at (4,0.25) {vehicle (SPV)};
\node at (7.2,0.35) {Investors};
\node at (4,-1.25) {Collateral account};
\node at (2.4,0.5) {Reimbursement};
\node at (5, -0.45) {Liquidation of assets};
\node at (5, -0.65) {+ Investment return};
\node at (7.6, -0.3) {= cash flow};
\end{tikzpicture}
\end{scaletikzpicturetowidth}
\caption{Cashflows in the presence of a triggering event (full exhaustion).}\label{fig:ILSevent} }
\end{figure}

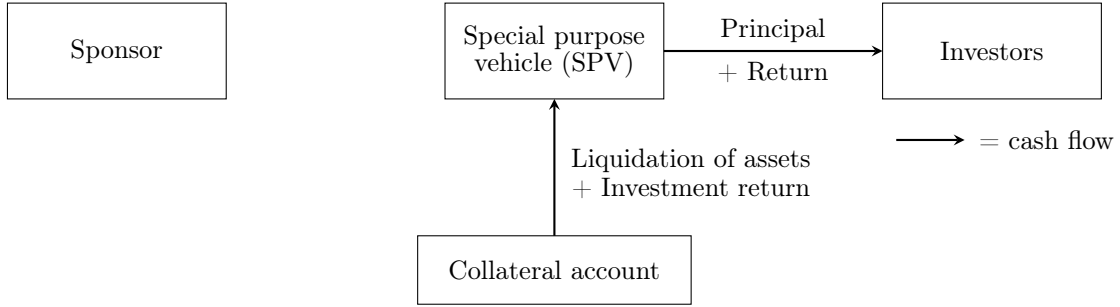
\begin{figure}[ht!]
{\centering
\begin{scaletikzpicturetowidth}{\textwidth}
\begin{tikzpicture}[scale=\tikzscale]
\tikzstyle{arrow} = [thick,->,>=stealth]
\draw (0,0) rectangle (1.6,0.7);
\draw (3.2,0) rectangle (4.8,0.7);
\draw (6.4,0) rectangle (8,0.7);
\draw (3,-1.5) rectangle (5,-1);
\draw [arrow] (4.8,0.35) -- (6.4,0.35);
\draw [arrow] (4,-1) -- (4,-0);
\draw [arrow] (6.5,-0.3) -- (7,-0.3);
\node at (0.8,0.35) {Sponsor};
\node at (4,0.45) {Special purpose};
\node at (4,0.25) {vehicle (SPV)};
\node at (7.2,0.35) {Investors};
\node at (4,-1.25) {Collateral account};
\node at (5.6,0.5) {Principal};
\node at (5.6,0.2) {+ Return};
\node at (5, -0.45) {Liquidation of assets};
\node at (5, -0.65) {+ Investment return};
\node at (7.6, -0.3) {= cash flow};
\end{tikzpicture}
\end{scaletikzpicturetowidth}
\caption{Cashflows in the absence of a triggering event.}\label{fig:ILSnoevent} }
\end{figure}

A cat bond pays off contingent upon a suitably predefined catastrophe event occurring during the risk period, such as a North American hurricane, a European windstorm or a Japanese earthquake. The bond is triggered in case some (proxy) quantity related to the loss and depending on the type of trigger, see the discussion below, exceeds some threshold called the attachment point.
In some instances a second (higher) threshold is specified, called the exhaustion point, which is the level at which investors would face a full investment loss. The payout structure can then take several forms, e.g., step-wise, linear or binary. In case a triggering event occurs during the lifetime of the contract, the proceeds from the collateral account are used to cover the sponsor's losses. In the event of a full loss the principal is not returned to the investors, see Figure \ref{fig:ILSevent}. In case of a partial loss, i.e., a loss between the attachment and the exhaustion point, only the impacted part of the principal is disbursed to the sponsor.
If there are no losses over the lifetime of the contract that are above the attachment point, investors receive the entire principal alongside the earned investment return, see Figure \ref{fig:ILSnoevent}.

Cat bonds have a proven track record of delivering higher returns\footnote{This can however be questionable since in some instances, the risk is not fully understood and what is observed as excess return may just be a margin to compensate for uncertainty. This has been the case for some recent wildfire transactions.} than other asset classes with a similar level of risk and of bringing diversification to portfolios due to their low correlation with other instruments, see, e.g., \cite{swissre11}. Given their multi-year duration, generally up to 3-4 years, insurers and reinsurers holding these bonds can achieve a partial decoupling from the pricing cycle in the traditional reinsurance market, reducing earnings volatility.
In addition, booking technical provisions for ILS allows a sponsor to reduce its minimum capital requirement.
Moreover, unlike reinsurance contracts, cat bonds are tradeable over-the-counter in secondary markets. Finally, they allow accessing the much larger capital markets and provide much needed capacity during hard stages of the reinsurance cycle, i.e., when reinsurance is more expensive than average.
On the downside, the costs associated with structuring and modeling can be quite high but they should always be put into perspective with brokerage fees and commissions in reinsurance. In addition, in terms of capacity, investor capital is not permanent as other asset classes with higher yields might be more attractive.

In the case of a cat bond, there is usually a modeling agency (most often among RMS, Verisk or CoreLogic) providing an independent modeling assessment of the underlying risk and reward structure of the bond: it computes the attachment probability (probability that the loss reaches the attachment point), the exhaustion probability (probability of a full loss) and the expected loss. On top of the floating rate generated from the risk-free collateral account, a risk spread is paid to investors.
The so-called multiple is the ratio of this risk spread over the modeled expected loss.

A cat bond can be structured to provide per-occurrence cover, meaning per-event exposure, or aggregate cover, that is, exposure to multiple events over the course of each annual risk period. 
Let us note that other types of risks can be covered through ILS structures, for instance, mortality, longevity, mortgage or casualty risks. In the sequel we focus on natural catastrophe risks and specifically on aggregate covers.

Cat bonds come with various triggering schemes, determining whether the entire or part of the principal is used to cover losses to the sponsor following an event \citep[see, e.g.,][]{ils}:
\begin{itemize}
    \item Industry loss index: payout based on the total loss to the insurance industry, for example, using Property Claims Services (PCS) or PERILS data.
    \item Pure parametric index: payout depends on hazard characteristics such as earthquake magnitude, storm central pressure or wind speed.
    \item Modeled loss: recovery depends on loss modeled by agencies (RMS, Verisk, CoreLogic) using the event characteristics, exposure data and the vulnerability module of the cat model. 
    \item Indemnity: recovery is entirely based on the loss experienced by the sponsor.
\end{itemize}
Parametric index triggers constitute a refinement of pure parametric triggers. They allow the trigger to be more flexible by introducing additional variables such as the radius of the storm or the geographical coordinates of the insured risk, \citep[see][]{swissre11}. They are equally transparent and additionally reduce basis risk, that is, the risk of a mismatch between the cat bond payout and the losses to the sponsor. Indemnity remains the most common trigger type in the market, followed by industry loss index \citep[see][]{naic}.
On the other hand, moral hazard could arise from the lack of transparency of the bond. Indeed, with an indemnity trigger, although there are contractual limitation factors to control exposure growth and yearly resets to adjust the attachment point and the expected loss \citep[see][]{ils}, an insurer may submit an inaccurate loss report or could decide to underwrite more contracts in regions covered by the bond. 
Sponsors and investors thus face a trade-off between transparency and basis risk.
Investors will generally favor the transparency and reduction in moral hazard offered by parametric covers. The latter additionally provide a faster settlement due to contract terms and the simplicity to determine whether the threshold has been met or not. The sponsor may also have an incentive for risk transfer products to pay out very fast. This is the case for World Bank transactions where funds need to be disbursed as quick as possible to rebuild in countries that get hit by a natural disaster.
On the other side of the spectrum, sponsors will tend to prefer indemnity triggers as these minimize basis risk. 
Finally, modeled loss triggers partially offset the moral hazard associated with indemnity triggers, while displaying a level of basis risk which is lower than for parametric triggers. However, the industry models, albeit constantly progressing in terms of data capture, accuracy and coverage, might still exhibit some model risk. Moreover, they are based on stochastic event catalogs which are difficult to challenge due to intellectual property and their lack of transparency.

A new type of product, known as hybrid cat bonds, is proposed in \cite{barrieu} to protect investors from a stock market crash on top of the exposure to catastrophe risk. 
Hybrid cat bonds can also be understood as bonds where the trigger is a mixture between different types of triggers, such as parametric and industry loss index. 
Various clustering algorithms are tested in terms of their classification accuracy for a binary trigger in \cite{calvet}, with a case study based on a catalog of stochastic earthquake events in Costa Rica. The results are benchmarked with the optimal trigger obtained from a genetic algorithm presented in \cite{franco}. In both cases, the trigger is parametric; however, it is elaborated on synthetic data and not on observations. This could raise the issue of complexity with regards to investors.
A distinction in the literature is set between so-called first generation or "cat-in-a-box" triggers \citep[see][]{franco} and second-generation triggers \citep[see][]{goda}, where in addition the reduction in basis risk is highlighted when using step-wise payouts instead of binary payouts.
First-generation triggers consider the physical characteristics of the covered event (e.g., for an earthquake, the depth, magnitude and coordinates of the epicenter). The attachment can be made dependent on the event magnitude and the location by defining a polygonal grid over the covered region. In this way, varying magnitudes within each polygon need to be attained to trigger a payout. 
In turn, second generation triggers are built from spatially distributed features, e.g., stations reporting wind speed or spectral acceleration, with weights that can be selected at each station in order to reflect the sponsor's exposure in the region, see, for example, the Bosphorus 2015 transaction covering earthquakes in Turkey\footnote{\url{https://www.artemis.bm/deal-directory/bosphorus-ltd-series-2015-1/}}.
Building on this work, \cite{pucciano} and \cite{goda19} study the interplay between the required complexity of the trigger formulation and the distribution of the reporting stations in terms of their vicinity with respect to insured assets.

Arguably, the most complex parametric trigger was the Compass Re II bond issued on behalf of AIG in 2015 for covering hurricanes in the US\footnote{\url{https://www.artemis.bm/deal-directory/compass-re-ii-ltd-series-2015-1/}}. The trigger was a functional of the latitude and longitude at landfall, the radius of the storm and the maximum sustained wind speed. The inclusion of the landfall coordinates allowed the sponsor to calibrate the trigger to be more responsive to areas with peak exposure. We could not find any successor transaction of this Compass Re deal. Among European windstorm transactions, Pylon II Capital was an innovative cover for protecting French utility Électricité de France\footnote{\url{https://www.artemis.bm/deal-directory/pylon-ii-capital-ltd/}}. 
Blue Fin Re was another parametric bond covering European windstorm and was issued in 2007 on behalf of Allianz\footnote{\url{https://www.artemis.bm/deal-directory/blue-fin-ltd/}}. There is, however, no publicly available information on the exact formulation of the trigger. 

Parametric triggers for outstanding cat bonds have been used for certain World Bank sponsored transactions (first generation trigger for Mexico earthquake and hurricane, and for Jamaica hurricane), and public authorities such as the Metropolitan Transportation Agency in New York, with the 2020 MetroCat bond covering storm surge and earthquake using a second-generation trigger with a binary payout\footnote{\url{https://www.artemis.bm/deal-directory/metrocat-re-ltd-series-2020-1/}}.

This shows that some interesting triggering mechanisms have been elaborated in the industry but remain confidential, and in the case where information is available in the public domain, for example, on the website \url{https://www.artemis.bm}, the trigger structure is not presented in detail. We attempt to address this gap with a parametric trigger for which we lay out all modeling assumptions and which can be compared in terms of basis risk to trigger formulations which have been used in practice.

\subsection{Spread and bond price}\label{sec:price}
As opposed to econometric models which regress historical bond spreads on observable variables such as the expected loss, 
covered territory, sponsor, spreads on corporate bonds with comparable ratings, and latent variables such as the stage of the reinsurance cycle \citep[see][]{braun16} in the present approach we consider the spread as exogenously given and playing the role of an input variable in the formulation of the bond price.
Under this scenario, cat bonds are priced in an incomplete market framework. Since the underlying process (natural catastrophe occurrence) triggering the event is not a tradeable security, it is not possible to construct a replicating portfolio of assets to match the payoff of the bond.
 
Suppose a sponsor wishes to issue a cat bond with maturity $T>0$ years. We discretize the time interval in yearly units, and we say that a catastrophe occurs in year $n$ if it takes place in the time interval $(n-1,n],$ for $n=1,\ldots,T.$ We denote the attachment point of the bond as $u_a$ and the exhaustion point by $u_e>u_a,$ these two quantities being contractually determined before inception. The investors receive in exchange for the risk they bear a coupon, which consists of a fixed spread $s>0$ given in percentage per annum of the face value $FV$, on top of a floating short rate (which is usually the 3-month US money market funds rate).

Assume all random variables defined in the sequel live in the filtered probability space $(\Omega,\mathcal{F},\mathbb{F},\QQ),$ for a filtration $\mathbb{F}=\left(\mathcal{F}_n\right)_{n=1,\ldots,T}.$ As in \cite{cox}, under our risk-neutral pricing measure $\QQ$, we assume that financial market (short rate) variables and insurance technical (catastrophe) variables are independent, which leads to a decoupling of $\mathbb{F}$ under $\QQ$ into
\begin{align*}
\mathbb{F}^{ins}&=\left(\mathcal{F}^{ins}_n\right)_{n=1,\ldots,T}\quad \textnormal{filtration of insurance technical events},\\
\mathbb{F}^{fin}&=\left(\mathcal{F}^{fin}_n\right)_{n=1,\ldots,T}\quad\textnormal{filtration of financial events},
\end{align*}
where for all $n=1,\ldots,T,$ it is assumed that $\mathcal{F}_n=\sigma\left(\mathcal{F}^{ins}_n,\mathcal{F}^{fin}_n\right)$ is the smallest $\sigma$-field containing all events of $\mathcal{F}^{ins}_n$ and $\mathcal{F}^{fin}_n,$ see \cite{mcav}. We assume that the filtrations $\mathbb{F}^{ins}$ and $\mathbb{F}^{fin}$ are independent under $\QQ.$ 

We consider at this stage a general model for the loss process $(S_n)_{n\geq 1}$ with finite time horizon $T>0$, where $S_n$ is the loss in year $n$, for $n=1,\ldots,T.$ We assume that the components $S_n$ are independent and identically distributed under $\QQ,$ and that the loss process $(S_n)_{n\geq 1}$ is 
$\mathbb{F}^{ins}$-adapted.
This general model could be either a cat model or a standard frequency-severity model. 
In Section \ref{sec:contrib}, we will present a loss model that is tailored to our problem. For the purpose of expressing the stream of cash flows, we define the stopping time 
$\tau^*:\Omega \to \mathbb{N}$ as
\begin{equation*}
\tau^*=\inf \{n \in \{1,\ldots,T\};\quad S_n\geq u_a \},
\end{equation*}
where we follow the convention $\tau^*=\infty$ if $S_n< u_a$ for all
$n\in\{1,\ldots,T\}.$

Losses below the attachment point $u_a$ are fully assumed by the sponsor, whereas a loss above the exhaustion point $u_e$ leads to a full loss of the principal for investors. Losses between the attachment and the exhaustion points trigger the bond and entail only a partial repayment of principal to investors.
In case a triggering event occurs before maturity, investors receive the amount $\left(1-p_{\tau^*}\right)FV \mathds{1}_{\{\tau^*\leq T\}}$,
where $p_{\tau^*}$ corresponds to the proportion of the face value $FV>0$ which is lost at time $\tau^*.$ Typically, a piecewise linear form in the loss is specified, that is, for $n\in\{1,\ldots,T\},$
\begin{equation*}
p_n=\left(\frac{(S_n-u_a)\vee 0}{u_e-u_a}\right)\wedge 1.
\end{equation*}
In the limiting case where $u_a=u_e,$ we obtain a binary loss, that is, a full loss to the investors when the bond is triggered and a full recovery of principal otherwise. Hence, this formulation also covers cat bond deals without exhaustion points.
Let us note that other formulations for the payoff can be equally considered, such as a stepwise constant structure which is often found in parametric covers, where the level depends on some physical characteristic of the peril, e.g., the earthquake magnitude, as in \cite{zimbidis}.
Assume that coupons are paid at the end of each year, contingent on no loss above the attachment point. 
The total cashflow is obtained by taking the sum of the principal redemption and the coupons. The former can be written at time $n$ as
\begin{equation*}
P_n=\begin{cases}
\left(1-p_n\right)FV \mathds{1}_{\{\tau^*=n\}},  &\mbox{if } n < T, \\
\left(1-p_T\right)FV \mathds{1}_{\{\tau^*=T\}}+ FV\mathds{1}_{\{\tau^*>T\}},  &\mbox{if } n = T,
\end{cases}
\end{equation*}
where for all $n=1,\ldots,T,$ $P_n$ is $\mathcal{F}^{ins}_n$-measurable since the loss process $(S_n)_{n\geq 1}$ is $\mathbb{F}^{ins}$-adapted.
In other words, we assume that the residual principal is paid at the end of the year in which the loss occurred.
We will assume Vasicek dynamics for the short rate, although any other stochastic model that is analytically tractable can be used without having to rely on simulation. 
The short rate process denoted by $r=(r_t)_{t\geq 0}$ is thus governed by
\begin{equation*}
\mathrm{d} r_t = \alpha(\mu-r_t)\mathrm{d}t + \sigma \mathrm{d} W_t^\QQ,
\end{equation*}
with initial value $r_0>0,$ where $\alpha>0,$ $\mu>0$ and $\sigma>0$ denote the speed of mean reversion, the level of mean reversion and the volatility, respectively, and where $W^\QQ$ is a Brownian motion under $\QQ$ that is $\mathbb{F}^{fin}$-adapted.
To derive the distribution of the loss variable under $\QQ,$ we introduce the Wang transform, see \cite{wang96,wang00,wang02,wang04}
\begin{equation}\label{eq:wang}
g(q)=\Phi\left(\Phi^{-1}(q)-\lambda\right),   
\end{equation}
where $g:[0,1]\to[0,1]$ is a distortion function which is non-decreasing and right-continuous, with $g(0)=0,$ $g(1)=1,$ and such that $g(q)\leq q$ for $q\in[0,1],$ $\lambda>0$ is a skewness parameter and $\Phi$ is the standard normal distribution function. The parameter $\lambda$ can be calibrated so as to ensure the bond is valued at par, see \cite{tang_yuan_2019}, and can be interpreted as the catastrophe risk premium as it reflects the extent to which the distribution of the catastrophe risk variable needs to be skewed towards the right for risk-neutral pricing purposes.
Assume that the loss random variables are also i.i.d. under $\PP$ with $S_1,\ldots,S_T\sim F_S.$ We define the (distorted) distribution of $S$ under $\QQ$ as
$F_S^\QQ(x)=g\circ F_S(x)=g(F_S(x)).$ 

The coupon at time $n$ is given by
$(r_n + s)FV \mathds{1}_{\{\tau^*>n\}},$ and this quantity is $\mathcal{F}_n$-measurable for all $n=1,\ldots,T.$    
This leads to the following expression for the price of the cat bond at $t=0$:
\begin{equation}\label{eq:priceT}
V_0=\sum_{n=1}^T \E_{\QQ}\left[B_n^{-1}\left(\left(r_n+s\right)\mathds{1}_{\{\tau^*>n\}}+\left(1-p_n\right)\mathds{1}_{\{\tau^*=n\}}\right)FV\right]+
\E_{\QQ}\left[B_T^{-1}\mathds{1}_{\{\tau^*>T\}}FV\right],
\end{equation}
where $B_n=\exp\left\{\int_0^n r_u du\right\}$ denotes the value of the bank account at time $n=1,\ldots,T.$ Since $(r_t)_{t\geq 0}$ is an affine process, one can write, see, e.g., \cite{brigo},
\begin{equation}\label{eq:bondpricevasicek}
\E_{\QQ}\left[e^{-\int_0^n r_u du}\right]
=e^{A_1(0,n)-A_2(0,n)r_0},
\end{equation}
where the exact forms of the functions $A_1$ and $A_2$ are given in \cite{brigo}.
In order to simplify the computation of the cat bond price, we consider the $n$-forward probability measure $\QQ^n$ which is equivalent to $\QQ$ and is defined by the Radon-Nikodym derivative
\begin{equation}\label{eq:measurechange}
\frac{d\QQ^n}{d\QQ}\biggr\lvert_{\mathcal{F}_n}= \frac{\exp\left(- \int_0^n r_u du\right)}{\E_{\QQ}\left[\exp\left(- \int_0^n r_u du\right)\right]}.
\end{equation}
We note that under this measure change, the independence between insurance technical and financial variables is preserved.
The dynamics of the short rate process under $\QQ^n$ for 
$0\leq t\leq n$ are then given by
\begin{equation*}
dr_t=\left(\alpha(\mu-r_t)-\sigma^2 A_2(t,n)\right)dt + \sigma dW_t^{\QQ,n},
\end{equation*}
where $W^{\QQ,n}$ is a $\QQ^n$-Brownian motion that is $\mathbb{F}^{fin}$-adapted and satisfies
\[
dW_t^{\QQ,n}=dW_t^\QQ+\sigma A_2(t,n)dW_t^\QQ,
\]
for $0\leq t\leq n,$ see \cite{brigo}.
Solving this stochastic differential equation, we obtain the mean of the Vasicek process at time $n$ under $\QQ^n$:
\begin{equation}\label{eq:vasicekmean}
\E_{\QQ^n}[r_n]=r_0 e^{-\alpha n}+\left(\mu-\frac{\sigma^2}{\alpha^2}\right)\left(1-e^{-\alpha n}\right)+\frac{\sigma^2}{2\alpha^2}\left(e^{-\alpha n}-e^{-2\alpha n}\right).
\end{equation}
The independence between financial and insurance technical variables allows to decouple the expression in
(\ref{eq:priceT}) as follows
\begin{multline}\label{eq:priceTdev}
V_0=\sum_{n=1}^T\E_{\QQ}\left[B_n^{-1}\left(r_n+s\right)\right]\QQ[\tau^*>n]FV\\
+\sum_{n=1}^T \E_{\QQ}\left[B_n^{-1}\right]\E_{\QQ}\left[\left(1-p_n\right)\mathds{1}_{\{\tau^*=n\}}\right]FV +\E_{\QQ}\left[B_T^{-1}\right]\QQ[\tau^*>T]FV.
\end{multline}
We can then compute $\E_{\QQ}\left[B_n^{-1}\left(r_n+s\right)\right]$ using the $n$-forward measure defined in (\ref{eq:measurechange}), and (\ref{eq:vasicekmean}).
We derive the expectations $\E_{\QQ}\left[B_n^{-1}\right]$ and $\E_{\QQ}\left[B_T^{-1}\right]$ using (\ref{eq:bondpricevasicek}).

\begin{figure}[!ht]
    \centering
    \begin{subfigure}[b]{0.48\textwidth}
        \centering
       \includegraphics[width=\textwidth]{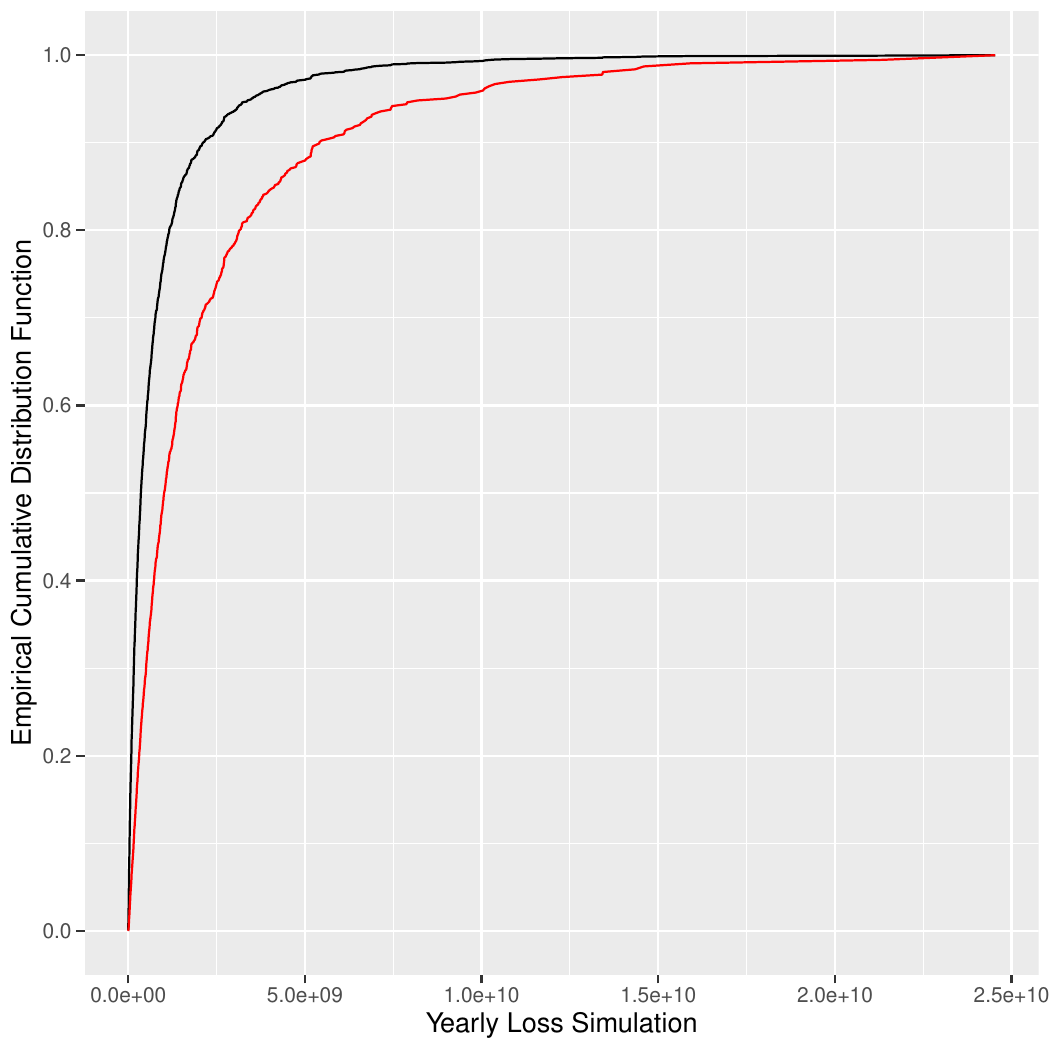}
        \end{subfigure}
        \hfill
        \begin{subfigure}[b]{0.48\textwidth}  
            \centering 
            \includegraphics[width=\textwidth]{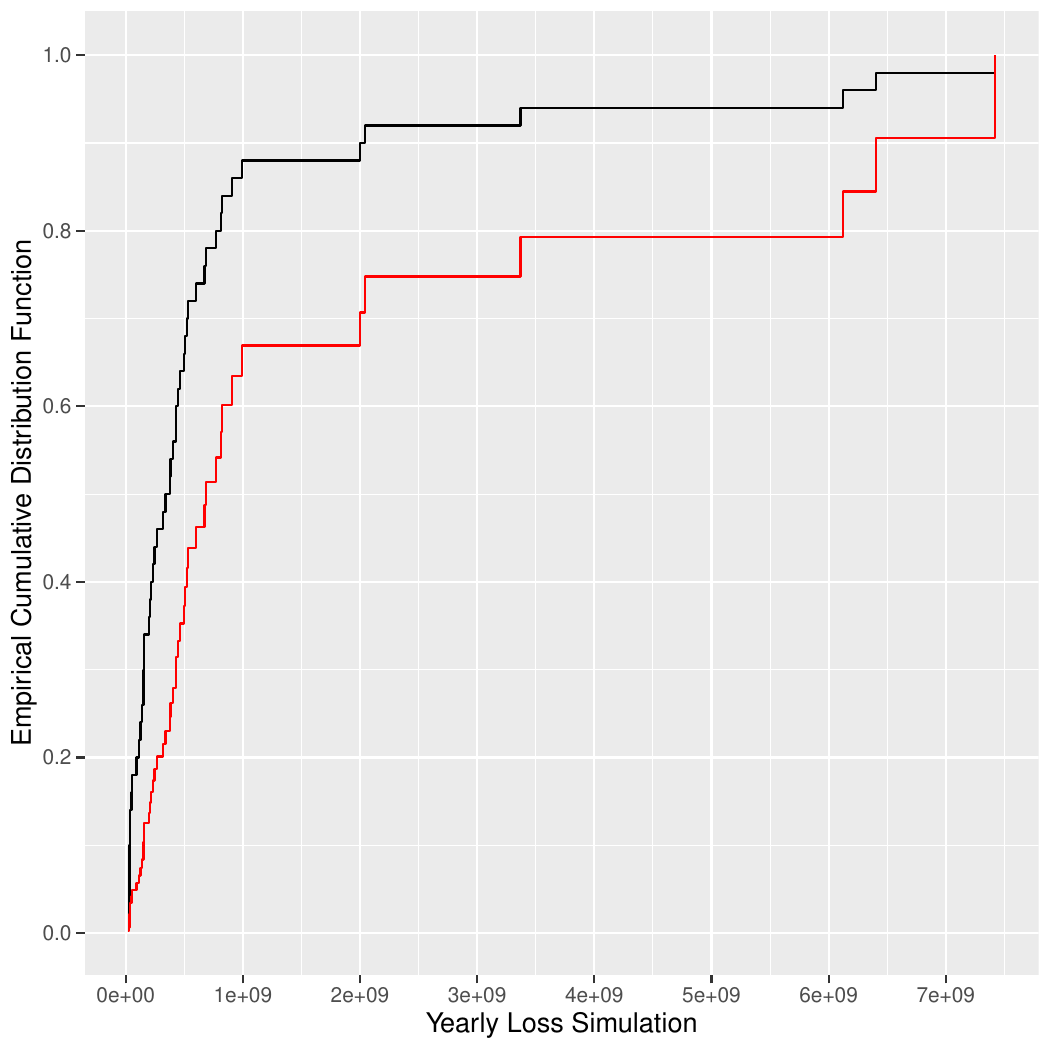}
        \end{subfigure}
        \caption{Empirical cumulative distribution function (black) of the yearly loss $S$ and Wang-transformed counterpart (red) for a sample of 2000 (left) and 50 simulations (right), respectively.}
\label{fig:emp_cdf}
\end{figure}

Under the assumption of independent components $S_n$ of the loss process, the stopping time decouples similar to a geometric distribution, i.e.,
\begin{equation*}
\QQ[\tau^*>n]=\QQ[S_1<u_a,\ldots,S_n<u_a]=\prod_{u=1}^n\QQ[S_u<u_a].
\end{equation*}
Under the latter assumption, we obtain analogously
\begin{equation*}
\E_{\QQ}\left[\left(1-p_n\right)\mathds{1}_{\{\tau^*=n\}}\right]=
\prod_{u=1}^{n-1} \QQ[S_u<u_a]\E_{\QQ}\left[(1-p_n)\mathds{1}_{\{S_n\geq u_a\}}\right].
\end{equation*}
One can consider the case $T=1$ without loss of generality as $S_1,\ldots,S_T\overset{\text{i.i.d.}}{\sim} S.$ In this case, the expression above reads
\begin{equation*}
\E_{\QQ}\left[\left(1-p_1\right)\mathds{1}_{\{\tau^*=1\}}\right]=
\E_{\QQ}\left[(1-p_1)\mathds{1}_{\{S_1\geq u_a\}}\right].
\end{equation*}
In Section \ref{sec:results}, we will evaluate the price at \eqref{eq:priceTdev} from a sample $\{S_j, \mbox{ } j=1,\ldots,J\}$ of $J$ realizations of $S$ based on empirical estimates of the attachment probability $\PP[S\geq u_a]$ and the exhaustion probability $\PP[S\geq u_e].$ 
The empirical distribution $\hat{F}_S^\QQ(x)=g\left(\hat{F}_S(x)\right)$ of $S$ under $\QQ$ is obtained by applying the Wang transform in \eqref{eq:wang} to the empirical distribution $\hat{F}_S$ of $S$ under $\PP$. 
The empirical counterpart of the probability $\QQ[S<u_a]$ is given by $$\hat\QQ[S<u_a]=g\left(\hat\PP[S<u_a]\right)=\Phi\left(\Phi^{-1}\left(\frac{1}{J}\sum_{j=1}^J \mathds{1}_{\{S_j<u_a\}}\right)-\lambda\right).$$ 
By definition of $p_1,$ 
$$(1-p_1)\mathds{1}_{\{S_1\geq u_a\}}=\left(1-\frac{S_1-u_a}{u_e-u_a}\right)\mathds{1}_{\{u_a\leq S_1 < u_e\}}.$$ 
We consider the reordered sample so that
$$S_{(1)}\leq \ldots \leq S_{(J)}.$$
The empirical counterpart of the expectation $\E_{\QQ}\left[(1-p_1)\mathds{1}_{\{S\geq u_a\}}\right]$ is then obtained using the Wang-transformed probability weights 
$g\left(\hat{F}_S\left(S_{(j)}\right)\right)-g\left(\hat{F}_S\left(S_{(j-1)}\right)\right):$
$$\hat{\E}_{\QQ}\left[(1-p_1)\mathds{1}_{\{S\geq u_a\}}\right]=\sum_{j=1}^J \left(g\left(\hat{F}_S\left(S_{(j)}\right)\right)-g\left(\hat{F}_S\left(S_{(j-1)}\right)\right)\right)\cdot\left(1-\frac{S_{(j)}-u_a}{u_e-u_a}\right)  \mathds{1}_{\{u_a\leq S_{(j)}<u_e\}},
$$
where $S_{(0)}=0.$
Note that the probability of attachment and the probability of exhaustion are higher under $\QQ$ due to the distortion introduced by the Wang transform. In Figure \ref{fig:emp_cdf} (left panel), we show how the empirical cumulative distribution of $S$ gets distorted when switching from $\PP$ to $\QQ.$ The right panel in Figure \ref{fig:emp_cdf} is based on a smaller sample of simulations to visualize how the Wang transform distorts the probability weights $1/J$ under $\PP$ to $g\left(\hat{F}_S\left(S_{(j)}\right)\right)-g\left(\hat{F}_S\left(S_{(j-1)}\right)\right)$ under $\QQ$. 

\subsection{Max-stable random fields}\label{sec:max-stable}
Throughout, $\bigvee$ denotes the supremum.
Moreover, $\overset{d}{=}$ and $\overset{d}{\rightarrow}$ stand for equality and convergence  in  distribution,  respectively. In the case of random fields, distribution has to be understood as the set of all finite-dimensional distributions.

A random field $X$ on $\Mbb{R}^d$ is said to be max-stable if there exist continuous functions $a_n(\cdot)$ and $b_n(\cdot)$ on $\Mbb{R}^d$ such that if $X_1, \ldots, X_n$ are independent replications of $X$, then
$$\bigvee_{i=1}^{n} \frac{X_i - b_n}{a_n} \overset{d}{=} X,$$
for all $n\geq 1.$
Moreover, it is called simple if it has standard Fréchet margins, i.e., for all $\bm{x} \in \Mbb{R}^d$, $\Mbb{P}(X(\bm{x}) < x)=\exp \left( -1/x \right), \mbox{ }x>0$.

Max-stable random fields are ideally suited to the modeling of pointwise maxima. 
Let $V_i,\mbox{ } i=1, \dots, n,$ be independent replications of a random field $V$ on $\Mbb{R}^d$ (for instance, the field of wind speed values in the case $d=2$)  and let $c_n(\cdot)>0$ and $d_n(\cdot)$ be sequences of functions. If there exists a non-degenerate random field $G$ on $\Mbb{R}^d$ such that
$$
\bigvee_{i=1}^n \frac{V_i -d_n}{c_n}\overset{d}{\rightarrow} G, \mbox{ for } n \to \infty,
$$
then $G$ is necessarily max-stable \citep[e.g.,][]{haan1984spectral}. More generally, as noted in, e.g., \citet[][Section 2.3]{huser2014space}, max-stable fields provide appropriate models for extremes of individual observations.

We now provide a probabilistic construction  of simple max-stable fields
that can be leveraged to build parametric max-stable models.
Let $(U_i)_{i \geq 1}$ be the points of a Poisson point process on $(0, \infty)$ with intensity function $u^{-2} \mathrm{d}u$ and let $Y_i,\mbox{ }i\geq 1$, be independent replications of a random field $Y$ which are independent of the $(U_i)_{i \geq 1}$ and  such that, for all $\bm{x} \in \mathbb{R}^d$,
$\mathbb{E}[Y(\bm{x})]=1$. Then the field $Z$ defined on $\Mbb{R}^d$ by
\Beq
\label{Eq_Spectral_Representation_Stochastic_Processes}
Z = \bigvee_{i=1}^{\infty} U_i Y_i
\Eeq
is simple max-stable,  see \cite{haan1984spectral}.

This so-called spectral representation has been exploited to define the Brown--Resnick, extremal-$t$, Schlather, and Smith max-stable models that we will use in Section \ref{sec:case study}. Recall that a random field $W$ has stationary\footnote{Throughout the paper, stationarity refers to strict stationarity.} increments if the distribution of the field $\{ W(\bm{x}+\bm{x}_0)-W(\bm{x}_0) \}$, $\bm{x} \in \Mbb{R}^d$, does not depend on the choice of $\bm{x}_0 \in \Mbb{R}^d$. If the increments of $W$ have a finite second moment, the semivariogram of $W$ is
$$ \gamma_W(\bm{x})=\frac{1}{2} \mathrm{Var}(W(\bm{x})-W(\bm{0})), \quad \bm{x} \in \Mbb{R}^d,$$
where $\mathrm{Var}$ stands for the variance. Let $W(\bm{x}),\mbox{ }\bm{x} \in \mathbb{R}^d$, be a centered Gaussian random field with stationary increments and with semivariogram $\gamma_W$. Then the field $Z$ defined by \eqref{Eq_Spectral_Representation_Stochastic_Processes} with 
$$Y(\bm{x})=\exp \left( W(\bm{x})-\frac{\mathrm{Var}(W(\bm{x}))}{2} \right),\quad \bm{x} \in \Mbb{R}^d,$$ 
is referred to as the Brown--Resnick random field associated with the semivariogram $\gamma_W$ \citep{brown,kabluchko}. It is stationary with a distribution fully described by $\gamma_W$ \citep[][Theorem 2 and Proposition 11, respectively]{kabluchko}. A commonly used semivariogram is 
\Beq
\label{Eq_Power_Variogram}
\gamma_W(\bm{x})= \left( \| \bm{x} \|/\kappa \right)^{\psi}, \quad \bm{x} \in \mathbb{R}^d,
\Eeq
where $\|\cdot\|$ denotes the Euclidean norm, and $\kappa>0$ and $0<\psi \leq 2$ are the range and smoothness parameters, respectively. 

Let $\varepsilon(\bm{x})$, $\bm{x} \in \mathbb{R}^d$, be a stationary standard Gaussian random field, let $\Gamma$ denote the gamma function. Then taking in \eqref{Eq_Spectral_Representation_Stochastic_Processes}
$$ Y(\bm{x}) = \frac{\sqrt{\pi} 2^{1 - \nu/2}}{\Gamma([1-\nu]/2)} \max \{ \varepsilon(\bm{x}), 0\}^{\nu}, \quad \bm{x} \in \mathbb{R}^d,$$ 
where $\nu >0$ is the number of degrees of freedom,
yields the extremal-$t$ model and one can verify that $\mathbb{E}[Y(\bm{x})]=1$ for all $\bm{x} \in \mathbb{R}^d$ \citep{opitz2013extremal}. The fact that any correlation function can be used for $\varepsilon$ offers flexibility. In the following, we will consider the Whittle-Matérn, powered exponential and Cauchy functions. 

The Cauchy function is
\begin{equation}
\label{Eq_CauchyCorrFunction}  
\rho(h)=\left( 1 + \left( h/\kappa \right)^2 \right)^{-\psi}, \quad h>0,
\end{equation}
where $h>0$ is the Euclidean distance between two locations, $\kappa >0$ and $\psi>0$ are the range and smoothness parameters, respectively, and it is common practice to add a nugget parameter $a_0\in [0, 1)$ to the correlation function as follows
$$ \rho^*(h) = (1 - a_0) \rho(h).$$
The case $\nu=1$ leads to the so-called Schlather model \citep{schlather2002models}. Under some conditions, the extremal-$t$ random field converges to the Brown--Resnick field as $\nu \to \infty$.

The Whittle--Mat\'ern correlation function is
\begin{equation*}
 \rho(h) = \frac{2^{1-\psi}}{\Gamma(\psi)}\left(\frac{h}{\kappa}\right)^\psi K_\psi\left(\frac{h}{\kappa}\right),
\end{equation*}
for range and smoothness parameters $\kappa>0$ and $\psi>0,$ respectively, and where $K_\psi$ is the modified Bessel function of the third kind with order $\psi.$

The powered exponential correlation function is
\begin{equation*}
\rho(h) =  \exp\left[-\left(\frac{h}{\kappa}\right)^\psi\right],
\end{equation*} for $\kappa>0$ and $0<\psi\leq 2.$


The Smith random field with covariance matrix $\Sigma$ \citep{smith1990max} corresponds to the Brown--Resnick field associated with the variogram $\gamma_W(\bm{x})=\bm{x}' \Sigma^{-1} \bm{x}/2$, $\bm{x} \in \mathbb{R}^d$, where $'$ designates transposition; see, e.g., \cite{huser2013composite}.

A well-known measure of spatial dependence for max-stable fields, which is also frequently used for goodness-of-fit assessments, is the bivariate extremal coefficient function $\theta$ \citep[e.g.,][]{schlather03}. It satisfies, for all $u >0$,
$$\mathbb{P}\left( Z(\bm{x}_1) \leq u, Z(\bm{x}_2) \leq u \right) = \exp ( -\theta(\bm{x}_1, \bm{x}_2)/u ), \quad \bm{x}_1, \bm{x}_2 \in \mathbb{R}^d,$$ where $Z$ is simple max-stable. 

In practice, max-stable fields are not necessarily simple but have generalized extreme-value (GEV) univariate marginal distributions. If $X$ is a max-stable field with location, scale and shape functions $\eta(\bm{x}) \in \mathbb{R}$, $\tau(\bm{x}) >0$ and $\xi(\bm{x}) \in \mathbb{R}$, we can write
\Beq
\label{Eq_Link_Maxstb_Simple_Maxstab}
X(\bm{x}) = 
\left \{
\begin{array}{ll}
\eta(\bm{x})-\tau(\bm{x})/\xi(\bm{x})  + \tau(\bm{x}) Z(\bm{x})^{\xi(\bm{x})}/\xi(\bm{x}), & \mbox{ } \xi(\bm{x}) \neq 0, \\
\eta(\bm{x}) + \tau(\bm{x}) \log Z(\bm{x}), & \mbox{ } \xi(\bm{x}) = 0,
\end{array}
\right.
\Eeq
where $Z$ is simple max-stable, see \cite{koch2022}.

\section{Main contributions}\label{sec:contrib}
In Section \ref{sec:overview}, we have outlined the types of parametric triggers which have been used for previous cat bond issuances, as well as proposals from the academic literature.
In this section, we present our main contributions: a new trigger type which can take a flexible form between parametric and modeled loss, and an approach to compute the price of the cat bond.

Consider a cat bond covering the insured claims for a specific type of natural disaster (e.g., extra-tropical or tropical cyclones) on a given region $A$ of interest and during a one-year period. The covered area can be an entire country, several countries or a particular subregion within a country where the sponsor is seeking coverage.

We consider a discretized region $A$ in the style of second generation parametric triggers, that is, we assume that we have $K$ sites $\bm{x}_1, \ldots, \bm{x}_K \in A\subseteq\mathbb{R}^3$ that are representative of $A$ in terms of insured claims. Each site is characterized by its longitude, latitude and altitude, that is, $\bm{x}_k=(lon_k,lat_k,alt_k)',$ for $k=1,\ldots,K.$
In the ideal scenario, each site $\bm{x}_k$ would correspond to a policy as this would allow to consider the exposure at each insured location and enable a closer calibration of the trigger to actual losses. However, this is not realistic in practice as exposure data is usually available only on an aggregated level, for example, administrative regions. Therefore, sites must be chosen to reflect the exposure in each region. We will discuss various modeling options in Section \ref{sec:windexpo}.

Let $N$ denote the number of events of the considered type occurring during the  one-year period under study. We propose to model the spatial field of insured claims triggered by the $i$-th event, for $i=1, \ldots, N$, using a field $\{C_i(\bm{x}) \}_{\bm{x} \in A}$ that is constructed solely from transformations of publicly available hazard-related physical fields (e.g., maximal wind speed) and, possibly, freely accessible industry exposures. Otherwise, predefined exposures in the neighborhood of each $\bm{x}_k$ can be used. The mentioned transformations are damage functions that map the physical features of the hazard to destruction percentages.

Mimicking the structure of cat models, we approximate the actual cost associated with event $i$ by the cost field
\begin{equation}\label{eq:cost field}
C_i(\bm{x})=D(Z_i(\bm{x})) E(\bm{x}),
\end{equation}
where $D:\mathbb{R}\mapsto [0,1]$ is a damage function, 
$\{Z_i(\bm{x})\}_{\bm{x}\in A}$ corresponds to a functional of the environmental variable characterizing the physical hazard for event $i$ (for example, the field of maximum wind speeds), and $\{E_i(\bm{x})\}_{\bm{x}\in A}$ represents the (deterministic) exposure field; see \cite{koch17}.

We assume that the fields \( Z_i \), for \( i = 1, \ldots, N \), are independent and identically distributed (i.i.d.), and that the damage function \( D \) is identical across events. Under these assumptions, the resulting cost fields \( C_i \) are also i.i.d. The assumption on $D$ is made for simplicity. However, in practice, one would expect that for two successive events occurring close in time and with similar wind speeds, the insured loss from the second event would be lower due to damage already caused by the first.
In the absence of loss data, the damage function is typically chosen based on physical and engineering principles and expert opinion. Otherwise, it can be calibrated empirically; see Section \ref{sec:vulnerability}. 
If exposure data are not available, a constant exposure field may be specified for simplicity.
Otherwise, several approaches can be used to model the exposure component $\{E(\bm{x})\}_{\bm{x}\in A}$ of the cost field. It can take, for instance, a parametric form $E(\bm{x})=F(\bm{x}),$ where $F$ is a function 
of latitude and longitude only. Alternatively, it may depend on a set of socio-economic variables 
\[E(\bm{x})=F\left(n(\bm{x}),w(\bm{x}),p(\bm{x}),o(\bm{x})\right),
\]
where $n(\bm{x}),$ $w(\bm{x}),$ $p(\bm{x})$ and $o(\bm{x})$ are the number of inhabitants, the mean wealth per inhabitant, the penetration rate of insurance and the orographic exposure at site $\bm{x},$ respectively. The latter variable captures the protection effect for a specific site: for example, if buildings at $\bm{x}$ are surrounded by very high buildings, there is a high protection effect. On the other hand, if $\bm{x}$ is on top of a hill, it will be much more exposed to wind (but also much less exposed to, e.g., flood). 
This alternative specification for the exposure component could be the subject of further research.

Note that, for each event, we assume the same damage function. In Section \ref{sec:vulnerability}, we will consider yearly exposure fields which take into account the evolution of sums insured. The iid assumption for event costs is made for simplicity.  The cost field in \eqref{eq:cost field} considers one environmental variable. However, one could extend our approach to account for a multidimensional set of environmental variables.

We define our annual loss trigger as
\begin{equation}
S=\sum_{i=1}^{N} L_i=\sum_{i=1}^{N} \sum_{k=1}^K C_i(\bm{x}_k),\label{eq:lossregion}
\end{equation}
where $L_i=\sum_{k=1}^K C_i(\bm{x}_k)$ is the loss from event $i=1,\ldots,N.$ We will refer to realizations of $S$ and $L_i$ as annual loss trigger realizations and event loss trigger realizations, respectively. We assume that $L_1,\ldots,L_N$ are i.i.d. and are independent of $N$.
By construction, realizations of $S$ can be computed using only publicly available data from reputed and unbiased third parties (weather or geological agencies and insurance industry loss aggregators). Unlike the case of indemnity triggers, it does not require the knowledge
of real losses to the insured, and unlike modeled loss triggers, no access to vendor cat models is needed. This results in full transparency of the triggering mechanism and no moral hazard, as in the case of parametric bonds.
At the same time, the proposed trigger can be a better proxy for the real loss than triggers commonly used for parametric bonds which generally do not account for damage functions and exposure. Indeed, such triggers, being based on characteristics of the hazard such as moment magnitude for earthquakes or minimum central pressure for cyclones, may not correlate sufficiently enough with actual losses. 
Our trigger incorporates transformations that allow us to take exposure via the field $E(\bm{x})$ and vulnerability via the damage function $D$ into account. As a result, we expect basis risk to be reduced. In its spirit, the trigger we propose goes in the direction of a modeled loss trigger since vulnerability curves and potentially exposures can be used, without being one since no cat models are involved. Hence, our trigger potentially offers a good balance between moral hazard and basis risk by keeping the transparency of a parametric trigger while simultaneously reducing its basis risk. The latter stems from two sources: the choice of the sites and the specification of the cost field $C.$ Note that there is no basis risk from the number of events as one would consider the observed number of events during the period under consideration. Moreover, the trigger could be useful for new regions where there are no cat models at hand, in particular, for emerging markets. In such cases, there are typically no loss data available (and when it is, the calculation of claims can be opaque and payouts can take several months) and often only simple parametric solutions such as first generations triggers, see Section \ref{sec:overview}, are offered.

For indemnity-based bonds, moral hazard might appear as the sponsor may have incentives to compute the trigger in its favor by inadequately reporting the incurred loss. In case the information is not transferred in an open way, investors may not be able to challenge the trigger evaluation even if it is done properly by the sponsor.
Indeed, there are historical precedents of indemnity triggers which have gone to court, due to ambiguities in the loss classification and inadequate loss reporting, as in the case of cat bond Mariah Re which defaulted following the active U.S. tornado season in 2011\footnote{\url{https://www.artemis.bm/deal-directory/mariah-re-ltd-series-2010-1/}}.

Our pricing methodology also offers advantages in terms of transparency.
We need a good statistical model for the field $Z(\bm{x})$ in \eqref{eq:cost field} in order to mitigate model risk. Once this model has been calibrated, see Section \ref{sec:calibevents}, investors can simulate from the loss model and compute the bond price using \eqref{eq:priceTdev}. 
The loss model additionally allows investors to derive the attachment probability, the exhaustion probability and the expected loss of the bond, quantities which they would otherwise obtain from vendor models.
The main disadvantage of relying on model agencies is that their outputs often provide different results (e.g., between AIR and RMS), making the inherent model risk hard to quantify. Moreover, sponsors can make an arbitrage and select the model which would produce the most favorable pricing. Giving investors and the sponsor the same view of the risk and an unambiguous pricing mechanism reduces moral hazard. In our framework, we clearly define all modeling hypotheses and provide them to all parties in the transaction.
Our approach also adds value for perils where cat models do not have a long track record and hence model uncertainty is high. Indeed, transactions covering these perils usually comprise an uncertainty loading in the premium to compensate investors. This is the case of so-called secondary perils such as severe convective storm, flood or wildfire, which as opposed to peak perils such as named storm and earthquake are not so well modeled and are subject to more scrutiny from investors. Secondary perils make up an increasingly growing part of global insured losses, see \cite{sigma21}, and as such the need for their accurate modeling is garnering attention. To reduce the protection gap, that is, the difference between insured losses and economic losses, parametric solutions will be key and we believe our trigger proposal can contribute to solving this pressing societal need.

\section{Case study}\label{sec:case study}
In this section, we present a real case study covering windstorm risk in Germany. We explain the structure of the selected data for our analysis and discuss the methodology to fit a max-stable random field to extreme realizations in the wind speeds. Our case study covers Germany which is a country with a high insurance penetration.
This case study is relevant in the sense that extreme windstorms impact (re)insurer's balance sheets in a significant manner. It could further serve as a benchmark for any (re)insurer covering European windstorm, and could be extended not only to any other geographical region where wind-related risk is salient, but also to other perils where data are available.

\subsection{PERILS data}\label{sec:windexpo}
"Catastrophe Risk Evaluation and Standardizing Target Accumulations" (CRESTA)\footnote{\url{https://www.cresta.org/}} is the most widely used geographical data aggregation standard by the global (re)insurance industry. CRESTA zones usually correspond to administrative zones such as groupings of postal codes within a country. 

For the purpose of calibrating a max-stable random field, we consider the maximal wind speeds for all 95 CRESTA zones in Germany and for 100 
windstorm events having occurred between 
February 1999 and February 2020, a dataset originating from the UK Met Office (UKMO) and provided by PERILS\footnote{\url{https://www.perils.org/}}.
The UKMO defines maximum wind gust for an event as the highest value of the 3-second running-average wind speed at 10 metres above ground. Gust values are generated using successive short-range forecast runs using the Met Office Unified Model. 
PERILS then reports the peak gust over a 72 hour time period\footnote{This time window corresponds to the so-called hours clause for windstorm events which is usually present in natural catastrophe reinsurance contracts to specify how losses can be accumulated and events delimited, see, e.g., \url{https://www.air-worldwide.com/publications/air-currents/2019/Modeling-Fundamentals--Accounting-for-the-Hours-Clause/}.} for each CRESTA zone as a representation of the UKMO modeled grid values 
within the CRESTA zone weighted by population density.

Moreover, shapefiles for each German CRESTA zone were provided by GfK GeoMarketing\footnote{\url{https://geodaten.gfk.com/landkarten/}}. 
In our context, the shapefiles represent the geographical border of each CRESTA zone. 
For each CRESTA zone we consider the geographical centroid as representative grid point. 
Other choices such as the population weighted-centroid could be considered in case population data are available.
For non-convex regions, this can produce centroids which are not part of the region. 
This undesired behavior is limited to a handful of CRESTA zones and can thus be considered negligible.

\begin{figure}[!ht]
\begin{center}
\includegraphics[
width=0.55\textwidth]{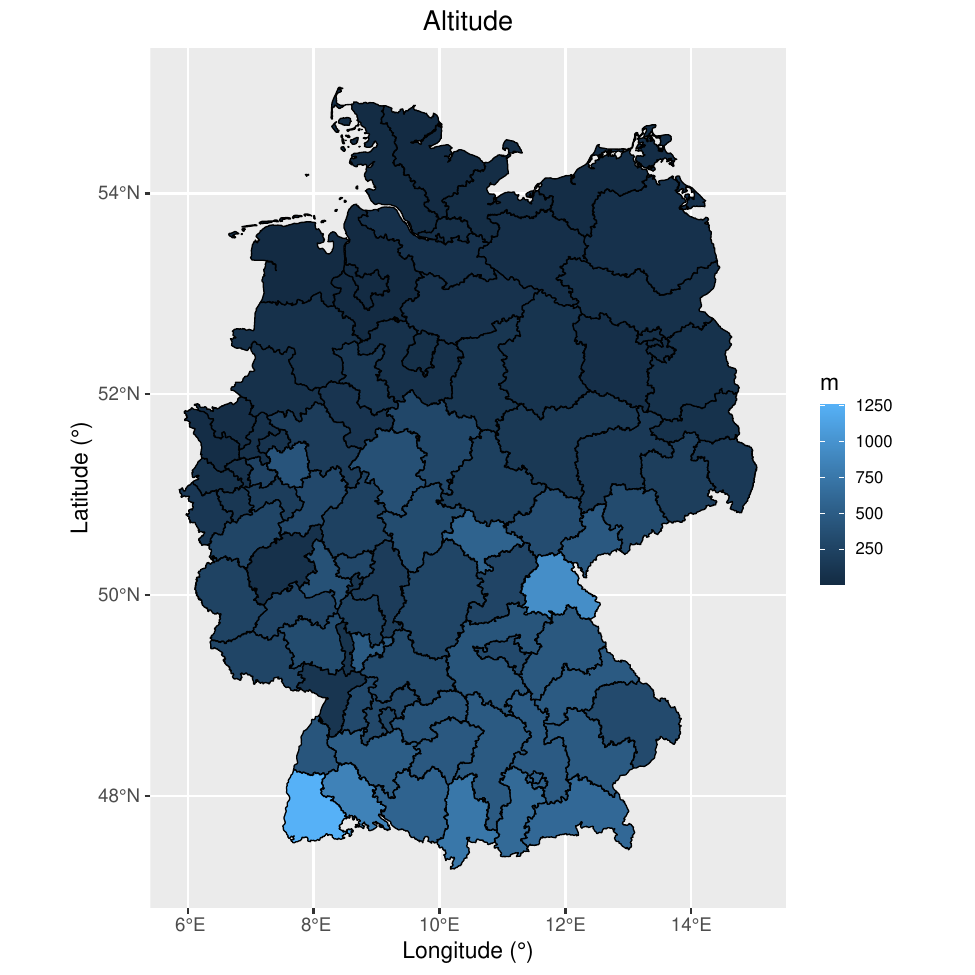}
\caption{Altitude of the representative grid point for each CRESTA zone in Germany.}\label{fig:CRESTAalt}
\end{center}
\end{figure}

We additionally collect altitude data at each CRESTA zone centroid using the Elevation API on the Google Maps platform\footnote{\url{https://developers.google.com/maps/documentation/elevation/start}}. 
We display in Figure \ref{fig:CRESTAalt} the altitude of the representative grid point for each CRESTA zone. Assigning the altitude of the geographical centroid as the representative altitude for each CRESTA zone instead of the altitude of the highest point within the CRESTA zone is a modeling choice and may lead to a fitted trend surface which is not optimal, see Section \ref{sec:calibevents}.

Industry insured losses are available for events having occurred since 1999 (when PERILS started reporting) and which had a total market loss across all impacted countries where data are collected (i.e., not only Germany) over EUR 200mn due to reporting guidelines set by PERILS. 
For such qualifying events before 2010, the industry loss is reported per country and not per CRESTA zone. This is not problematic as we are dealing with aggregate losses for Germany. Finally, the industry exposure database only starts in the year 2013. For all years between 2013 and 2020, this data corresponds to the evolution of exposure per CRESTA zone. In the sequel, we will consider the total PERILS exposure over residential, commercial and industrial lines, and over the building, contents and business interruption coverages, for all 95 German CRESTA zones. 
Due to these data limitations, we are unable to compute realizations of our trigger for historical storms before 2013 without any further assumption on industry exposure.
We thus project the exposure per CRESTA zone using linear trending for each year back to 1999 to use as much of the available information as possible. This yields a sample of 20 windstorm events having occurred between 1999 and 2020 for which we have industry loss data. These events form a subset of the sample of 100 historical windstorm events for which we have wind speed data per CRESTA zone and which we will use for the calibration of the hazard model. We denote the insured loss sub-sample as $\{\ell_i^{ins},i=1,\ldots,20\}.$
We provide a list of these 20 windstorm events reported by PERILS in Table \ref{event_names}, alongside the (unadjusted) as-is loss, the loss adjusted for inflation and exposure increase, the total exposure, the average damage ratio and the maximal wind speed observed for each event.

\begin{sidewaystable}
\begin{center}
\resizebox{\textwidth}{!}
{\begin{tabular}{c|c|r|r|r|r|r}
Event name & Event start date & As-is loss $L_i$ (mn EUR) & Adjusted loss (mn EUR) & Total exposure $E_i$ (tn EUR) & Damage ratio $L_i/E_i$ (\%) & Max wind speed $Z_i$ (m/s)\\
\hline
Anatol & 03.12.1999 & 135.279 & 509.311 & 4.494 &  0.003& 41.28\\
Lothar & 26.12.1999 & 606.668 & 2,284.042 & 4.494 & 0.014 & 43.39\\
Jeanett & 26.10.2002 & 739.694 & 2,031.300 & 6.161 &  0.012 & 36.67\\
Kyrill & 18.01.2007 & 2,326.540 & 4,339.593 & 9.070 & 0.026 & 47.65\\
Xynthia & 28.02.2010 & 480.809 & 737.962 & 10.821 & 0.004 & 37.26\\
Joachim & 15.12.2011 & 45.215 & 66.528 & 11.405 & 0.0004 & 26.87\\
Andrea & 04.01.2012 & 128.475 & 180.384 & 11.989 & 0.001 & 29.77\\
Christian & 27.10.2013 & 413.546 & 544.931 & 12.808 & 0.003 & 39.02\\
Xaver & 05.12.2013 & 151.621 & 200.753 & 12.808 & 0.001 & 36.97\\
Elon-Felix & 08.01.2015 & 157.547 & 196.845 & 13.615 & 0.001 & 31.79\\
Mike-Niklas & 29.03.2015 & 639.979 & 800.066 & 13.615 & 0.005 & 32.13\\
Egon & 12.01.2017 & 84.423 & 97.453 & 14.580 & 0.001 & 31.28\\
Thomas & 23.02.2017 & 75.780 & 87.669 & 14.580 & 0.001 & 30.27\\
Xavier & 05.10.2017 & 324.233 & 378.847 & 14.580 & 0.002 & 39.95\\
Herwart & 29.10.2017 & 197.858 & 229.756 & 14.580 & 0.001 & 33.77\\
Burglind & 02.01.2018 & 134.864 & 147.491 & 15.457 & 0.001 & 31.72\\
Friederike & 17.01.2018 & 1,151.504 & 1,264.908 & 15.457 & 0.007 & 37.23\\
Dragi-Eberhard & 09.03.2019 & 392.001 & 410.718 & 16.111 & 0.002 & 34.44\\
Sabine & 09.02.2020 & 709.192 & 709.192 & 16.918 & 0.004 & 35.41\\
Victoria & 15.02.2020 & 52.952 & 52.952 & 16.918 & 0.0003 & 28.74\\
\end{tabular}}
\end{center}
\caption{List of 20 windstorm events for which PERILS reported industry losses. Adjusted losses correspond to event losses adjusted to the 2020 industry exposure database (i.e., inflation- and exposure-adjusted). The last three columns correspond to the total exposure $E_i=\sum_{k=1}^K E_i(\bm{x}_k)$, the average damage ratio $L_i/E_i,$ and the maximal wind speed $Z_i=\max_k Z_i(\bm{x}_k),$ respectively, for $i=1,\dots,20.$}
\label{event_names}
\end{sidewaystable}

\subsection{Calibration and validation of the vulnerability function}\label{sec:vulnerability}
The vulnerability function maps a wind speed to a damage ratio, i.e., the percentage of exposure which suffers a loss following an event. A comparison of damage functions for windstorms is carried out in \cite{prahl15} 
for German data. 
The choice of a power damage function is rooted in physical considerations: indeed, structural damage is proportional to the dissipation rate of wind kinetic energy which scales with the third power of wind speed. However, other studies indicate a higher exponent for this relationship, see \cite{prahl12} for an empirical discussion of the link between German windstorms and insured losses.

\begin{figure}[!ht]
\begin{center}
\includegraphics[
width=0.6\textwidth]{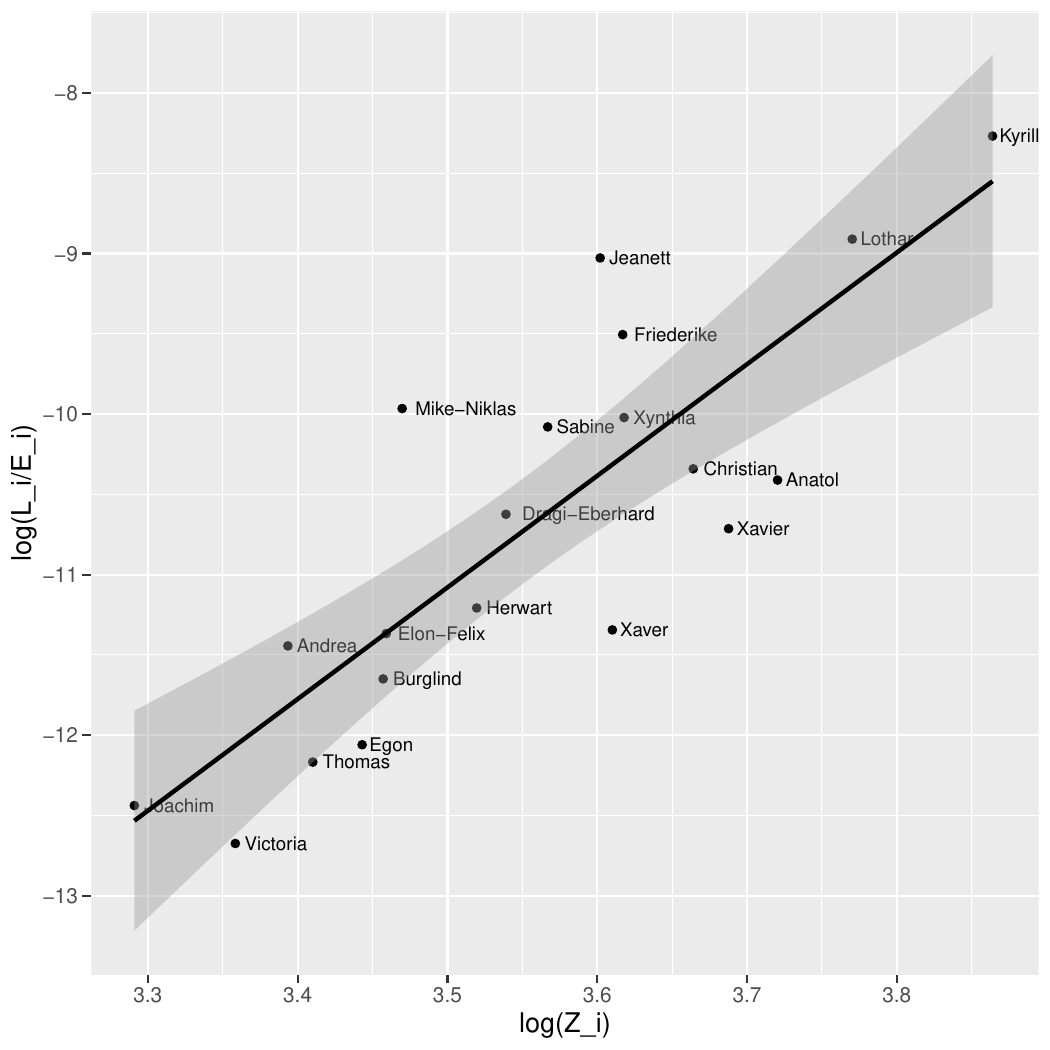}
\caption{log($L_i/E_i$) as a function of log($Z_i$) for the sample of 20 events with insured loss data.}\label{fig:logprahl}
\end{center}
\end{figure}

For a realization $z(\bm{x})$ of the field $Z(\bm{x})$ characterizing the physical hazard (extreme wind speeds), we assume a power law form $D(z(\bm{x}))=(z(\bm{x})/c_1)^\beta$ applied component-wise for each site $\bm{x}\in A$ and which is particularly adapted to the wind hazard, see \cite{prahl12} and \cite{prahl15}, where $\beta\in\mathbb{R}_+$ and $c_1>0$ is a scaling coefficient in order to retrieve as output of the vulnerability function a damage ratio between 0 and 100\%.
There is a natural link with the typical sigmoid function with steep initial increase and saturation at large wind speeds.
Indeed, in \cite{prahl12} a sigmoid function of the form $$D(y)=\frac{d_{\textnormal{max}}}{1+e^{-\beta y}},$$ is specified, where $d_{\textnormal{max}}>0$ is the asymptotic upper bound. Applying the transformation $y=\ln(z/b_z)$ to maximal wind speeds $z$ scaled by a local constant $b_z>0,$ one can approximate the damage function in the regime $d\ll d_{\textnormal{max}}$ (valid for Germany, see \cite{prahl12}) as
$$D(z)=\frac{d_{\textnormal{max}}}{1+(z/b_z)^{-\beta}}\approx \left(\frac{z}{c_1}\right)^\beta,$$ where $c_1\approx b_z d_{\textnormal{max}}^{-1/\beta}.$
Note that the specified vulnerability function does not depend on the site. This is a modeling choice which increases model error but which allows for a reduction of the estimation uncertainty. More granular information on the exposure mix at each site (e.g., variables such as construction class, occupancy type, year built, age of roof) would allow us to define a location-specific vulnerability function taking into account the local building stock and regional construction practices. One advantage (and one of the motivations) of our trigger lies in the fact that it involves a pretty simple vulnerability model, and thus it can be fitted even in cases where we have very few data.
The parameters $c_1$ and $\beta$ can be calibrated to industry insured loss data when the latter is available. This calibration yields a parameter $c_1$ which ensures that the damage function produces values between 0 and 100\%. In the absence of such data, the coefficients $c_1$ and $\beta$ be chosen based on expert knowledge, see, e.g., \cite{prahl12}.
In Figure \ref{fig:logprahl}, we plot the log of the average damage ratio $\log(L_i/E_i)$ in terms of the log of the maximal wind speed $Z_i,$ for $i=1,\ldots,20.$

\begin{figure}[!ht]
\begin{center}
\includegraphics[
width=0.6\textwidth]{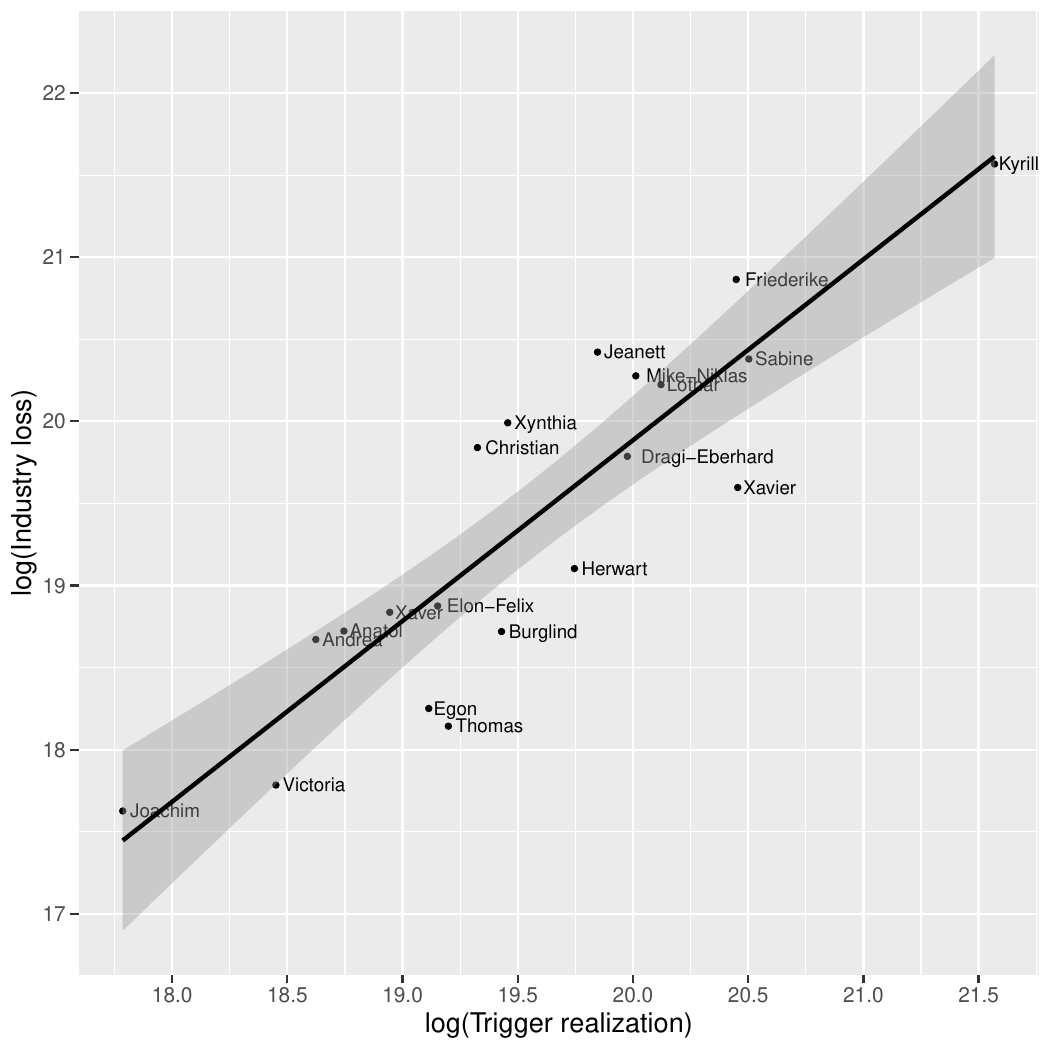}
\caption{Event trigger realizations versus industry insured losses for 20 selected historical events between 1999 and 2020 on the logarithmic scale, fitted regression line with 95\% confidence interval characterizing the uncertainty on the best regression line.}\label{fig:perils}
\end{center}
\end{figure}

Using industry loss data for our sample of 20 windstorm events between 1999 and 2020, we infer the vulnerability exponent $\beta$, as well as the parameter $c_1$ from the regression of industry losses $\ell_i^{ins}$ on event loss trigger realizations $\sum_{k=1}^K \left(z_i(\bm{x}_k)\right)^\beta E_i(\bm{x}_k),$ for $i=1,\ldots,20,$ and where for each event $i$ we consider the relevant exposure vector, that is, for an event in year $1999,\ldots,2020,$ we take the exposure vector from that year. 
For various integer values of $\beta$ between 5 and 13, where for $\beta$ between 7 and 9 we refine the grid to include all values with increments of 0.1, we choose the regression coefficient $\hat{\beta}_1$ such that the coefficient of determination $R^2$ of the following constrained linear regression (without intercept) is maximized:
\begin{equation*}
\argmin_{\beta_1}\sum_{i=1}^{20}\left(\ell_i^{ins}-\beta_1\sum_{k=1}^K \left(z_i(\bm{x}_k)\right)^\beta E_i(\bm{x}_k)\right)^2,    
\end{equation*}
where $\beta_1=c_1^{-1/\beta}.$
Considering our set of 20 observations, this yields a maximal $R^2=0.9307$ for $\beta=7.8,$ see Table \ref{Table_R2}, and associated regression coefficient 
$\hat{\beta}_1=1.442 \cdot 10^{-16}\mbox{ }(9.028 \cdot 10^{-18})$ (the standard error of the estimate is given in parentheses). 
\begin{table}
\begin{center}
\begin{tabular}{l|c|c|c|c|c|c|c|c|c|c}
$\beta$ & 5 & 6 & 7 & 7.8 &  8 & 9 & 10 & 11 & 12 & 13 \\
\hline
$R^2$ & 0.8145 & 0.8839 & 0.9222 & 0.9307 & 0.9301 & 0.9151 & 0.8865 & 0.8523 & 0.8178 & 0.7858
\end{tabular}
\end{center}
\caption{Coefficient of determination $R^2$ for various integer values of $\beta$ between 5 and 13 and for $\beta=7.8$ where $R^2$ is maximized. Note that we have ran the regression for all values between 7 and 9 with increments of 0.1. In order to not overload the table, only the $R^2$ value for $\beta=7.8$ is shown here.} 
\label{Table_R2}
\end{table}
For personal lines, \cite{prahl12} suggest an exponent between 8 and 12. 
The non-integer powers found in \cite{prahl15} stem from the consideration of insurance deductibles. 
For our choice of $\beta=7.8,$ we compare in Figure \ref{fig:perils} historical industry insured losses $\ell_i^{ins}$ with our event loss trigger realizations $\hat{\beta}_1\sum_{k=1}^K \left(z_i(\bm{x}_k)\right)^\beta E_i(\bm{x}_k)$. This allows us to assess the basis risk associated with our trigger. 
We observe that for each historical storm, the regression line displayed as the black line in Figure \ref{fig:perils} provides a rather good fit ($R^2=0.9307$), that is, for any industry loss size (with its corresponding return period on a German insurance market level), the trigger is able to capture the loss level solely based on wind speeds and an affine transformation of a power vulnerability function with $\beta=7.8$. The confidence bounds in Figure \ref{fig:perils} describe the uncertainty around the slope coefficient of the best regression line. It is thus expected that points fall outside of this interval. 

We perform a 10-fold cross-validation analysis where in each fold, 2 of the 20 storms are left out and a linear regression is conducted for all integer values of $\beta$ between 5 and 13. As above, we refine the grid for $\beta$ between 7 and 9 to include all values with increments of 0.1. This produces 10 estimates for the coefficients $\beta$ and $\beta_1$ and allows to assess the uncertainty in the parameter estimation of the damage function. For 4 of the 10 folds, the coefficient of determination is maximized for $\beta=7.8,$ for 3 of the 10 folds, the coefficient of determination is maximized for $\beta=7.6,$ for 2 of the 10 folds, the coefficient of determination is maximized for $\beta=7.9,$ whereas for one fold the maximal coefficient of determination is obtained for $\beta=8.1.$ We provide the fitted $\hat{\beta}_1$ parameter for each fold using $\beta=7.8$ in Table \ref{folds}. Over these 10 folds, the standard deviation of the fitted $\hat{\beta}_1$ parameter is $3.421\cdot 10^{-18},$ which is in line with the uncertainty around the fitted value using the full sample. One could refine this study in the presence of a larger sample of insured losses.

\begin{table}  
\begin{center}
\begin{tabular}{c|c|c|c|c|c|c|c|c|c}
Fold 1 & Fold 2 & Fold 3 & Fold 4 & Fold 5 & Fold 6 & Fold 7 & Fold 8 & Fold 9 & Fold 10\\
\hline
1.436 & 1.383 & 1.432 & 1.435 & 1.444 & 1.433 & 1.507 & 1.460 & 1.395 & 1.456 \\
\end{tabular}
\end{center}
\caption{Fitted regression coefficient $\hat{\beta}_1$ (expressed in multiples of $10^{-16}$) from 10-fold cross validation using $\beta=7.8$.} 
\label{folds}
\end{table}

We observe in Figure \ref{fig:perils} that two events stand out based on their distance to the regression line. Firstly, the losses associated with the event Friederike in January 2018 seem to be underestimated by our trigger. For this particular event, the actual losses are greater than suggested by the maximal wind speeds produced by the storm, see Table \ref{event_names}, which could be due to the track affecting larger parts of Germany. Our trigger would have to be further refined, for example including the storm size, in order to capture such effects. 

Furthermore, our trigger seems to overestimate the loss from the storm Xavier that crossed Northern Germany in early October 2017. The PERILS report for this event underlines the exceptional speed at which the storm traveled over the region under concern. Initiating over the Northern Atlantic, the depression traveled 3000km within 24 hours with winds of hurricane strength at several locations in Germany. The loss realization from our trigger, being based on maximal wind speeds, overestimates the industry loss from this event, which could be explained by the rapidity at which this particular winter storm traveled over Germany. Indeed, as for hurricanes, for a fixed maximal wind speed, the insured damage is usually more important when the storm travels slower over land. 
Capturing such exceptional events from a meteorological perspective would require adjusting the trigger to include more than just the wind speed as input variable in the physical hazard component.

Note that we calibrated our damage function only for events with losses above the PERILS threshold of EUR 200mn. We thus implicitly assume that our damage function is also suited for the range of wind speed values associated with lower losses. 

\begin{figure}[!ht]
\begin{center}
\includegraphics[
width=0.6\textwidth]{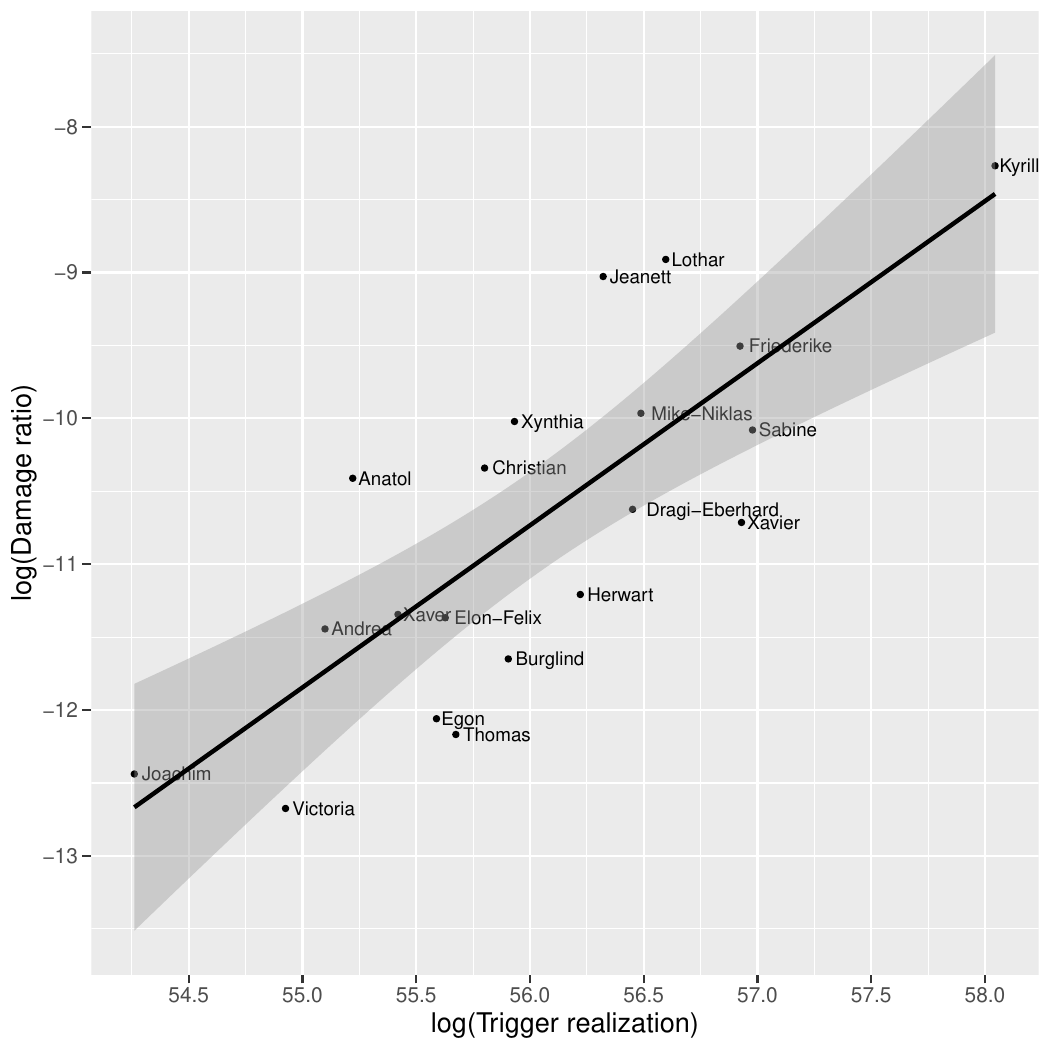}
\caption{(Log) average damage ratio versus (log) event trigger realizations for 20 selected historical events between 1999 and 2020, fitted regression line with 95\% confidence interval characterizing the uncertainty on the best regression line.}\label{fig:dmg}
\end{center}
\end{figure}

In Figure \ref{fig:dmg}, we plot the average damage ratio $L_i/E_i$ against the event trigger realizations for our sample of 20 historical events on the logarithmic scale. We observe that for the older events Anatol (1999), Lothar (1999) and Jeanett (2002), the damage ratios are underestimated, which might be due to the fact that the exposure extrapolation via backwards linear trending is too strong (and hence the real exposure should be higher). Another possibility is that more recent building codes were not in place at that time, implying that, ceteris paribus, an event occurring today would result in a lower damage ratio than an event occurring 20 years ago.

One can now benchmark this analysis with an event loss trigger depending only on wind speeds, and with other triggers additionally considering the exposure. We consider three alternative specifications for the event loss trigger in \eqref{eq:lossregion}:
\begin{enumerate}
    \item $L_i^1=\sum_{k=1}^K Z_i(\bm{x}_k)$,
    \item $L_i^2=\sum_{k=1}^K w_{k,i} Z_i(\bm{x}_k)$,
    \item $L_i^3=k_1 E_i (\max_k Z_i(\bm{x}_k))^{k_2}$,
\end{enumerate}
that is, the first choice only considers wind speeds at each CRESTA zone, the second choice introduces yearly weights $w_{k,i}>0$ such that $\sum_{k=1}^K w_{k,i}=1$ for all $i$, depending on the contribution of each CRESTA zone in terms of exposure to the total exposure in the given year, and the third choice is justified by the graphical inspection undertaken in Figure \ref{fig:logprahl}.

The first two proposals mirror previous European windstorm CAT bond transactions structured with parametric triggers such as Pylon (2003) and its successor Pylon II Capital (2011), Eurus II (2009), or Green Valley (2007) and its successor Green Valley 2 (2010), see \url{https://www.artemis.bm}. The parameters $k_1$ and $k_2$ from the third trigger specification are estimated by performing a linear regression of $\log(L_i/E_i)$ on $\log(\max_k Z_i(\bm{x}_k)),$ for $i=1,\ldots,20.$ This yields $\log(k_1)=-35.416\mbox{ }(3.885)$ and $k_2=6.953\mbox{ }(1.093).$ We thus obtain $k_1=4.158\cdot 10^{-16}$.

\begin{figure}[!ht]
    \centering
    \begin{subfigure}[b]{0.32\textwidth}
        \centering
       \includegraphics[width=\textwidth]{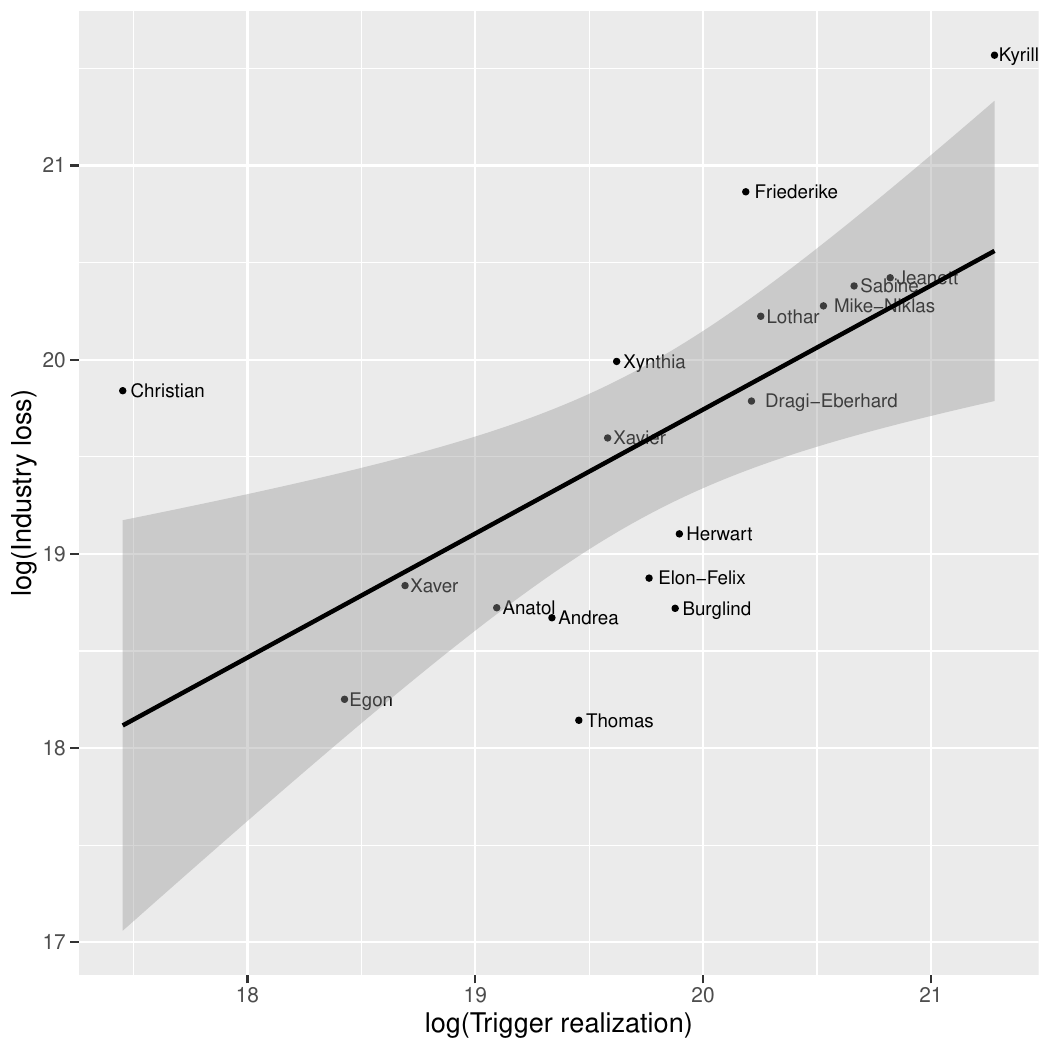}
        \end{subfigure}
        \hfill
        \begin{subfigure}[b]{0.32\textwidth}  
            \centering 
            \includegraphics[width=\textwidth]{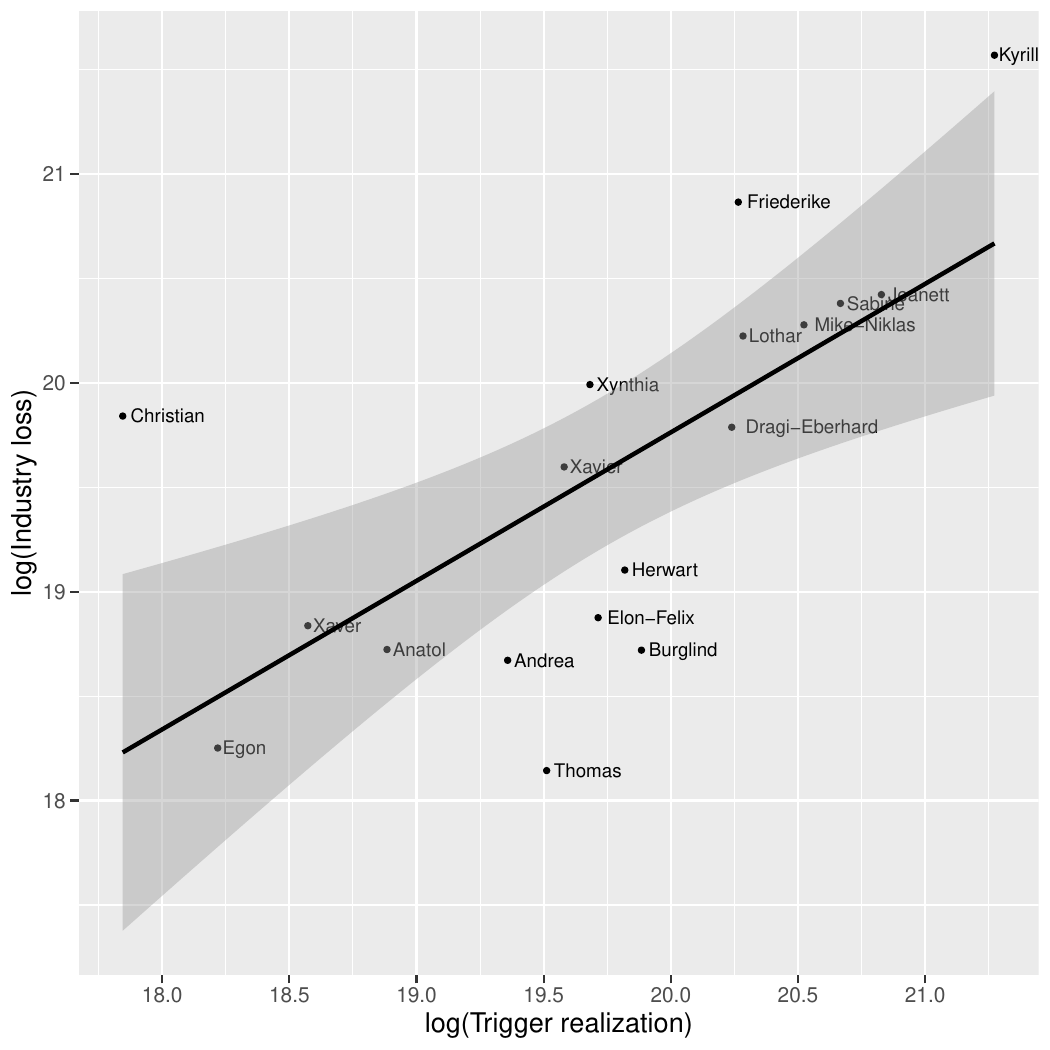}
        \end{subfigure}
                \hfill
        \begin{subfigure}[b]{0.32\textwidth}  
            \centering 
            \includegraphics[width=\textwidth]{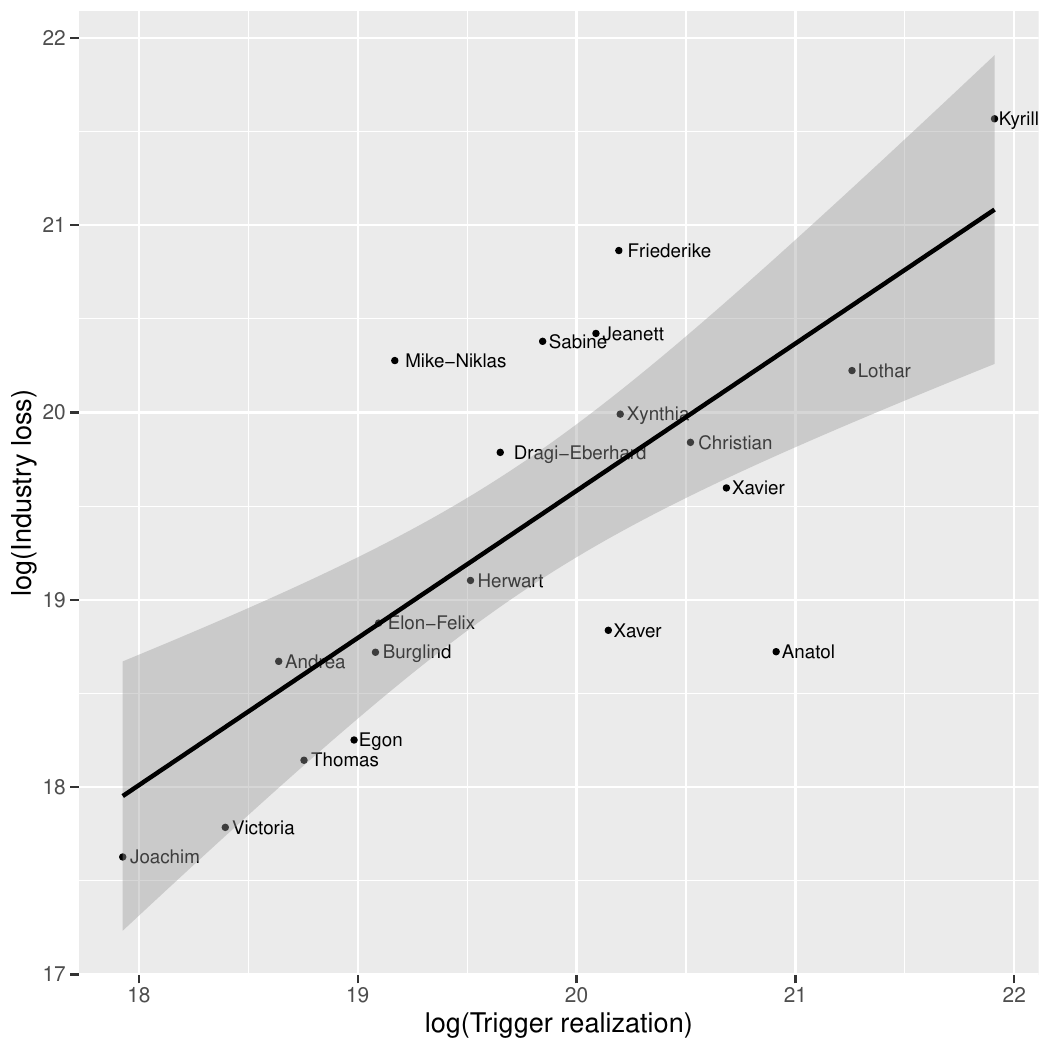}
        \end{subfigure}
        \caption{Event trigger realizations for alternatives 1 (left, $R^2=0.7293$), 2 (center, $R^2=0.7390$) and 3 (right, $R^2=0.6923$) versus industry insured losses for 20 selected historical events between 1999 and 2020 on the logarithmic scale, fitted regression line with 95\% confidence interval.
        }
\label{fig:alt_trigger}
\end{figure}

In Figure \ref{fig:alt_trigger}, we display historical industry insured losses versus realizations of $L^1$(left), $L^2$ (middle) and $L^3$ (right). This allows us to study basis risk for various trigger specifications. These plots highlight the reduction in basis risk as measured by the $R^2$ brought by our trigger choice compared to cat bonds with parametric triggers that have been issued in the past as well as with a trigger only depending on the maximal wind speed across all locations. We keep the third specification $L^3$ for benchmarking our results in Sections \ref{sec:modelrisk} and \ref{sec:results}, and we will refer to this trigger in the sequel as the benchmark trigger.

\subsection{Windstorm model calibration and validation}\label{sec:calibevents}
In this section, we discuss the calibration of our windstorm model to historical events under the CRESTA zone representation.
From the sample of 100 windstorm events, we fit a GEV distribution for each CRESTA zone centroid via maximum likelihood using the function \texttt{gev.fit} from the {\sf R} package \texttt{ismev}, see \cite{ismev}. 
We fit the Smith, Schlather, Brown--Resnick, and extremal-$t$ random fields for the dependence structure by assuming that the marginals are unit Fréchet, see Section \ref{sec:max-stable}, that is, once the dependency structure is calibrated, the marginals are transformed from Fréchet to their respective GEV distribution. In the case of the Schlather and extremal-$t$ models, we considered the powered exponential, Whittle--Matérn and Cauchy correlation functions. Parameter estimates are obtained using composite likelihood techniques, more precisely by maximizing the pairwise composite log-likelihood given at Equation (6) in \cite{padoan}. Model selection is done based on the composite likelihood information criterion (CLIC). We refer the reader to Section \ref{sec:fitmaxstable} in the Appendix for more details on the calibration procedure and the CLIC. According to Table \ref{Table_CLIC_Values_2step}, the extremal-$t$ model with Cauchy correlation function produced the lowest CLIC. This yields the estimates for the dependency parameters found in Table \ref{table_Param_Estimates_2step}. We also display the means of the GEV parameters across all locations alongside their respective standard deviations.
We note that when performing the calibration in two steps (separately fitting the marginal parameters and the dependency structure), uncertainty is not properly taken into account, see the discussion in Section \ref{sec:calibevents_apdx}, but it allows for a more accurate modeling of the marginals. 

In Section \ref{sec:calibevents_apdx}, we present another possible calibration of the windstorm model where we consider trend surfaces for the marginal parameters, and we perform a validation of the marginals and of the dependency structure. Indeed, when considering trend surfaces, one can separate the set of locations into a training sample and a validation sample, which is not the case for the two-step calibration presented here. However, trend surfaces present the disadvantage of introducing modeling errors in the wind speeds at each location. We display the 5 largest differences between the fit per location and the trend surface for each GEV parameter in Tables \ref{table_gev_delta_eta}, \ref{table_gev_delta_tau} and \ref{table_gev_delta_xi} for the location, scale and shape parameters, respectively. Moreover, these differences are amplified for CRESTA zones with high exposure. In particular, for regions such as Munich (DEU80, DEU81) and Würzburg (DEU97) with high total sums insured, the $\xi$ parameters suggested by the trend surface are much higher than the equivalent parameters fitted to these locations. Similarly, for highly exposed regions such as Freiburg (DEU79), Frankfurt (DEU60) and Berlin (DEU12), the $\eta$ parameters suggested by the trend surface are higher than the equivalent parameters fitted to these locations.

\begin{table}
\begin{center}
\begin{tabular}{l|c|c}
Model & CLIC & Number of model parameters \\
\hline
Extremal-$t$ Cauchy & 3,368,504 & 279 \\
Extremal-$t$ powered exponential & 3,368,592 & 279 \\
Extremal-$t$ Whittle--Matérn & 3,371,138 & 279 \\
Schlather powered exponential & 3,377,762 & 279 \\
Schlather Whittle--Matérn & 3,377,870 & 279 \\
Schlather Cauchy & 3,378,000 & 279 \\
Smith & 3,424,245 & 278 \\
Brown--Resnick & 3,432,668 & 277 

\end{tabular}
\end{center}
\caption{CLIC values for the different fitted models (in ascending order).}
\label{Table_CLIC_Values_2step}
\end{table}

\begin{table}[!ht]
    \centering
    \begin{tabular}{c|r}
    $a_0$ & 0.006 (0.001) \\ 
    $\kappa$ & 4.662 (0.962) \\ 
    $\psi$ & 0.392 (0.193) \\ 
    $\nu$ & 2.064 (0.429) \\ 
    $\eta$ & 16.916 (0.924) \\
    $\tau$ & 5.029 (0.499) \\
    $\xi$ & -0.116 (0.049)
    \end{tabular}
    \caption{Parameter estimates (standard errors inside parentheses) of the best model (extremal-$t$ with Cauchy correlation function). For the GEV parameters, the means across locations are given alongside their respective standard deviation.}
    \label{table_Param_Estimates_2step}
\end{table}

\begin{table}
\begin{center}
\begin{tabular}{l|c}
CRESTA zone & Difference in fitted $\eta$ parameter \\
\hline
DEU79 & 1.432 \\
DEU65 & 1.410 \\
DEU60 & 1.268\\
DEU68 & 1.097 \\
DEU12 & 0.899
\end{tabular}
\end{center}
\caption{Five largest differences between the fit of the location parameter $\eta$ from the trend surface and the fit per location.}
\label{table_gev_delta_eta}
\end{table}

\begin{table}
\begin{center}
\begin{tabular}{l|c}
CRESTA zone & Difference in fitted $\tau$ parameter \\
\hline
DEU17 & 0.535 \\
DEU83 & 0.530 \\
DEU02 & 0.440\\
DEU61 & 0.431 \\
DEU60 & 0.403 
\end{tabular}
\end{center}
\caption{Five largest differences between the fit of the scale parameter $\tau$ from the trend surface and the fit per location.}
\label{table_gev_delta_tau}
\end{table}

\begin{table}
\begin{center}
\begin{tabular}{l|c}
CRESTA zone & Difference in fitted $\xi$ parameter \\
\hline
DEU97 & 0.224 \\
DEU69 & 0.212 \\
DEU63 & 0.192 \\
DEU81 & 0.185\\
DEU80 & 0.182 
\end{tabular}
\end{center}
\caption{Five largest differences between the fit of the shape parameter $\xi$ from the trend surface and the fit per location.}
\label{table_gev_delta_xi}
\end{table}

Furthermore, following the graphical inspection in Figure \ref{fig:logprahl} and the benchmark trigger $L_i^3$ defined in Section \ref{sec:vulnerability}, we fit a univariate GEV model to the 100 wind speed maxima over all CRESTA zones. This produces the following fitted GEV parameters: $\tilde{\eta}=22.519\mbox{ }(0.672),$ $\tilde{\tau}=5.947\mbox{ }(0.482),$ $\tilde{\xi}=-0.078\mbox{ }(0.075).$ From this univariate model one can simulate realizations of the benchmark trigger, see Section \ref{sec:modelrisk}.

For cat bonds structured on aggregate basis, all covered events within the period must be considered. This justifies the specification of a frequency distribution for the number of events within a year, which will be the subject of the next section.

\subsection{Frequency distribution}\label{sec:freqdist}
In this section, we discuss the choice of the frequency distribution for the variable $N$ appearing in \eqref{eq:lossregion}.
For each year between 1999 and 2020, we count the number of German windstorm events. We present yearly event count data for the years between 1999 and 2020 in Table \ref{countevents}.

\begin{table}
\begin{center}
\begin{tabular}{l|c}
Year & Number of events\\
\hline
1999 & 6 \\
2000 & 6 \\
2001 & 2 \\
2002 & 5 \\
2003 & 0 \\
2004 & 7 \\
2005 & 4 \\
2006 & 4 \\
2007 & 8 \\
2008 & 5 \\
2009 & 7 \\
2010 & 4 \\
2011 & 7 \\
2012 & 2 \\
2013 & 9 \\
2014 & 12 \\
2015 & 3 \\
2016 & 0 \\
2017 & 4 \\
2018 & 2 \\
2019 & 1 \\
2020 & 2 
\end{tabular}
\end{center}
\caption{Yearly windstorm event counts. 
}
\label{countevents}
\end{table}

\begin{figure}[!ht]
\begin{center}
\includegraphics[
scale=0.55]{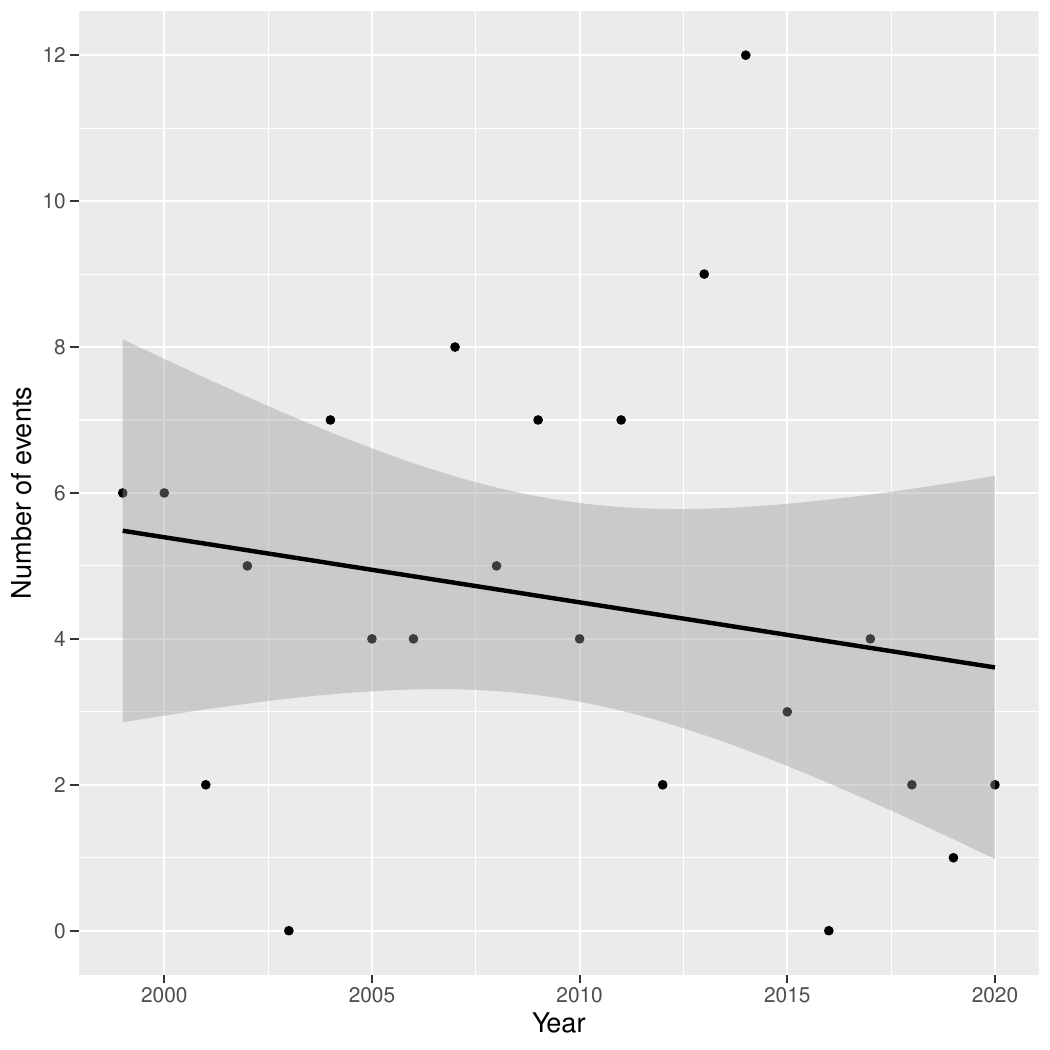}
\caption{Number of windstorm events per year between 1999 and 2020 with the trend line from the linear regression of the observed number of events against time and the associated 95\% confidence bounds on the regression estimates.}\label{fig:yearly_count}
\end{center}
\end{figure}

\begin{figure}[!ht]
\begin{center}
\includegraphics[scale = 0.8]{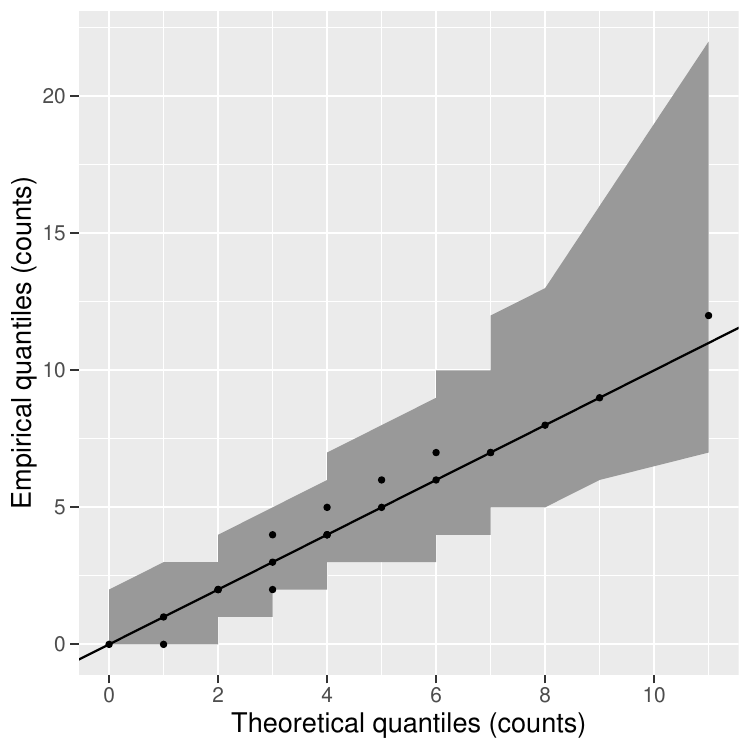}
\caption{Q-Q plot of the observed versus simulated yearly event counts. Overall envelopes at the 95\% confidence level are depicted in dark gray.}\label{fig:QQplotFreqModel}
\end{center}
\end{figure}

We consider the non-parametric Pettitt's test, see \cite{pettitt}, from the {\sf R} package \texttt{trend}, see \cite{trend}, to check whether there is a temporal trend in the yearly count data. This test does not reject the null hypothesis of no shift in the central tendency of the time series of event counts with $p$-value 0.1426. This is confirmed by a visual inspection of the time series of yearly events between 1999 and 2020 in Figure \ref{fig:yearly_count}. Moreover, we perform an overdispersion test in the Poisson GLM setting using the function \texttt{dispersiontest} from the {\sf R} package \texttt{AER}, see \cite{aer}. This test rejects the null hypothesis of a Poisson distribution with $p$-value 0.043 and suggests an over-dispersion factor of 1.935. The residual deviance of the Poisson GLM model is 48.596 on 21 degrees of freedom. We then fit a negative binomial distribution to the count data using maximum likelihood. This results in a frequency distribution with size parameter $3.955\mbox{ }(2.474)$ and mean $4.546\mbox{ }(0.666),$ where the standard errors of the respective estimates are given in parentheses. We thus obtain a fitted over-dispersion factor of 2.149.
To assess the goodness-of-fit of the frequency model, we show a two sample Q-Q plot between the observations and the simulated values, see Figure \ref{fig:QQplotFreqModel}.

\subsection{Full model assessment}\label{sec:modelrisk}
Following the calibration and validation of the three components appearing in our trigger definition, that is, the vulnerability function in Section \ref{sec:vulnerability}, the max-stable model for the hazard component in Section \ref{sec:calibevents} and the frequency distribution for the number of yearly events in Section \ref{sec:freqdist}, we perform a full model assessment.
For this purpose, we generate $J=2,000,000$ realizations from the frequency distribution $N$. This results in a sample of $M=9,090,933$ events that we generate from the extremal-$t$ model with Cauchy correlation function and parameters in Table \ref{table_Param_Estimates_2step} using the
\texttt{rmaxstab} function from the {\sf R} package~\texttt{SpatialExtremes}, see \cite{spatialextremes}, the vulnerability function with $\beta=7.8$ as discussed in Section \ref{sec:vulnerability} and PERILS exposure data per CRESTA zone for the year 2020. 

\begin{figure}[!ht]
    \centering
    \begin{subfigure}[b]{0.32\textwidth}
        \centering
       \includegraphics[width=\textwidth]{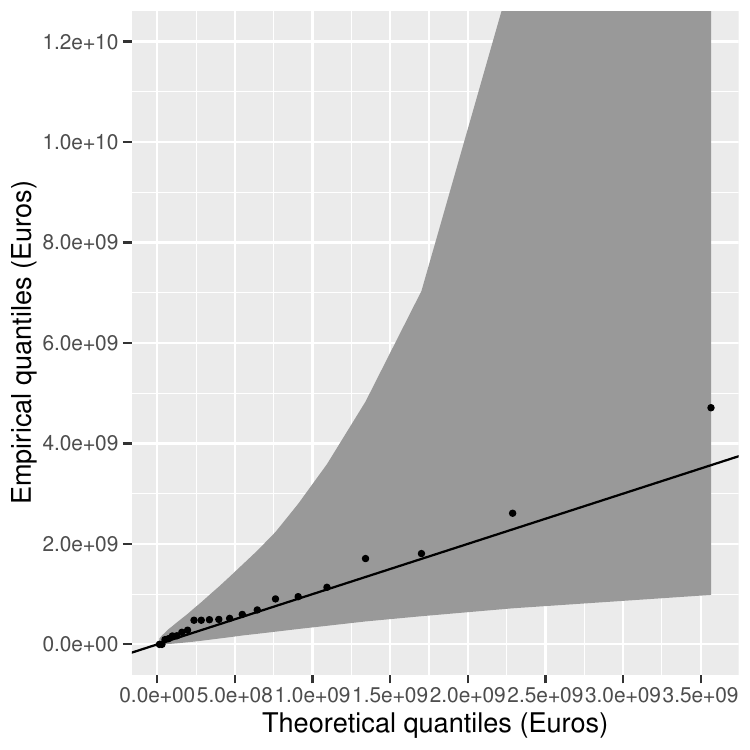}
        \end{subfigure}
        \hfill
        \begin{subfigure}[b]{0.32\textwidth}  
            \centering 
            \includegraphics[width=\textwidth]{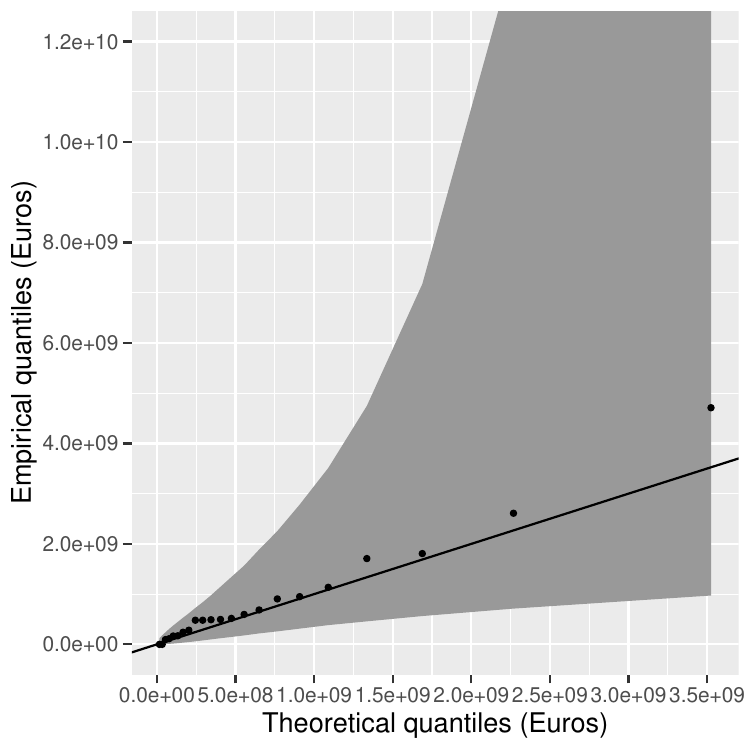}
        \end{subfigure}
                \hfill
        \begin{subfigure}[b]{0.32\textwidth}  
            \centering 
            \includegraphics[width=\textwidth]{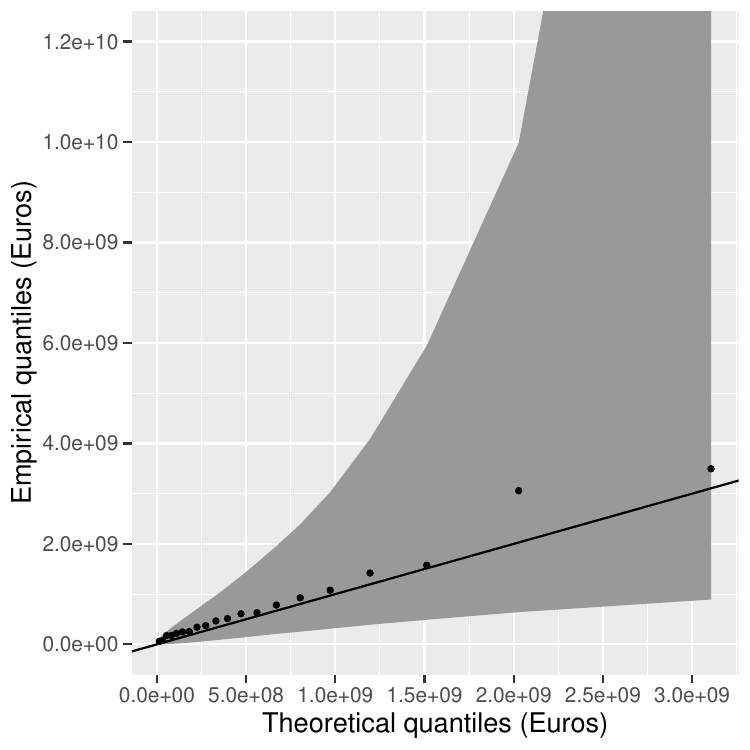}
        \end{subfigure}
               
        \caption{Q-Q plot of the observed versus simulated yearly trigger values obtained from the fitted extremal-$t$ random field using our trigger (left), the fitted Brown--Resnick random field using our trigger (center), and the univariate model using the benchmark trigger (right). Overall envelopes at the 95\% confidence level are depicted in dark gray.}
\label{fig:QQplotTrig}
\end{figure}

We thus obtain a sample of cost field realizations $\{C_m\left(\bm{x}\right), \mbox{ } m=1,\ldots,M\}_{\bm{x}\in A}$, from which we construct a sample of yearly losses $\{S_j, \mbox{ }j=1,\ldots,J\}$ by summing the losses from all simulated events within each simulated year as:
\begin{equation*}
S_j=\sum_{m=M_{j-1}+1}^{M_j}\sum_{k=1}^K C_m\left(\bm{x}_k\right),
\end{equation*}
where $M_j=\sum_{j'=1}^j N_{j'}$, $M_0=0,$ $N_j$ is the number of simulated events in year $j=1,\ldots,J,$ and 
$\sum_{j=1}^J N_j=M,$
that is, we allocate to each year the simulated events in increasing order
$$\{\underbrace{C_1(\bm{x}), \ldots,C_{N_1}(\bm{x})}_\text{year 1},\underbrace{C_{N_1+1}(\bm{x}),\ldots,C_{N_1 +N_2}(\bm{x})}_\text{year 2},\mbox{ }\ldots\mbox{ },\underbrace{C_{M_{J-1}+1}(\bm{x}),\ldots,C_M(\bm{x})}_\text{year $J$}\}.$$
We then compare the empirical distribution of the yearly losses as given by our simulations $S_j$ with the yearly trigger realizations. The latter are obtained by evaluating the annual loss trigger at \eqref{eq:lossregion} for the 100 events that occurred during the 22 years between 1999 and 2020.
As a benchmark, we also evaluate the trigger $L_i^3$ defined in Section \ref{sec:vulnerability}.
We show both comparisons by means of Q-Q plots in Figure \ref{fig:QQplotTrig}.

The theoretical quantiles on the $x$-axis in Figure \ref{fig:QQplotTrig} correspond to simulated yearly trigger values, whereas the empirical quantiles on the $y$-axis correspond to yearly trigger realizations. The confidence bounds in Figure \ref{fig:QQplotTrig} are obtained using the \texttt{envelope} function from the {\sf R} package~\texttt{boot} which computes the bounds using a bootstrap procedure. These confidence bounds describe the variability around the quantiles if the chosen model is the right one. Note that both our trigger (left) and the benchmark trigger (right) underestimate losses in the right tail of the distribution which would lead to investors not being adequately compensated for the risk they are taking. However, for the rest of the distribution, the Q-Q plots yield satisfactory results and this kind of model error when extrapolating into the far end of the distribution is not uncommon. 
Indeed, due to the small sample size, the confidence bounds are so large that the empirical quantiles still fall within the 95\% confidence interval. 
These results are compared with simulations from the fitted Brown--Resnick random field which produce a slightly stronger underestimation in the tail for our trigger, see Figure \ref{fig:QQplotTrig} (center). 

In this analysis, we are only evaluating the error from the wind speed model and from the frequency model. Indeed, since our trigger is by design parametric, we do not need to be penalized from the error on the vulnerability function. However, one can go a step further and look at the case of indemnity triggers by comparing the yearly loss simulations $S_j$ with the observed industry losses $\ell_i^{ins}$ that are aggregated per year to produce observed yearly industry losses.

From the 20 windstorm events between 1999 and 2020, we build yearly observed industry losses by aggregating the adjusted losses from Table \ref{event_names} for each year between 1999 and 2020. Note that during the years 2000, 2001, 2003, 2004, 2005, 2006, 2008, 2009, 2014 and 2016, no qualifying events (above an industry insured loss of EUR 200mn) were reported by PERILS and thus we obtain only 12 years with non-zero annual aggregate industry loss between 1999 and 2020. As trigger realizations were simulated based on 2020 industry exposure data, we perform an exposure adjustment to have losses on the same exposure basis (year 2020). This exposure adjustment is performed per CRESTA zone for the years 2010 to 2020 (with the exposure data extrapolated backwards for the period 2010 to 2012), and on an aggregate level (sum over all CRESTA zones) for the years 1999 to 2007 as the insured loss data is given per country and not per CRESTA zone for the years before 2010.

In order to provide a fair comparison with the censoring procedure employed by PERILS, we select a threshold under which we manually set simulated event losses to zero.
In Table \ref{prop_zeros}, we show the proportions of zeros obtained for various selected thresholds between EUR 50mn and EUR 200mn.
These proportions can be compared with the number of years for which PERILS did not report any event loss over EUR 200mn over the time period between 1999 and 2020, which yields a proportion of 45.5\% (10 out of 22 years).
\begin{table}
\begin{center}
\begin{tabular}{c|c}
Threshold & Proportion \\
\hline
50  & 18.1\% \\ 
100 & 30.9\% \\ 
150 & 40.3\% \\ 
200 & 47.7\%  \\ 
\end{tabular}
\end{center}
\caption{Proportion of zeros for various selected thresholds (in mn EUR).}
\label{prop_zeros}
\end{table}

\begin{figure*}
    \centering
    \begin{subfigure}[b]{0.48\textwidth}
        \centering
       \includegraphics[width=\textwidth]{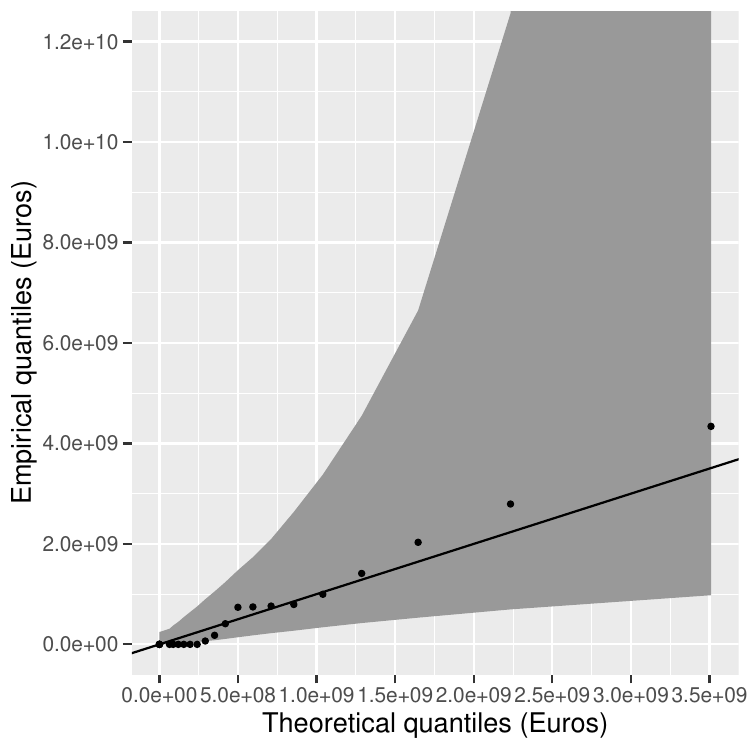}
        \end{subfigure}
        \hfill
        \begin{subfigure}[b]{0.48\textwidth}  
            \centering 
            \includegraphics[width=\textwidth]{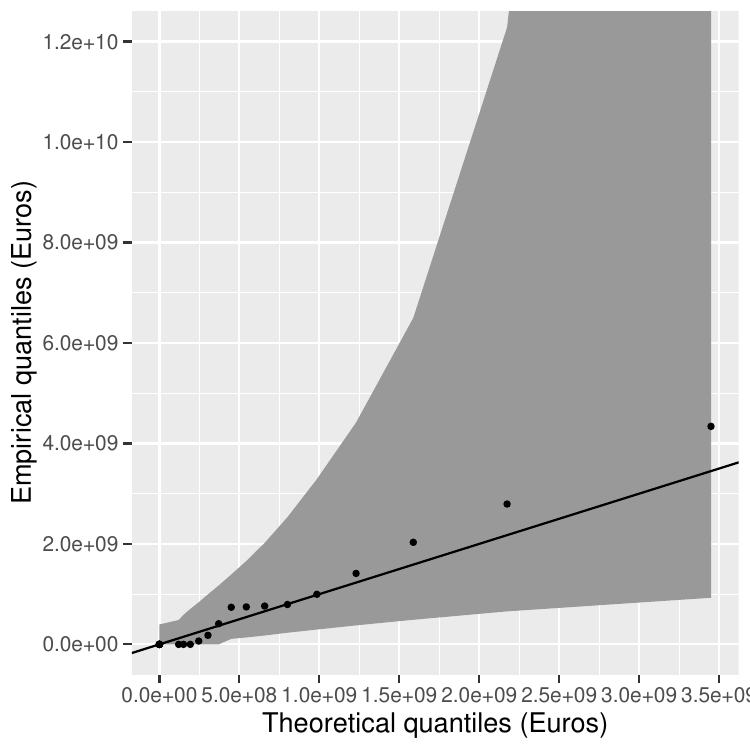}
        \end{subfigure}
        \hfill
        \vskip\baselineskip
        \vskip\baselineskip
        \begin{subfigure}[b]{0.48\textwidth}   
            \centering 
            \includegraphics[width=\textwidth]{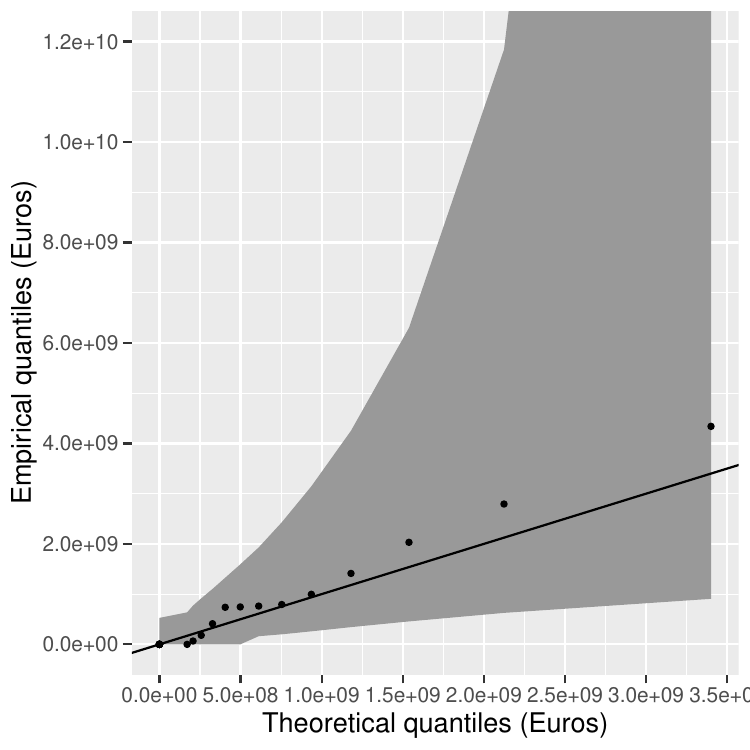}
        \end{subfigure}
        \hfill
        \begin{subfigure}[b]{0.48\textwidth}   
            \centering 
            \includegraphics[width=\textwidth]{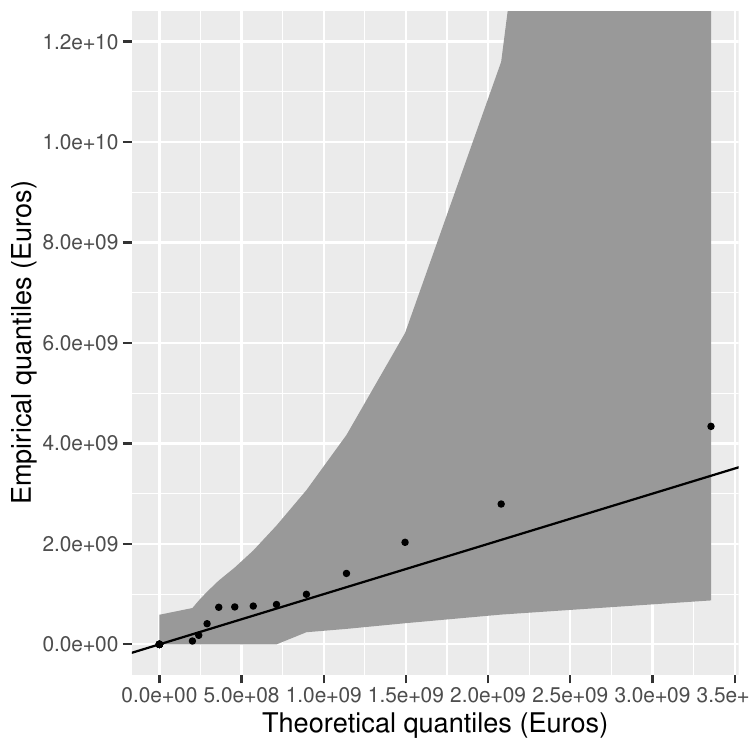}
        \end{subfigure}
        \hfill
        \caption{Q-Q plot of the observed versus conditional simulated losses for various thresholds: EUR 50mn (top left), EUR 100mn (top right), EUR 150mn (bottom left), EUR 200mn (bottom right). Overall envelopes at the 95\% confidence level are depicted in dark gray.}
            \label{fig:Fig_GoodnessFitFull}
    \end{figure*}

In Figure \ref{fig:Fig_GoodnessFitFull}, we display Q-Q plots for non-zero conditioned simulated values versus the PERILS observations, for various threshold levels between EUR 50mn and EUR 200mn. Increasing the threshold reduces the proportion of zeros, see Table \ref{prop_zeros}, but also leads to a minor deterioration in the Q-Q plot since it slightly increases underestimation of the yearly industry loss in the upper tail, see Figure \ref{fig:Fig_GoodnessFitFull}.
Note that owing to the lack of data, we were not able to perform an out-of-sample validation. 

The extremal-$t$ model with Cauchy correlation in combination with our trigger definition thus produce the best results when it comes to the full model assessment. Indeed, one has to jointly consider the basis risk implied by the trigger choice, see Figures \ref{fig:perils} and \ref{fig:alt_trigger}, and the quality of the model for the wind speeds, see Figure \ref{fig:QQplotTrig}. Considering both these elements together, we believe our trigger proposal alongside the extremal-$t$ model with Cauchy correlation for the wind speeds produce better results compared to the benchmark trigger derived from the univariate model. 

\subsection{Numerical results}\label{sec:results}
In this section, we present numerical results for the pricing of cat bonds based on our trigger proposal and the benchmark trigger. Using our trigger, we will also assess how bond prices vary when changing the model for the wind speeds.
From the sample of $J=2,000,000$ simulations of the one-year loss $S$, we can either empirically evaluate the attachment and exhaustion probabilities or fit a generalized Pareto distribution (GPD) to loss exceedances above a certain threshold which is chosen via a combination of graphical tests, i.e., a visual inspection of the mean excess plot and the stability plot for the GPD parameters. We explore the former option since the sample is large enough to ensure that empirical quantiles are close enough to their true counterparts. 

We present a pricing example of our cat bond with maturity $T=1$ year on two separate tranches, where the attachment points of these tranches are chosen to match various return periods of the one-year loss under the extremal-$t$ model with Cauchy correlation. 
Prices will be expressed per FV=100\euro{} of face value, using the formulation at \eqref{eq:priceTdev}. We consider a spread $s=7\%$ 
and the following parameters for the short rate model:
$\alpha=20\%,$ $\mu=2\%,$ $\sigma=10\%,$ and $r_0=1\%.$
We consider the same exhaustion point $u_e=11.008\cdot 10^{9}$ (modeled 200 year return period) for both tranches of the bond. For tranche A, we set $u_a=7.861\cdot 10^{9}$ (modeled 100 year return period) and for tranche B, we set $u_a=9.593\cdot 10^{9}$ (modeled 150 year return period). The average annual loss (that is, the expected loss in any given year) according to our model is $0.891\cdot 10^9$ (around 0.9 billion \euro{}). Cat bond layers typically sit at the top of reinsurance towers and solvency considerations require purchasing reinsurance up to the modeled 200 year return period, i.e., the 99.5\% value-at-risk. These examples allow to examine the behaviour of the different models further in the tail, as opposed to the Q-Q plots in Figure \ref{fig:QQplotTrig} which are based on a subdivision of the interval $[0,1]$ in 22 data points, and hence the largest quantile considered is the 95.65\% quantile.

\begin{figure}[!ht]
\begin{center}
\includegraphics[
width=0.6\textwidth]{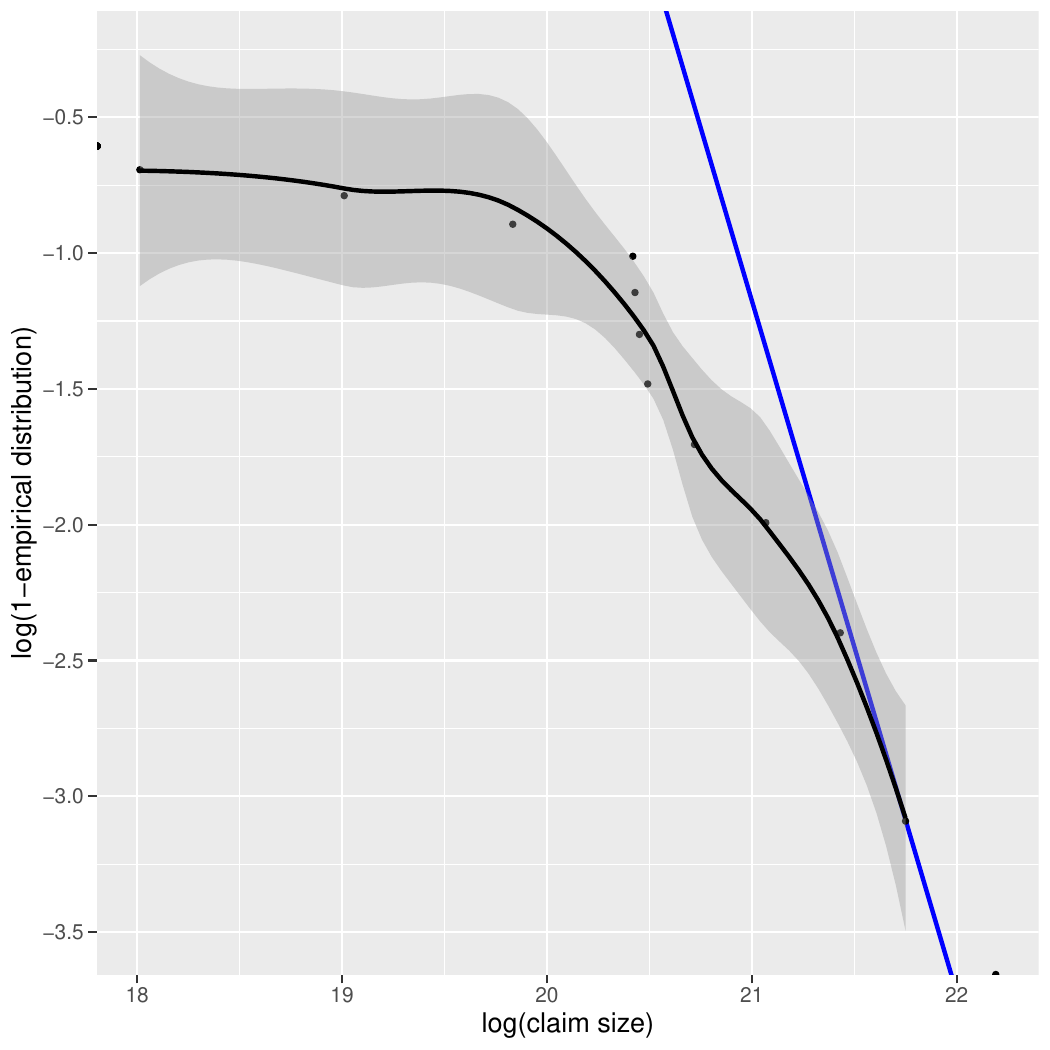}
\caption{Empirical log-log plot for the sample of observed yearly losses with local regression line (in black) and line with slope $-2.55$ (in blue).}\label{fig:logalpha}
\end{center}
\end{figure}

A comparison of the largest historical (adjusted) loss over the sample given by $4.340\cdot 10^9$ (with a corresponding return period of 22 years) with the simulated 100 year return period loss of $7.861\cdot 10^9$ (attachment point for tranche A) suggests a Pareto-$\alpha$ of 2.55 (rule of thumb derived from the quantiles of a Pareto($\alpha$) distribution). In Figure \ref{fig:logalpha}, we verify this property for the empirical sample of yearly losses. Due to the small sample size, there is high uncertainty in the estimation of the tail parameter $\alpha.$ 

For tranche A, we plot the evolution of the bond price for various values of the parameter $\lambda$ between 0.25 and 1.25 with steps of 0.05. Recall that $\lambda$ corresponds to the catastrophe risk premium. We select $\lambda$ such that the market price of tranche A is equal to 100. This results in $\lambda=0.737.$ 
This parameter is tailored to the specification of tranche A. In \cite{tang_yuan_2019}, this value is 1.23 based on a case study on Californian earthquakes between 1769 and 2000 using the earthquake magnitude as the variable describing the hazard and a generalized Pareto distribution to model threshold exceedances, whereas in \cite{galeotti}, this value is 0.7 based on a regression of historical spreads on observable variables using all cat bonds issued between 1999 and mid 2009.
\begin{figure}[!ht]
\begin{center}
\includegraphics[
width=0.6\textwidth]{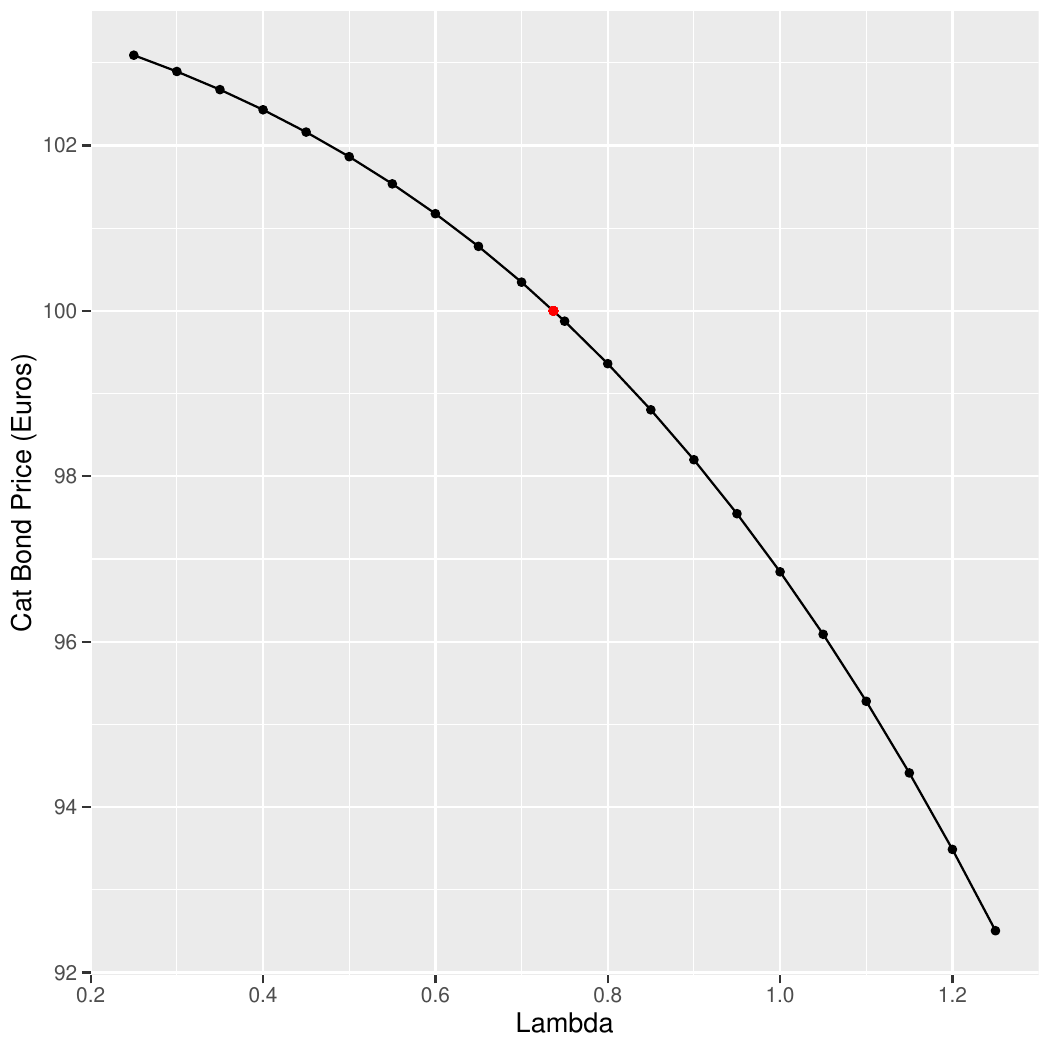}
\caption{Evolution of the price of the cat bond for various values of $\lambda$ (catastrophe risk premium) between 0.25 and 1.25.}\label{fig:kappa}
\end{center}
\end{figure}
We obtain a price for tranche A of 100.00\euro{} and for tranche B of 100.69\euro{} (in other words, the spread for tranche B has to be set lower to obtain a market price of 100\euro{} - we compute this spread to be 6.28\%). As expected, the price for tranche B is higher as it is less likely to be triggered. 
In actual cat bond transactions, the spread is determined as a multiple of the expected loss. This multiple typically depends on qualitative factors such as the state of the market (capital supply and demand mechanics), the sponsor's track record (sponsors with a longer presence in the market and with robust underwriting practices could be rewarded by paying a lower spread), the region/peril covered (European windstorm bonds tend to have lower spreads due to their diversifying profile compared to US wind or US earthquake bonds) and the coverage basis (bonds on an aggregate basis tend to pay higher spreads due to stronger modeling uncertainty for perils such as severe thunderstorms or wildfire). The value chosen for the spread is in line with the spread for outstanding cat bonds covering European windstorm, see \url{https://www.artemis.bm}. 

Keeping all pricing parameters equal, we compare these results with those obtained with the fitted Brown--Resnick random field.
We obtain a price for tranche A of 100.11\euro{} and for tranche B of 100.79\euro{}. The Brown--Resnick random field produces slightly lower losses as the corresponding return periods of interest are lower:
the modeled 100 year return period loss is $7.737\cdot 10^{9}$, the modeled 150 year return period loss is $9.428\cdot 10^9,$ and the modeled 200 year return period loss is $10.809\cdot 10^9$ and hence the prices for both tranches are higher than with the extremal-$t$ model with Cauchy correlation. 
The corresponding catastrophe risk premium in the Brown--Resnick model is $\lambda=0.749.$

The univariate model for the maxima over all locations, see Section \ref{sec:calibevents}, produces a price for tranche A of 101.26\euro{} and for tranche B of 101.87\euro{}, which shows that this model produces even lower losses than the Brown--Resnick model in the tail of the yearly loss distribution.

\begin{figure}[!ht]
\begin{center}
\includegraphics[
width=0.6\textwidth]{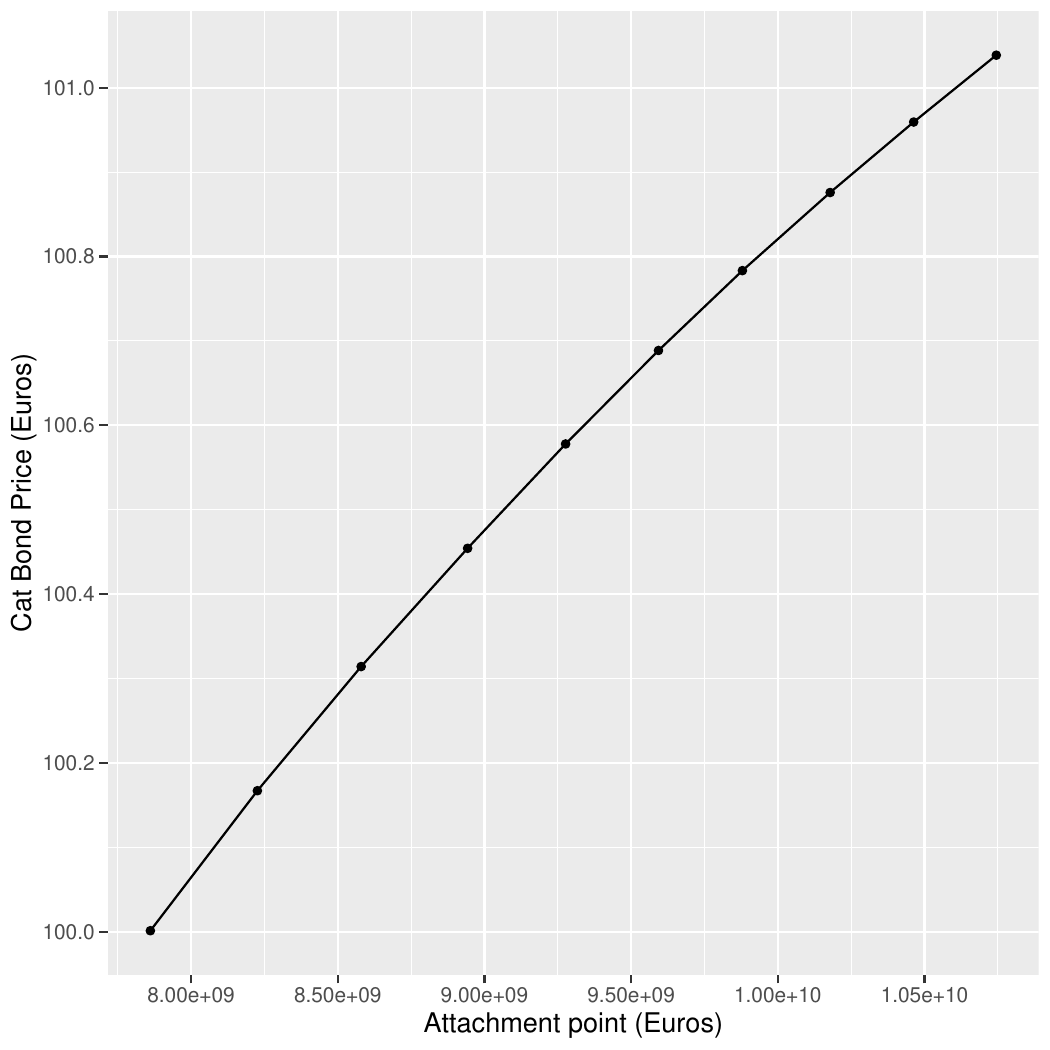}
\caption{Evolution of the price of the cat bond with increasing attachment points.}\label{fig:attach}
\end{center}
\end{figure}

Keeping the exhaustion point $u_e$ and all other parameters fixed, we plot the price of the bond for increasing attachment points corresponding to increasing return periods between 100 and 190 years with steps of 10 years in Figure \ref{fig:attach}. We observe a concave behavior of the price which can be explained by the fact that the higher the attachment point, the smaller the gap with the exhaustion point.

Note that $u_a$ and $u_e$ were chosen based on return periods of losses over the whole German territory, for all lines of business (residential, commercial and industrial) and for all types of coverage (buildings, contents and business interruption). This is essentially an aggregate industry loss warranty (ILW) cover. For industry loss index triggers, the sponsor would typically specify payout factors for CRESTA zone, peril and line of business depending on their market share in that region for that specific peril and line of business. For indemnity triggers, the sponsor could simply specify its exposure in each CRESTA zone. 

\section{Conclusion}\label{sec:conclu}
Cat bonds are often considered as the most popular product in the family of ILS, due to their historical track record and liquidity in the secondary market. The choice of the trigger when issuing a cat bond involves striking a balance between transparency and basis risk. Indemnity triggers are generally favored by sponsors but they carry moral hazard and come with additional duties such as reporting and reputational risk, whereas parametric triggers are usually preferred by investors for their transparency, at the expense of introducing basis risk on the side of the sponsor. We propose a flexible trigger and a statistical loss model displaying a high level of transparency due to our modeling assumptions. As a result, we obtain a blended parametric-modeled loss trigger which can be easily evaluated and used for pricing not only existing cat bonds, but also to build innovative covers for regions where there are no historical loss data and/or no available or reliable cat models. 
As a complementary tool to standard frequency-severity models from the actuarial literature or vendor cat models from modeling agencies, we propose a framework inspired by the cost field laid out in \cite{koch17, koch2024correlation}, and depending only on observable environmental variables and industry exposure. In doing so we bring tools from the theory of spatial extremes to model catastrophe risk. 
We believe this new trigger and pricing methodology could increase the attractiveness of the cat bond product offering and convince more investors and sponsors to enter this asset class. Going further, we could use $r$-Pareto processes as an alternative to max-stable random fields. 
One could also study the future evolution of cat bond prices by considering forecasts for the socio-economical variables appearing in the exposure component of the cost field, or by devising stress scenarios on the physical hazard to take into account the impact of climate change. 

\newpage
\section*{Acknowledgements}{Both authors kindly thank Prof. Patrick Cheridito, Prof. Mario V. Wüthrich, Dr. Philipp Arbenz, Dr. Robert Salzmann and Dr. Maurizio Savina for their insightful comments and careful readings of previous versions of the manuscript. We kindly thank PERILS AG for providing us with wind speed, exposure and insured loss data for German historical windstorm events, and GfK GeoMarketing for providing us with shape file data for the German CRESTA zones.
\\As SCOR Fellow, John Ery thanks SCOR for financial support. Erwan Koch would like to thank the Expertise Center for Climate Extremes (ECCE) at UNIL for financial support.}

\bibliographystyle{apalike}
\bibliography{bibl}

\newpage
\appendix
In this appendix we provide supplementary material on the calibration on events and the calibration on monthly maxima.

\section{Calibration on events}
\subsection{Windstorm model calibration and validation}\label{sec:calibevents_apdx}
\begin{figure}[!ht]
\begin{center}
\includegraphics[scale=0.7]{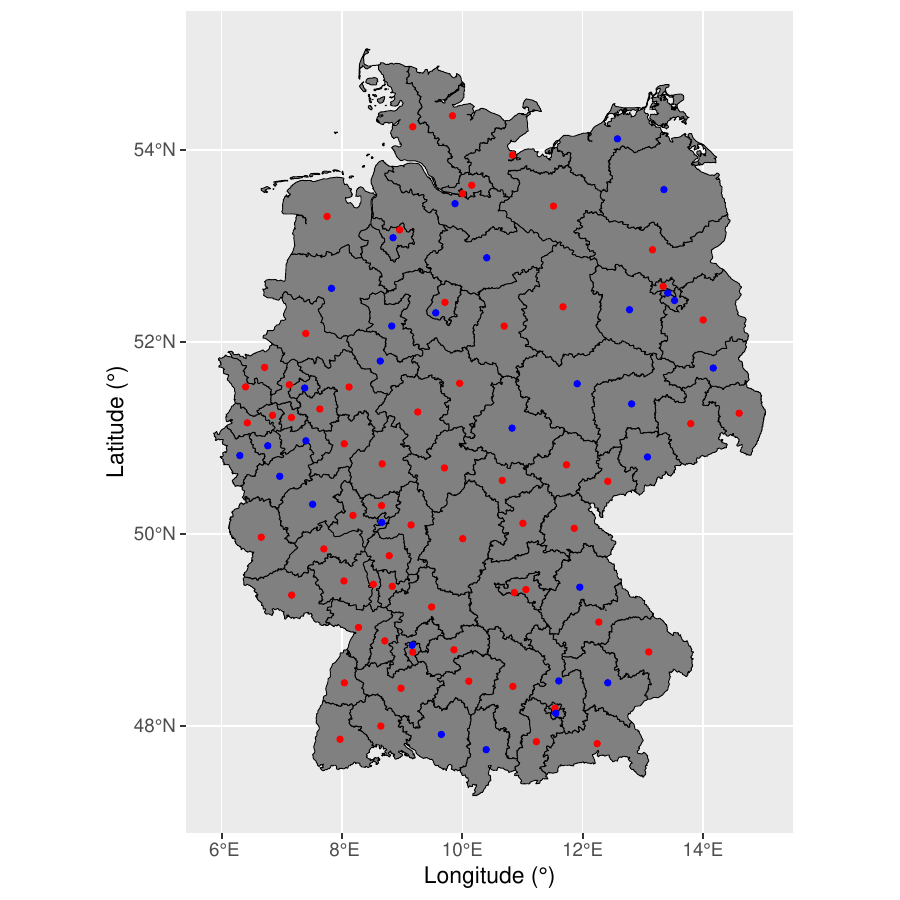}
\caption{Map showing the 95 CRESTA zone centroids used for our analysis. We consider 64 centroids for calibration (red points) and 31 for validation (blue points).}\label{fig:mapStations}
\end{center}
\end{figure}

In this section, we discuss the calibration and validation of our model to historical events under the CRESTA zone representation using trend surfaces for the marginals.
We consider a split of the CRESTA zone centroids into two groups: we randomly choose 64 centroids for calibration and the remaining 31 are used for validation. 
We display the map of the CRESTA zones in Figure \ref{fig:mapStations}, where the red points correspond to calibration centroids and the blue points to validation centroids. 

\begin{figure}[!ht]
    \centering
    \begin{subfigure}[b]{0.32\textwidth}
        \centering
       \includegraphics[width=\textwidth]{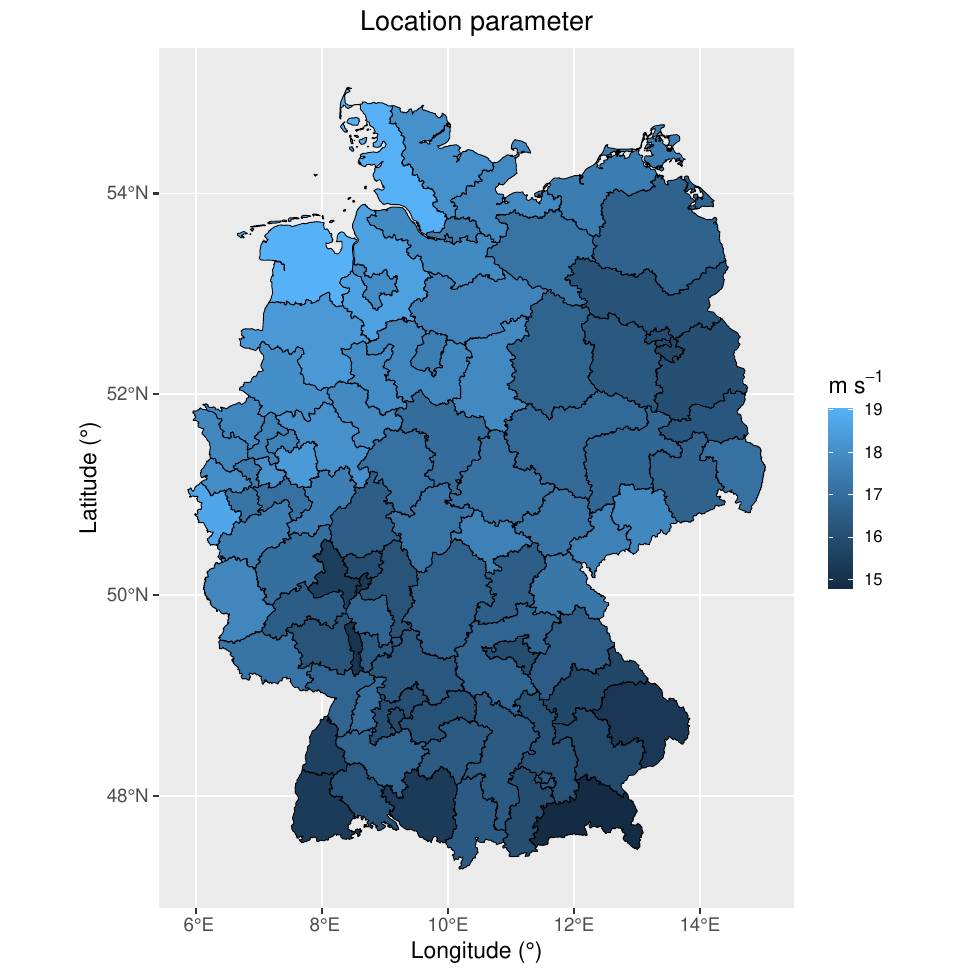}
        \end{subfigure}
        \hfill
        \begin{subfigure}[b]{0.32\textwidth}  
            \centering 
            \includegraphics[width=\textwidth]{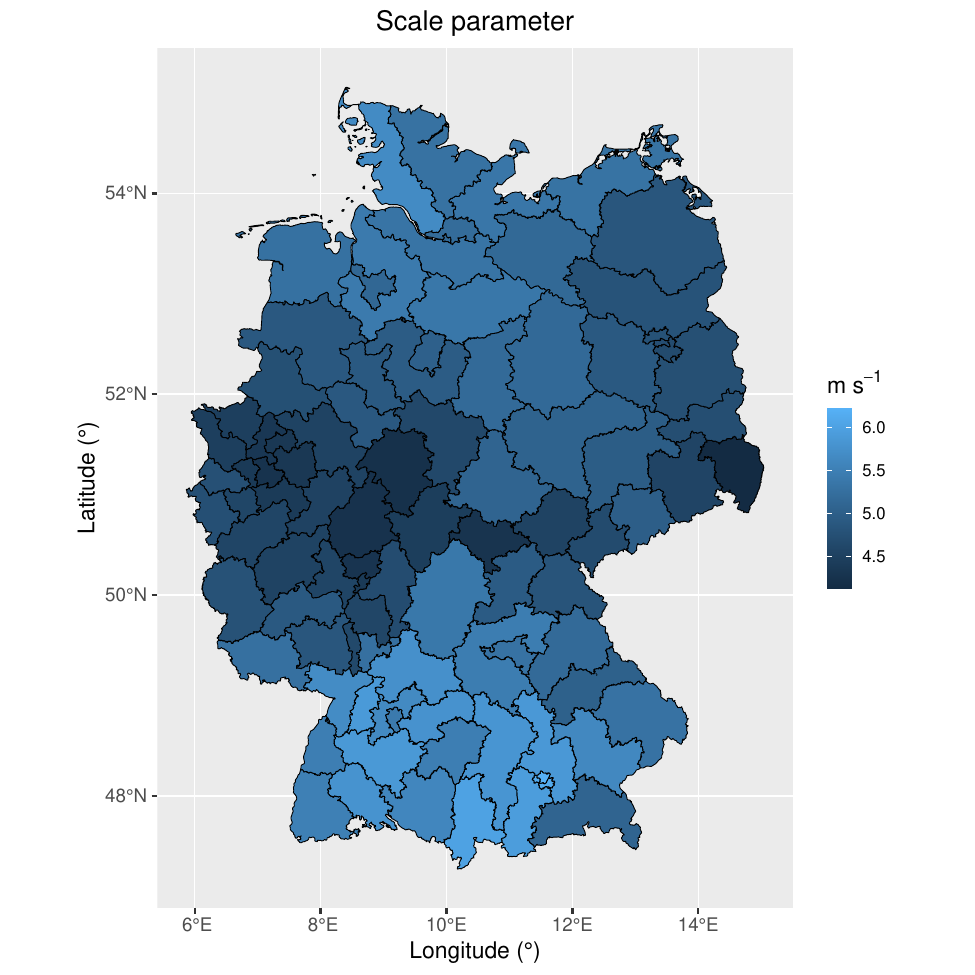}
        \end{subfigure}
        \hfill
        \begin{subfigure}[b]{0.32\textwidth}  
            \centering 
            \includegraphics[width=\textwidth]{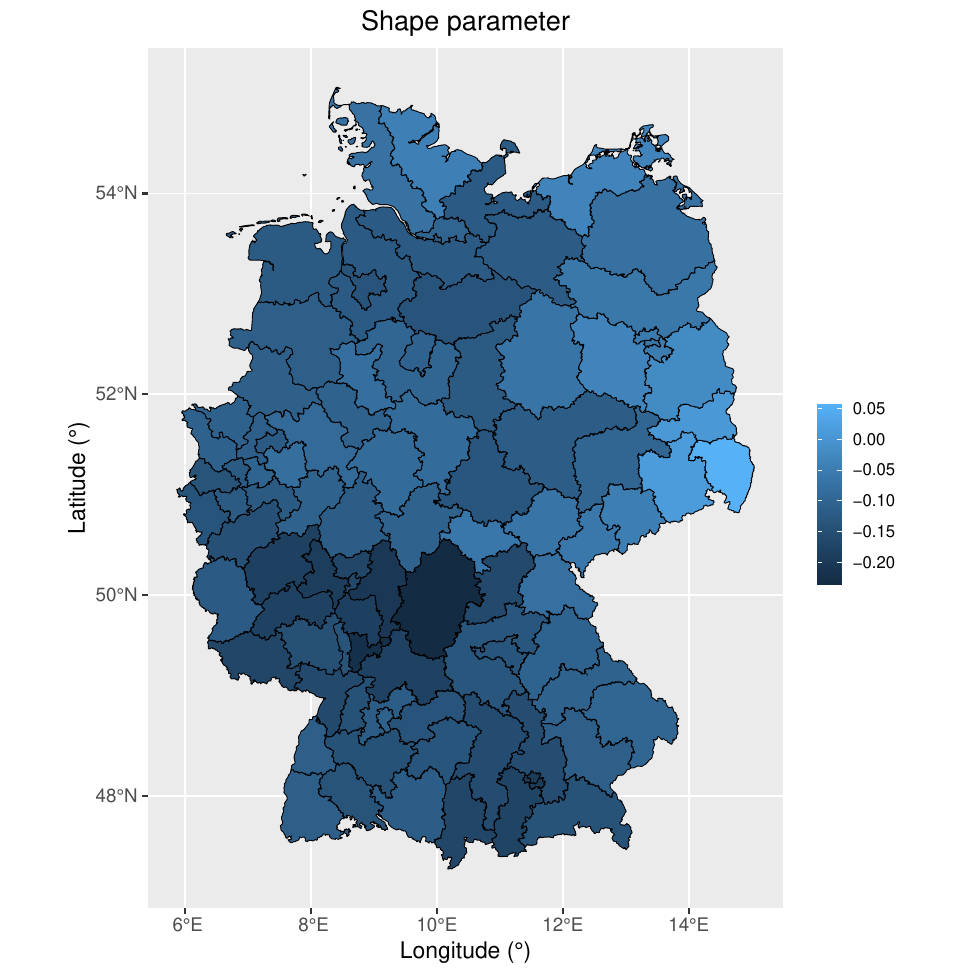}
        \end{subfigure}
        \caption{Map of the fitted GEV parameters for the 95 CRESTA zones.}
\label{fig:mapsGEV}
\end{figure}

Using trend surfaces for the GEV parameters rather than fitting them separately at each site as in Section \ref{sec:calibevents} is fairly common as it reduces parameter uncertainty, allows a joint estimation of all marginal and dependence parameters in a reasonable amount of time (the optimization is based 
on the pairwise likelihood method which is very time-consuming, see Section \ref{sec:fitmaxstable} in the Appendix) and
enables prediction at sites where no observations are available.
For the GEV parameters, we choose the following trend surfaces:
\begin{align*}
\eta(\bm{x}) &= \eta_0 + \eta_1  \mathrm{lon}(\bm{x}) + \eta_2 \mathrm{lat}(\bm{x}) + \eta_3 \mathrm{alt}(\bm{x}), \\
\tau(\bm{x}) &= \tau_0 + \tau_1 \vert \mathrm{lat}(\bm{x}) - 51 \vert, \\
\xi(\bm{x}) &= \xi_0,
\end{align*}
where $\mathrm{lon}(\bm{x})$, $\mathrm{lat}(\bm{x})$ and $\mathrm{alt}(\bm{x})$ denote the longitude, latitude and altitude of location $\bm{x}$, respectively. The presence of longitude and latitude in the location parameter $\eta$ and the form of the scale parameter $\tau$ are natural when looking at the maps of the GEV parameters at each location, see Figure \ref{fig:mapsGEV}. Indeed, in the middle plot of Figure \ref{fig:mapsGEV}, we observe that the scale parameter is minimal at latitude $51^\circ$ and increases north and south from this level.

Adding the altitude in $\eta$ will lead to an improvement in the fit. For the shape parameter $\xi$, we choose a constant, as is commonly done in the literature.
We fit the Smith, Schlather, Brown--Resnick, and extremal-$t$ random fields for the dependence structure on the sample of 100 windstorm events. In the case of the Schlather and extremal-$t$ models, we considered the powered exponential, Whittle--Matérn and Cauchy correlation functions. Parameter estimates are obtained using composite likelihood techniques, more precisely by maximizing the pairwise composite log-likelihood given at Equation (6) in \cite{padoan}, and model selection is done based on the composite likelihood information criterion (CLIC). We refer the reader to Section \ref{sec:fitmaxstable} in the Appendix for more details on the calibration procedure and the CLIC.
According to Table \ref{Table_CLIC_Values_event}, the extremal-$t$ model with Cauchy correlation function produced the lowest CLIC.
\begin{table}
\begin{center}
\begin{tabular}{l|c|c}
Model & CLIC & Number of model parameters \\
\hline
Extremal-$t$ Cauchy & 2,328,464 & 11 \\
Extremal-$t$ Whittle--Matérn & 2,328,778 & 11 \\
Extremal-$t$ powered exponential & 2,330,744 & 11 \\
Schlather Whittle--Matérn & 2,335,478 & 10 \\
Schlather powered exponential & 2,335,526 & 10 \\
Schlather Cauchy & 2,336,119 & 10 \\
Brown--Resnick & 2,337,706 & 9 \\
Smith & 2,341,105 & 10
\end{tabular}
\end{center}
\caption{CLIC values for the different fitted models (in ascending order).}
\label{Table_CLIC_Values_event}
\end{table}

\begin{figure*}
    \centering
    \begin{subfigure}[b]{0.32\textwidth}
        \centering
       \includegraphics[width=\textwidth]{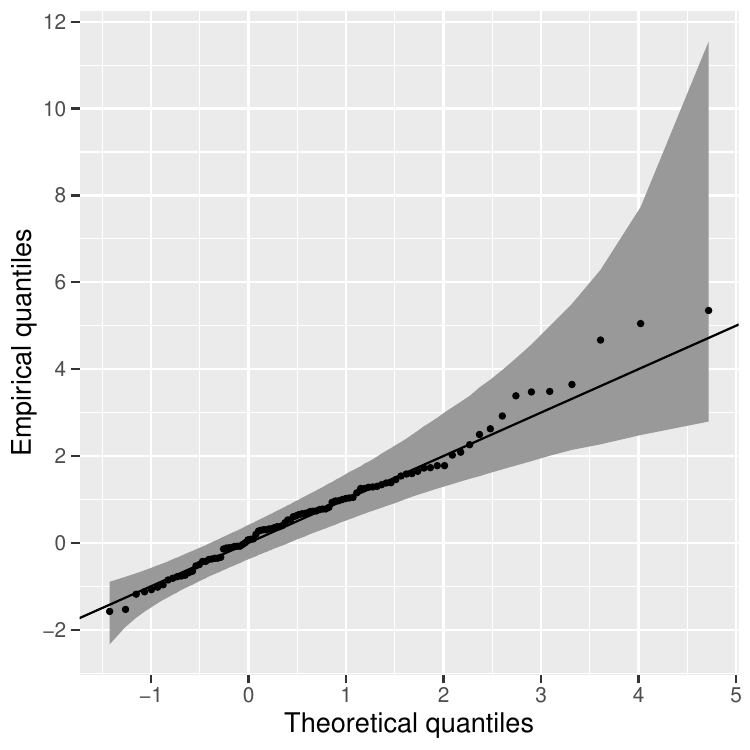}
        \end{subfigure}
        \hfill
        \begin{subfigure}[b]{0.32\textwidth}  
            \centering 
            \includegraphics[width=\textwidth]{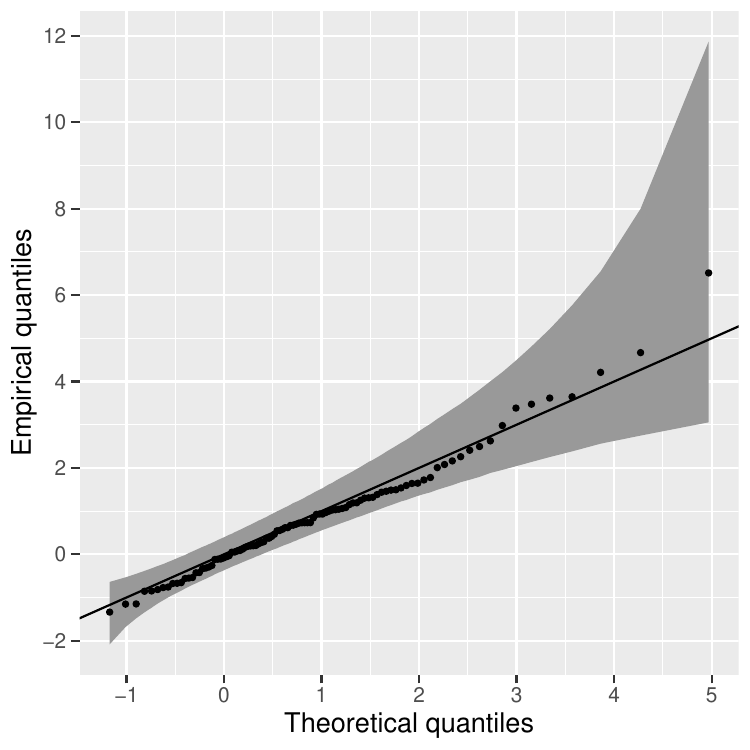}
        \end{subfigure}
        \hfill
        \begin{subfigure}[b]{0.32\textwidth}  
            \centering 
            \includegraphics[width=\textwidth]{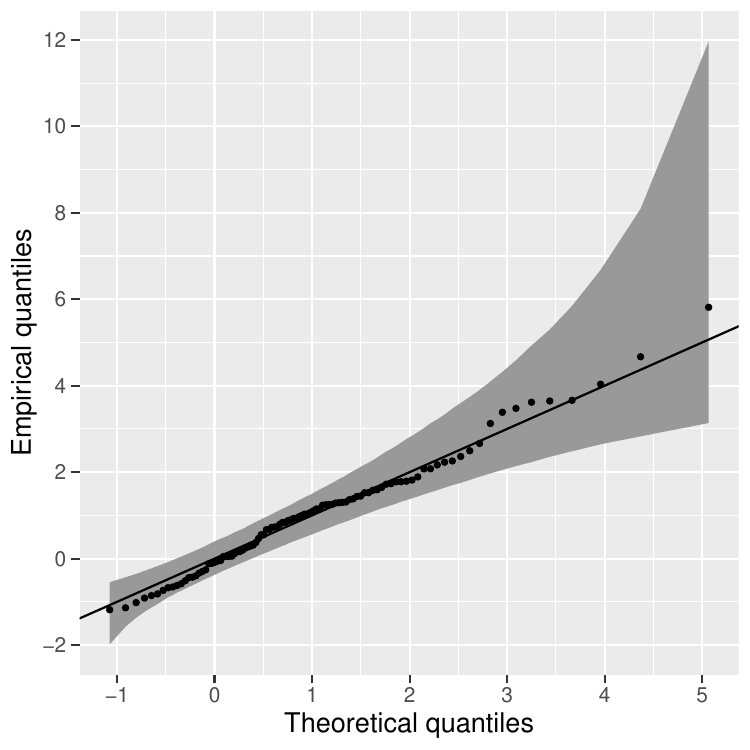}
        \end{subfigure}
        \vskip\baselineskip
        \begin{subfigure}[b]{0.32\textwidth}   
            \centering 
            \includegraphics[width=\textwidth]{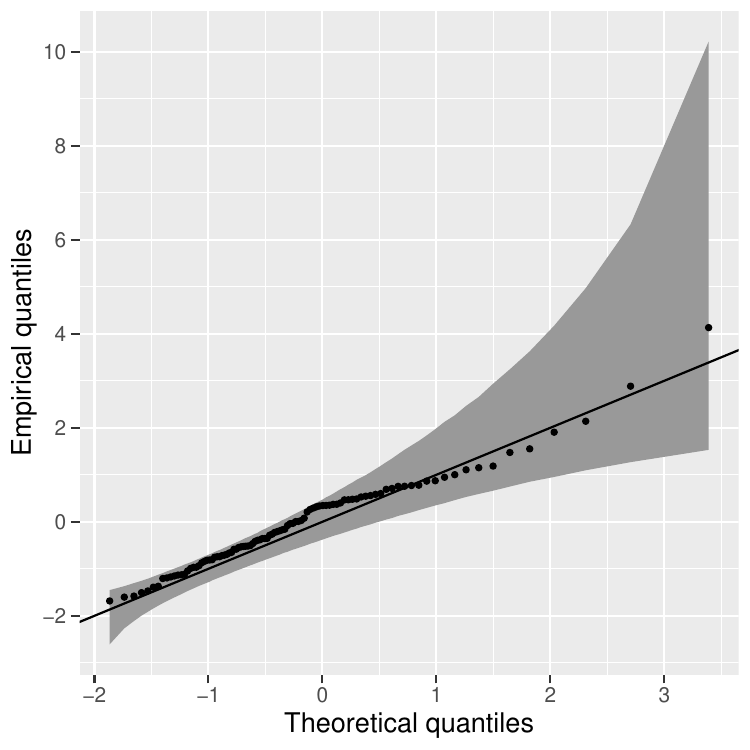}
        \end{subfigure}
        \hfill
        \begin{subfigure}[b]{0.32\textwidth}   
            \centering 
            \includegraphics[width=\textwidth]{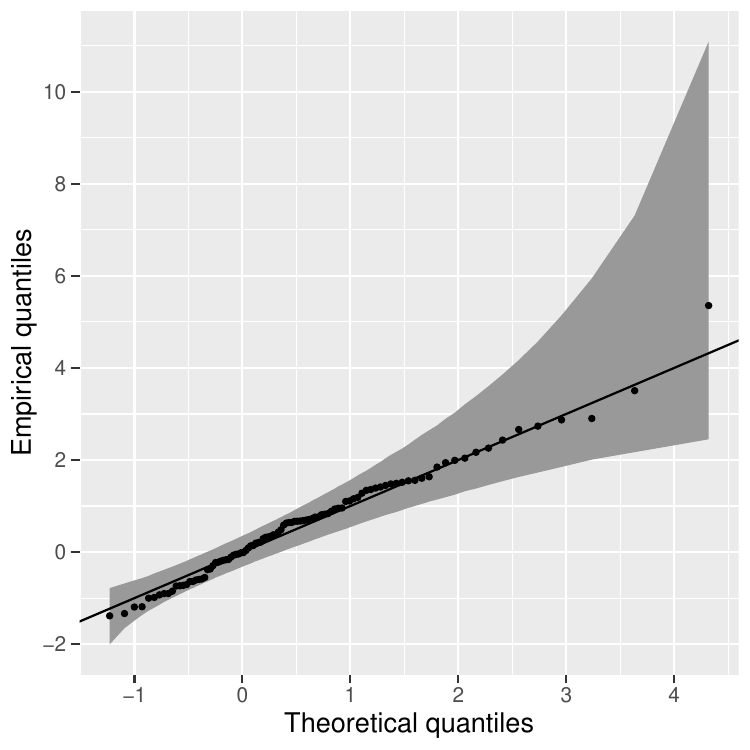}
        \end{subfigure}
        \hfill
        \begin{subfigure}[b]{0.32\textwidth}   
            \centering 
            \includegraphics[width=\textwidth]{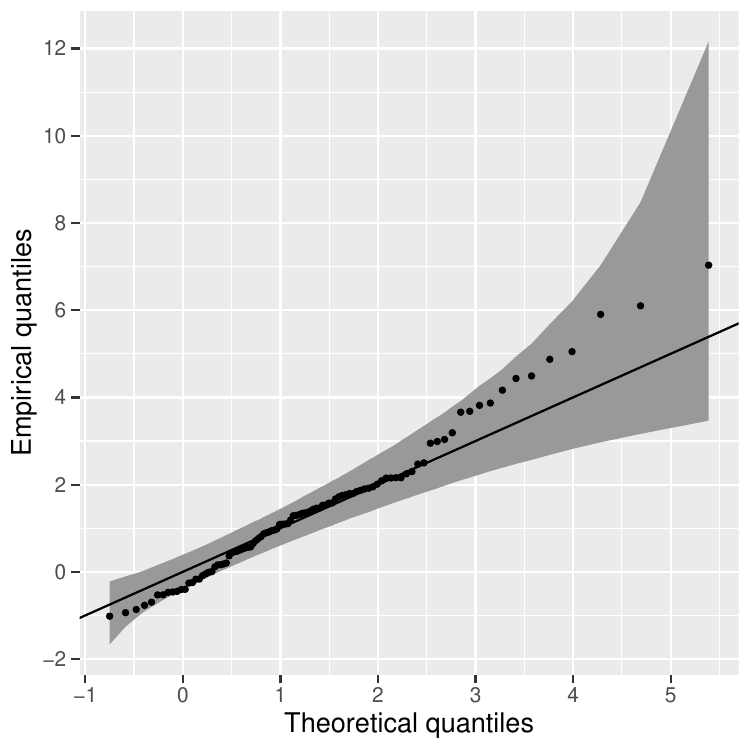}
        \end{subfigure}
        \vskip\baselineskip
        \begin{subfigure}[b]{0.32\textwidth}   
            \centering 
            \includegraphics[width=\textwidth]{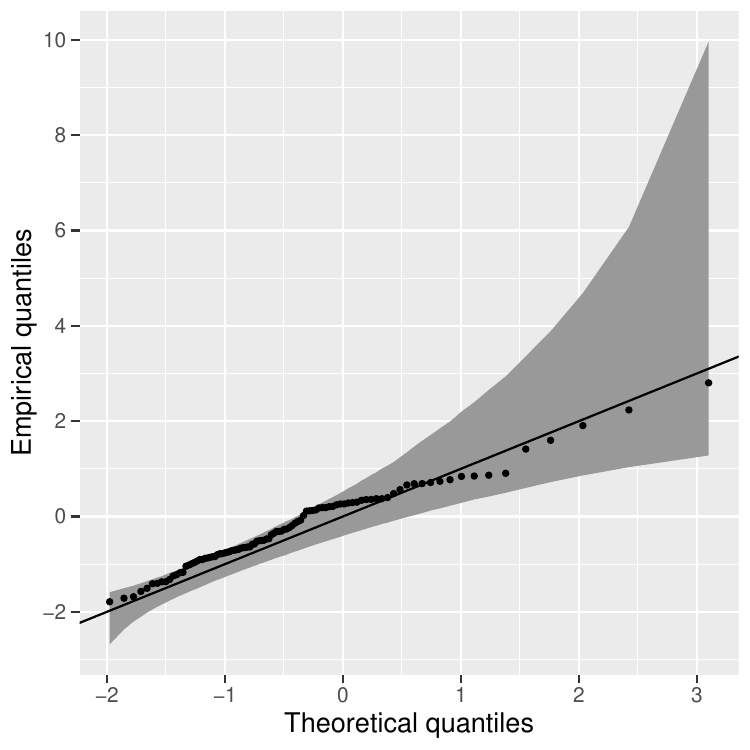}
        \end{subfigure}
        \hfill
        \begin{subfigure}[b]{0.32\textwidth}   
            \centering 
            \includegraphics[width=\textwidth]{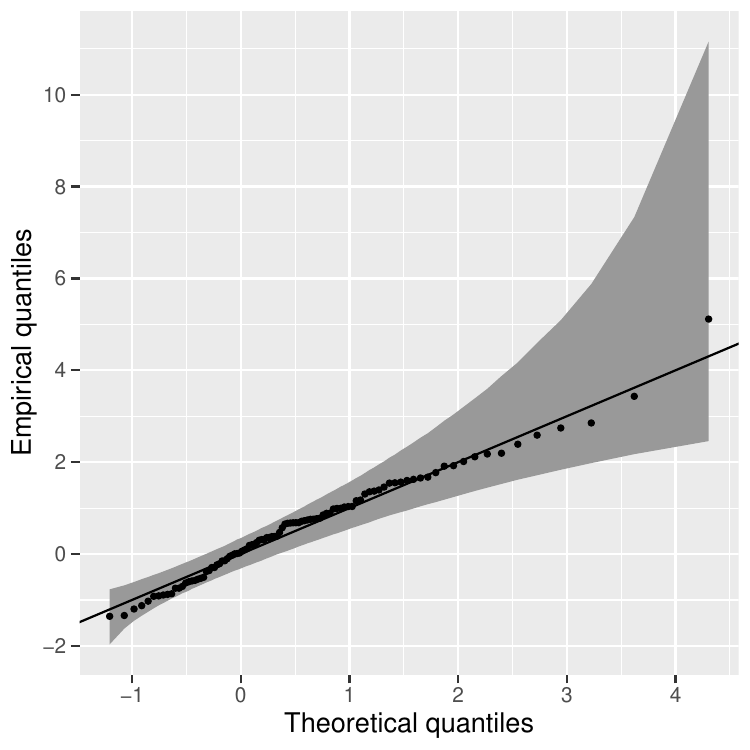}
        \end{subfigure}
        \hfill
        \begin{subfigure}[b]{0.32\textwidth}   
            \centering 
            \includegraphics[width=\textwidth]{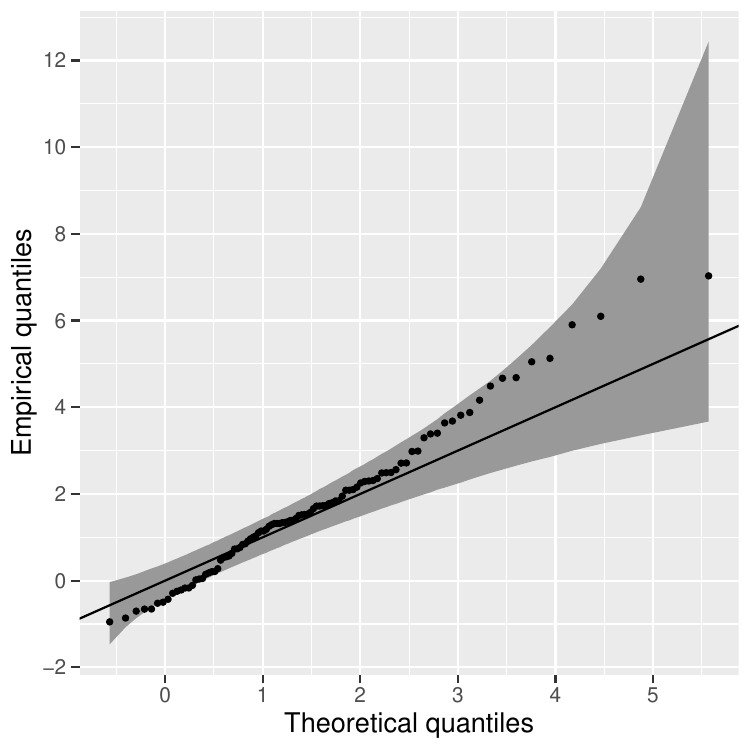}
        \end{subfigure}
        \caption{Performance of the chosen extremal-$t$ model with Cauchy correlation on the validation centroids. The top row concerns maxima for pairs of validation centroids separated by a low (left), moderate (middle) and a long (right) distance. The middle row focuses on minima (left), mean (middle) and maxima (right) for a group of $10$ validation centroids chosen randomly. The bottom row concerns  minima (left), mean (middle) and maxima (right) for all $31$ validation centroids. Overall envelopes at the 95\% confidence level are depicted in dark gray.}
        \label{fig:Fig_GoodnessFitQQPlotExtCauchyEve}
    \end{figure*}

\begin{figure}[!ht]
    \centering
    \begin{subfigure}[b]{0.48\textwidth}
        \centering
       \includegraphics[width=\textwidth]{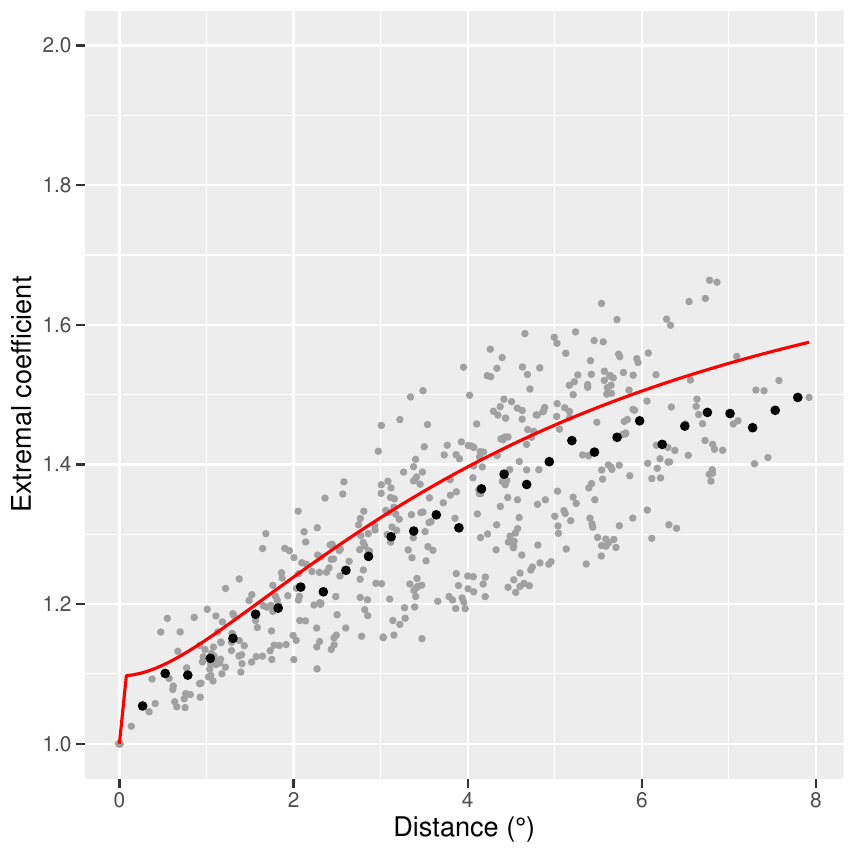}
        \end{subfigure}
        \hfill
        \begin{subfigure}[b]{0.48\textwidth}  
            \centering 
            \includegraphics[width=\textwidth]{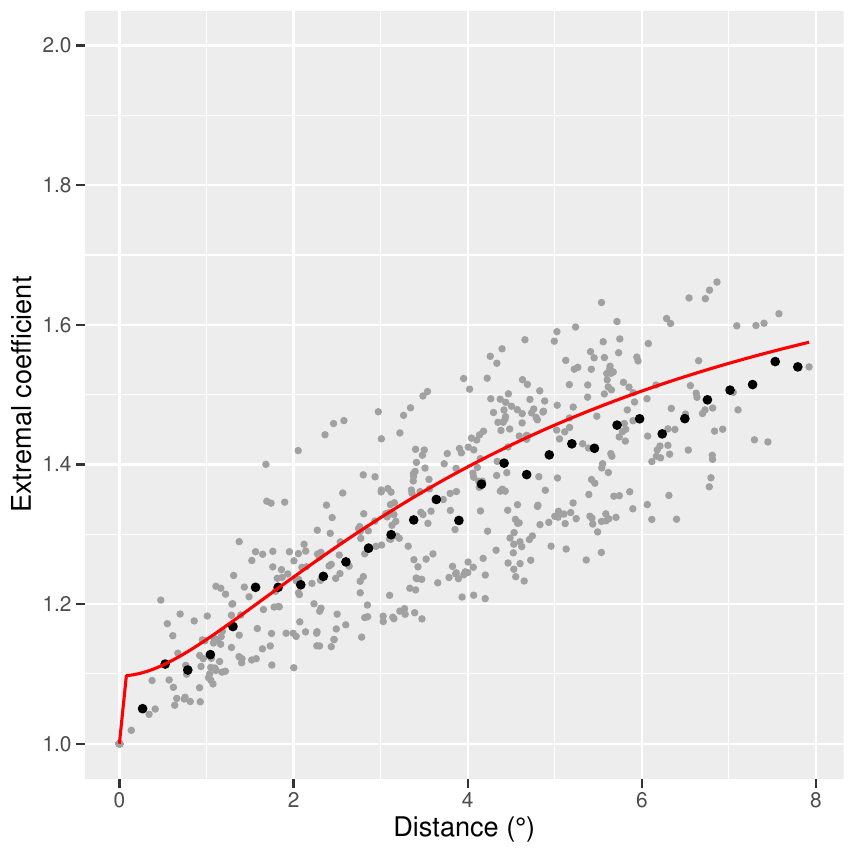}
        \end{subfigure}
        \caption{Model performance on the validation centroids. Theoretical pairwise extremal coefficient function from the extremal-$t$ Cauchy model (red line), and empirical pairwise extremal coefficients (dots). The gray and black
dots are pairwise and binned estimates, respectively. The empirical extremal coefficients have been computed using the empirical distribution functions (left) and the obtained GEV parameters (right).}
\label{fig:Fig_GoodnessFitExtCoeffExtCauchyEve}
\end{figure}

\begin{figure*}[!ht]
    \centering
    \begin{subfigure}[b]{0.48\textwidth}
        \centering
       \includegraphics[width=\textwidth]{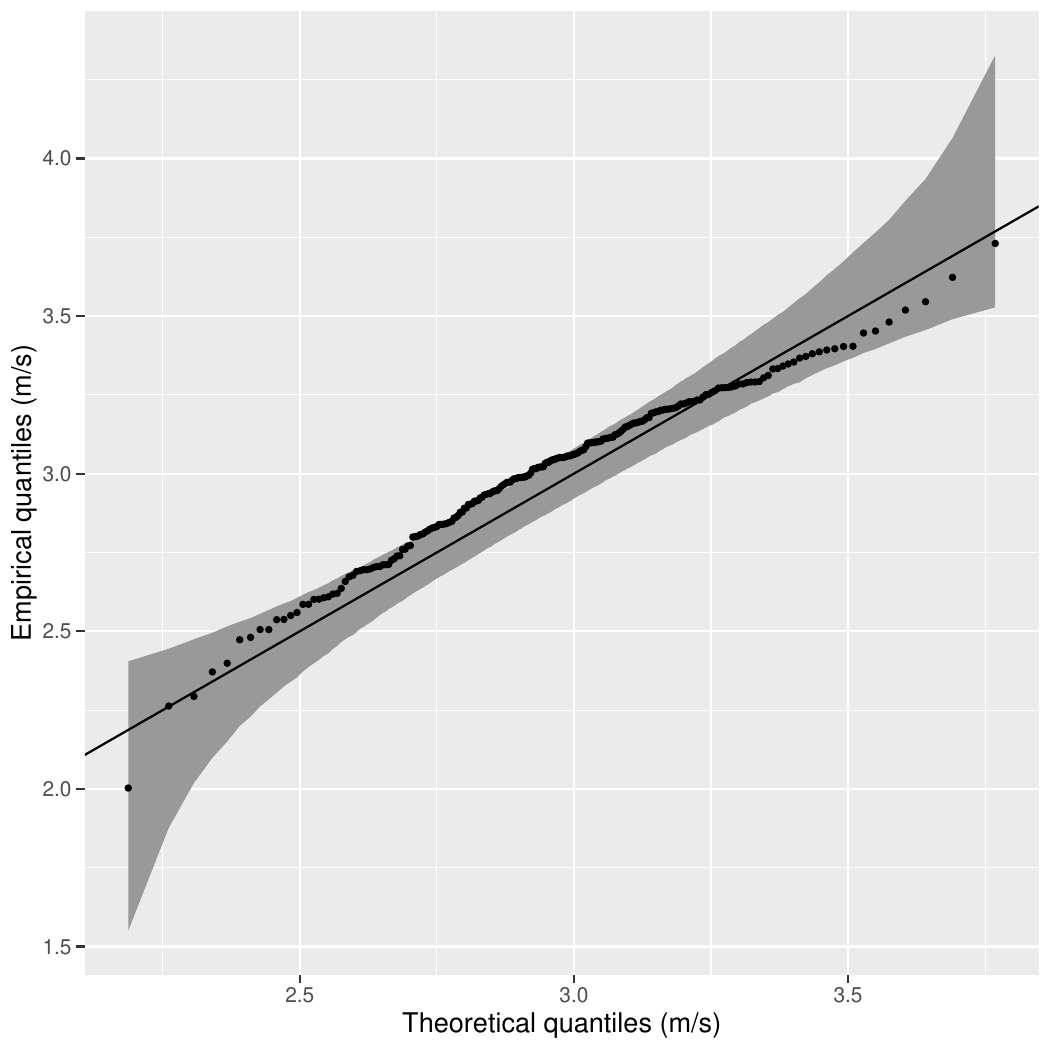}
        \end{subfigure}
        \hfill
        \begin{subfigure}[b]{0.48\textwidth}  
            \centering 
            \includegraphics[width=\textwidth]{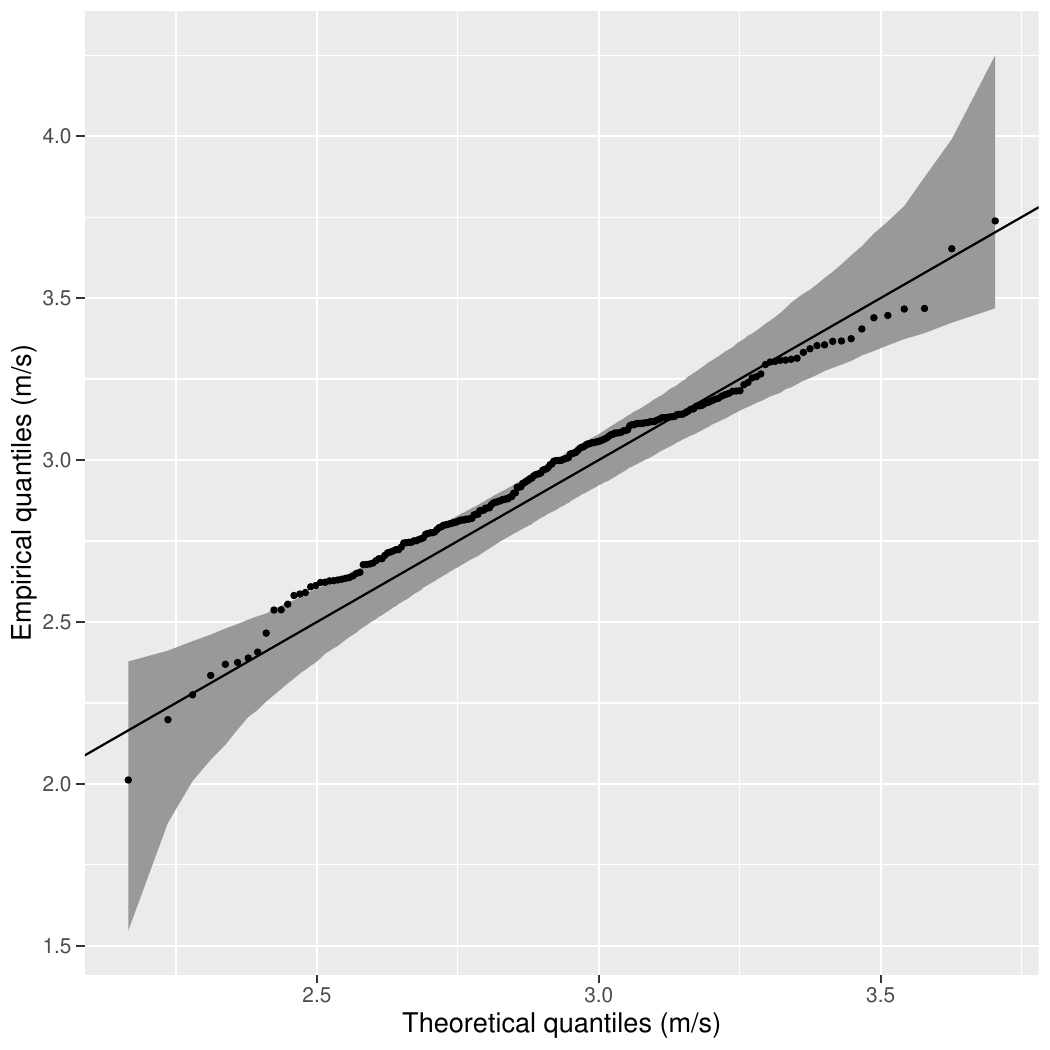}
        \end{subfigure}
        \hfill
        \vskip\baselineskip
        \vskip\baselineskip
        \begin{subfigure}[b]{0.48\textwidth}   
            \centering 
            \includegraphics[width=\textwidth]{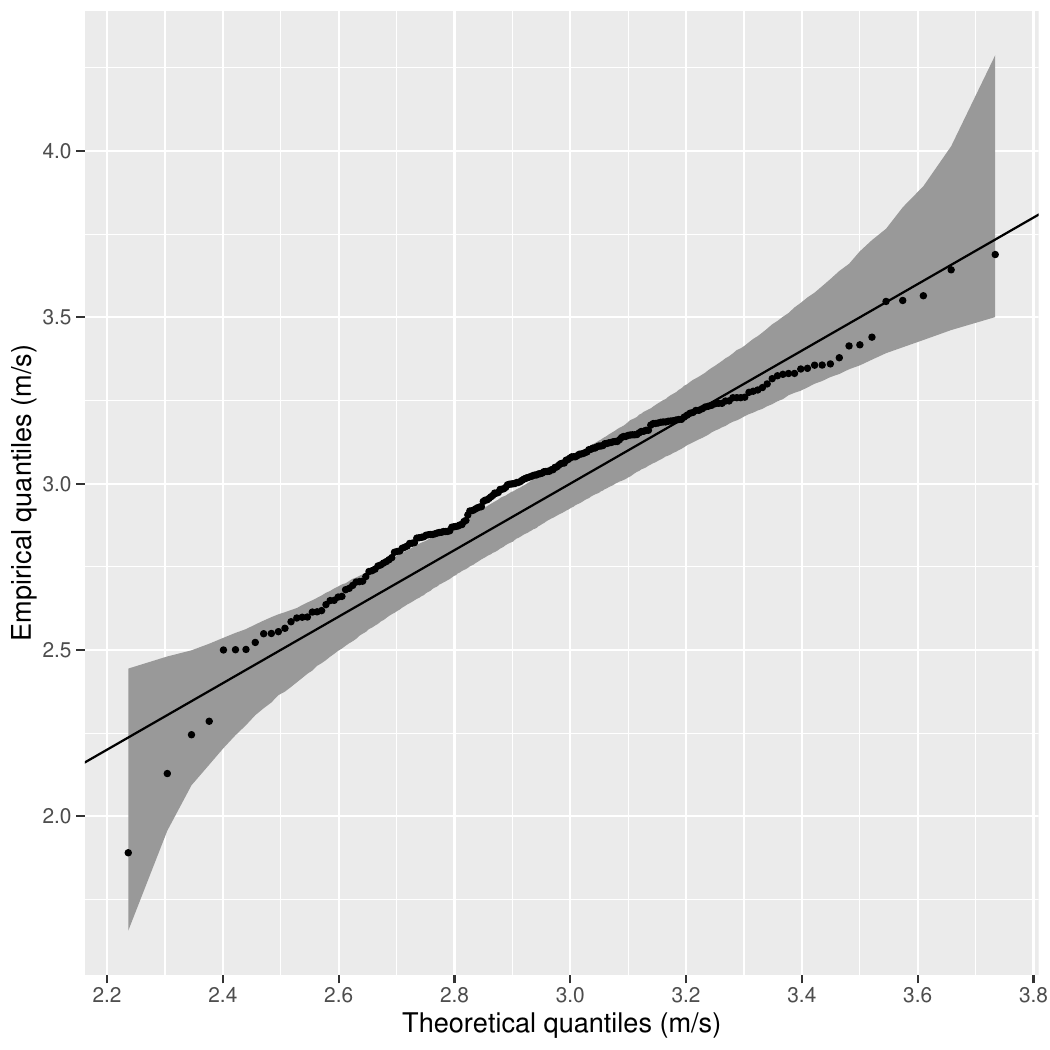}
        \end{subfigure}
        \hfill
        \begin{subfigure}[b]{0.48\textwidth}   
            \centering 
            \includegraphics[width=\textwidth]{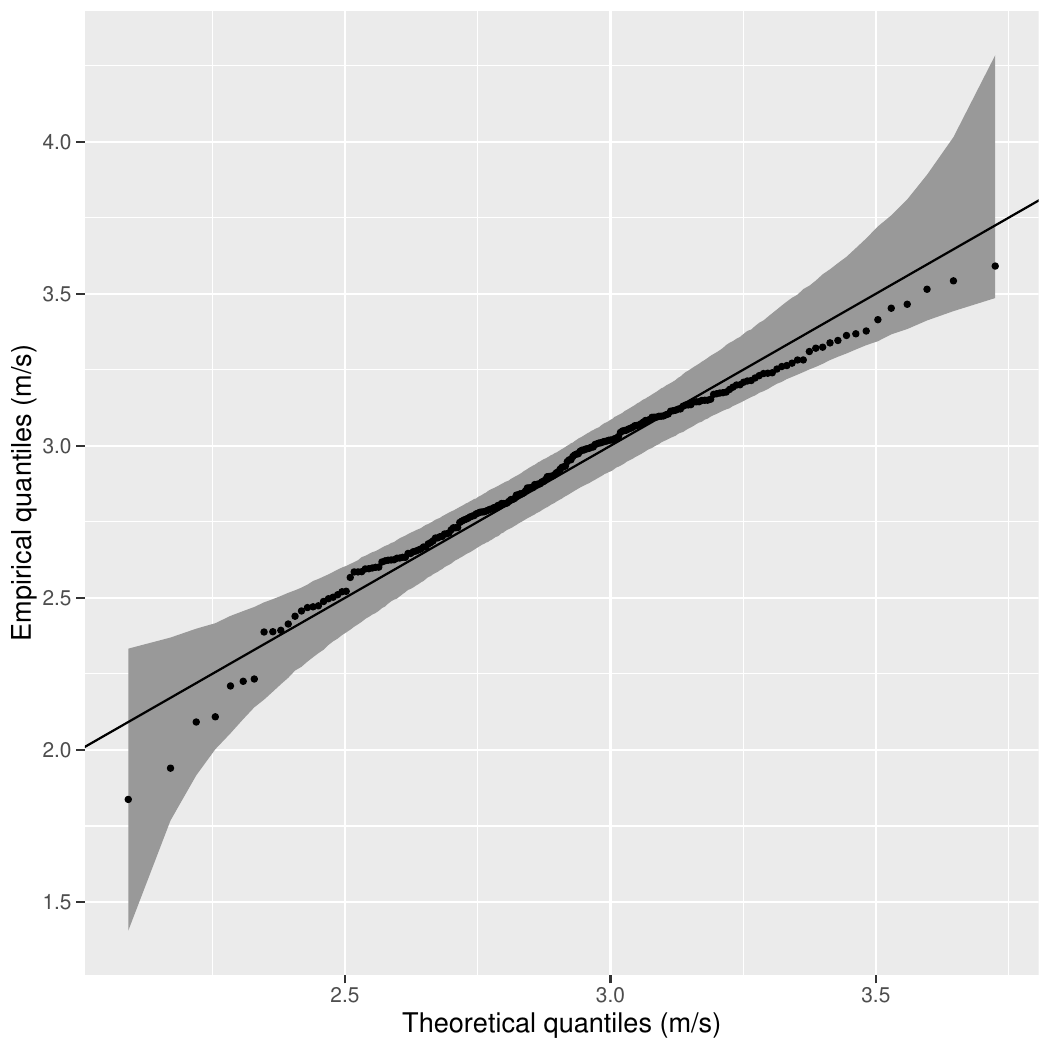}
        \end{subfigure}
        \hfill
        \caption{log-log plot of the observed versus simulated wind speeds from the fitted extremal-$t$ model for four randomly chosen CRESTA zones from the validation set. Overall envelopes at the 95\% confidence level are depicted in dark gray.}
        \label{fig:windval}
    \end{figure*}

Moreover, we perform max-stability tests analogously to \cite{davison} for the extremal-$t$ model with Cauchy correlation function fitted above. We present several Q-Q plots in Figure \ref{fig:Fig_GoodnessFitQQPlotExtCauchyEve} for various groupwise maxima based on the validation locations. We observe from the plots in the first row that pairwise dependencies seem to be modeled properly, no matter the distance between the two locations under consideration. The last two rows confirm that the higher dimensional properties are also correctly taken into account, using different summary statistics.
The plot for the fitted extremal coefficient on the validation locations is shown in Figure \ref{fig:Fig_GoodnessFitExtCoeffExtCauchyEve}. Even though it seems that we slightly underestimate dependence with the selected extremal-$t$ model, we deem this fit to be satisfactory. Indeed, we are fitting an extremal-$t$ random field to events and not to block maxima, see Section \ref{sec:appendixcalib} in the Appendix. This provides supplementary evidence that the dependence structure is adequately captured. Moreover, we provide log-log plots for four randomly chosen CRESTA zones from the validation set to validate the GEV marginals in Figure \ref{fig:windval}.

Refitting the model using all locations produces the estimates for the model parameters found in Table \ref{table_Param_Estimates_Final}. 
\begin{table}[!ht]
    \centering
    \begin{tabular}{c|r}
    $a_0$ & 0.007 (0.002) \\ 
    $\kappa$ & 4.064 (0.589) \\ 
    $\psi$ & 0.213 (0.099) \\ 
    $\nu$ & 2.974 (0.638) \\ 
    $\eta_0$ & -4.297 (4.878) \\ 
    $\eta_1$ & -0.155 (0.059) \\ 
    $\eta_2$ & 0.440 (0.094) \\ 
    $\eta_3$ & 0.001 (0.0001) \\ 
    $\tau_0$ & 4.485 (0.324) \\ 
    $\tau_1$ & 0.354 (0.063) \\ 
    $\xi_0$ & -0.011 (0.035) 
    \end{tabular}
    \caption{Parameter estimates (standard errors inside parentheses) of the best model when using all centroids for the fit.}
    \label{table_Param_Estimates_Final}
\end{table}

\subsection{Composite likelihood for the fitting of max-stable random fields}\label{sec:fitmaxstable}
In this section, we provide a brief summary of the general procedure employed to fit max-stable random fields to data.
The first step is to fit a trend surface (spatial GEV model) for the marginals. This is done using the function \texttt{fitspatgev} from the {\sf R} package \texttt{SpatialExtremes}. This function performs maximum likelihood estimation under the (wrong) assumption that locations are mutually independent. Indeed, the log-likelihood function which is maximized is given by
\begin{equation*}
\ell(\bm{z};\bm{\theta})=\sum_{m=1}^M \sum_{k=1}^K \log f_{GEV}(z_{m,k};\bm{\theta}_k),
\end{equation*}
where $M$ is the number of observations, $\bm{\theta}_k\in\Theta\subseteq\R^{q_k}$ is the vector of model parameters at location $k$, $f_{GEV}$ is the GEV density and $z_{m,k}$ is the $m$-th observation at location $k.$
To account for model misspecification, standard errors for model parameters are estimated with their sandwich estimates, see \cite{white82}.

Fitting a trend surface instead of calibrating a GEV distribution at each location enables the evaluation of the model at a new location using interpolation. We select the trend surface which yields the lowest composite likelihood information criterion (CLIC) value, see \eqref{eq:clic}.
Other non-parametric forms such as splines could be specified, see for example \cite{chavez05}.
The calibration on event data from Section \ref{sec:calibevents_apdx} yields the following trend surfaces for the GEV parameters:
\begin{align*}
\eta(\bm{x}) &= \eta_0 + \eta_1  \mathrm{lon}(\bm{x}) + \eta_2 \mathrm{lat}(\bm{x}) + \eta_3 \mathrm{alt}(\bm{x}), \\
\tau(\bm{x}) &= \tau_0 + \tau_1 \vert \mathrm{lat}(\bm{x}) - 51 \vert, \\
\xi(\bm{x}) &= \xi_0,
\end{align*}
where $\mathrm{lon}(\bm{x})$, $\mathrm{lat}(\bm{x})$ and $\mathrm{alt}(\bm{x})$ denote the longitude, latitude and altitude of location $\bm{x}$, respectively. It is common to assume a constant $\xi_0$ for the shape parameter.

We then fit the dependence structure for various max-stable random fields (Smith, Schlather, Brown--Resnick and extremal-$t$) on top of the trend surfaces obtained above. This step is performed using the function \texttt{fitmaxstab} from the {\sf R} package \texttt{SpatialExtremes}. For the Schlather and extremal-$t$ models we consider the powered exponential, Whittle--Matérn and Cauchy correlation functions.
Since there is no closed form distribution for max-stable processes in greater than $K=2$ dimensions (the density involves a sum taken over the set of all possible partitions of $\{1,\ldots,K\}$ and the dimension of this set equals the Bell number of order $K$), we rely on a composite-likelihood approach, see \cite{lindsay88} and \cite{varin08}.
Availability of the bivariate density $f_{k,k'}$ at locations $k$ and $k'$ leads to the formulation of a pairwise composite log-likelihood as a sum of log-likelihoods corresponding to each bivariate contribution:
\begin{equation*}
\ell_P(\bm{z};\bm{\varphi})=\sum_{m=1}^M \sum_{k=1}^{K-1}\sum_{k'=k+1}^K \ell_{k,k'}^m(\bm{\varphi}),    
\end{equation*}
where $M$ is the number of observations, $\bm{z}_m$ is the vector of wind speeds for observation $m$, $\bm{\varphi}\in\Phi\subseteq\R^q$ is the vector of model parameters, and each $\ell_{k,k'}^m(\bm{\varphi})=\log f(z_{m,k},z_{m,k'};\bm{\varphi})$ is the bivariate marginal log-likelihood based on data at locations $k$ and $k'$.
The bivariate density can be derived from the cumulative distribution of the extremal-$t$ process 
\begin{multline*}
\PP\{Z(\bm{x}_1)\leq z_1,Z(\bm{x}_2)\leq z_2\}=\exp\Bigg[-\frac{1}{z_1}T_{\nu+1}\left\{-\frac{\rho(\|\bm{x}_1-\bm{x}_2\|)}{b}+\frac{1}{b}\left(\frac{z_2}{z_1}\right)^{1/\nu}\right\} -
\\ -\frac{1}{z_2}T_{\nu+1}\left\{-\frac{\rho(\|\bm{x}_1-\bm{x}_2\|)}{b}+\frac{1}{b}\left(\frac{z_1}{z_2}\right)^{1/\nu}\right\}
\Bigg],   
\end{multline*}
where $T_\nu$ is the cumulative distribution function of a Student-$t$ random variable with $\nu$ degrees of freedom and $b^2=\frac{1-\rho(\|\bm{x}_1-\bm{x}_2\|)^2}{\nu+1}.$
In practice, one can include only closely located pairs in the composite likelihood to improve the statistical efficiency of the estimator, see \cite{padoan}. Model selection is then performed via the CLIC, see \cite{varin}, on the basis of the expected Kullback--Leibler divergence between the true unknown model and the adopted model:
\begin{equation}\label{eq:clic}
\textnormal{CLIC}=-2\left\{\ell_P\left(\bm{z};\hat{\bm{\varphi}}\right)-\textnormal{tr}\left[J\left(\hat{\bm{\varphi}}\right)H\left(\hat{\bm{\varphi}}\right)^{-1}\right]\right\},
\end{equation}
where $H(\hat{\bm{\varphi}})$ and $J(\hat{\bm{\varphi}})$ are the analogues of the expected information and the covariance matrix of the score vector, respectively, see \cite{padoan}.
This is equivalent to the Akaike Information Criterion (AIC) under model misspecification, see \cite{davison03}.

\section{Calibration on monthly maxima}\label{sec:appndx}
In this appendix we present an alternative calibration to our previous approach using block maxima of wind speeds. The wind speed data we use for the calibration is generated using an approach which is analogous to that used for the UKMO data in Section \ref{sec:windexpo}. 
\subsection{Reanalysis data}\label{sec:appendixdata}
Reanalysis data present spatial and temporal homogeneity for a large-scale study of European windstorm phenomena. 
They provide a global and realistic picture of the weather and climate by combining historical data with up-to-date climate models. The data are generated from a combination of data assimilation systems leveraging state-of-the-art global forecasting models and real observations which are consistent with so-called station or in situ observations.
In contrast, in situ observations are too coarse in space and/or short and inhomogeneous in time, see \cite{ceppi}.

In the following, we will use data from the European centre for medium range
weather forecasting (ECMWF), see \cite{ecmwf}, under the ERA5 (ECMWF Re-Analysis) atmospheric model\footnote{The dataset used in this study is freely accessible at \url{https://cds.climate.copernicus.eu/cdsapp\#!/home}}. The data consist of hourly observations of wind gust speeds since the 1st of January 1979 to the 31st of December 2020 with a spatial resolution of 0.25$^{\circ}$. 
The chosen variable is called "10m wind gust since previous post-processing" in the ECMWF terminology, which represents at each time stamp the maximal wind gust in the past hour, and where wind gusts are measured over 3 seconds at a height of 10 meters, consistently with the UKMO wind speed variable definition, see Section \ref{sec:windexpo}.
We must take specific care of the first day as the observations start only at 7am coordinated universal time (UTC). 
The choice of this variable is backed by \cite{dellamarta} which states that wind gust is ideal to describe wind peaks at the surface, further leading to storm damages. There is no abrupt change in observations and additionally the data coverage is satisfactory for most problems.

However it is also known that wind gust provides unrealistic measurements for coastal regions and sites with steep orographic gradient. It is further discussed in \cite{ceppi} that reanalysis wind speed data are a parametrized quantity, display biases in the wind speed magnitude and may not necessarily be representative of local wind speed data due to the coarse resolution of the grid. One therefore needs to rely on in-situ wind data in order to capture the local dynamics and obtain accurate wind speed estimates.
We can however ignore this aspect as we are interested in aggregate risk over a relatively large region. 
On the other hand, wind speed data are sometimes available through their $u$ and $v$ components which represent instantaneous measurements. This results in confounding issues and unreliable estimates of quantities such as return periods, see \cite{dellamarta}.
For these reasons we decide to base our analysis on wind gust data.

A subregion of interest can be selected by specifying geographical coordinates.
We specify the smallest rectangle encompassing the entire German territory and retrieve wind speed data at all grid points with a $0.25^\circ$ resolution. This rectangle covers a region with latitudes between 47.25$^\circ$ and 55.25$^\circ$ and longitudes between 5.75$^\circ$ and 15.25$^\circ$. This results in a grid with $33\cdot 39=1287$ sites. 
Moreover, the spatial resolution of the data being 0.25$^{\circ}$, two adjacent grid points are approximately 27km apart from each other. To be more precise, 0.25$^{\circ}$ in terms of latitude represent approximately 27.8km and based on the sphericity of the earth and the cosine of the latitude, 0.25$^{\circ}$ in terms of longitude at our level represent approximately 25.1km.
By doing this, we are splitting the region into rectangles, with each grid point at the bottom-left corner of a given rectangle acting as the representative grid point for this rectangle.
In order to select only the grid points lying within the German border, we take the intersection with the extent of the shape file for Germany, resulting in 730 grid points. 

When exposure data is available, it is usually not under this gridded resolution. It is typically submitted to players in the reinsurance and ILS market on a regional or CRESTA zone basis, see Section \ref{sec:windexpo}. The exposure data would then have to be mapped to the gridded representation. One solution would be to allocate proportionately the exposure of a CRESTA zone to all grid points within this CRESTA zone.
Due to the $0.25^\circ$ resolution of the grid, there are six CRESTA zones with a small geographical footprint which do not contain any grid point: these are CRESTA zones DEU12, DEU13, DEU20, DEU60, DEU80 and DEU81. 

\subsection{Model calibration and validation}\label{sec:appendixcalib}
We consider only the data between the months of October and March, which apart from a few exceptions correspond to the months with the highest wind speeds. Moreover, during these months, insured losses are mainly due to wind and not to other physical hazards such as hail, see \cite{prahl12} for a study using insured loss data from 1997 to 2007 on a German administrative district level.
From a meteorological perspective, extreme wind speeds occurring during this period are due to extra-tropical cyclones which cause storms that are more stationary in space and live on a larger scale; that is, if a point on our grid is hit, it is very likely that a lot of other points will be hit, this effect dampening with the distance between points. Conversely, severe convective storms which occur during the summer lead to so-called straight-line winds, which are more random and localized, and thus, more difficult to model.
We thus consider the maximal wind speeds at each grid point over each month in the extended winter period, and fit a GEV distribution to these time series. Here we are implicitly assuming that these maxima are i.i.d. and by considering monthly blocks, we are significantly reducing seasonality effects in the underlying series. Indeed, the wind speeds tend to be lower during the spring and summer months. The independence of the maxima and the stationarity of the time series of monthly maxima also seem to be reasonable assumptions for our dataset. The underlying motivation to consider monthly maxima is due to an observed average number of events per year between 4 and 5, see Section \ref{sec:freqdist}. 
To ensure a fair comparison with the calibration on events in Section \ref{sec:calibevents}, we take an average of the monthly maximal wind speeds for each grid point within the 89 CRESTA zones containing at least one grid point. 
On top of ensuring a fair comparison between both approaches it reduces the computational effort as the fit would be very long when calibrating a max-stable random field to 730 grid points. 

\begin{figure}[!ht]
\begin{center}
\includegraphics[scale=0.7]{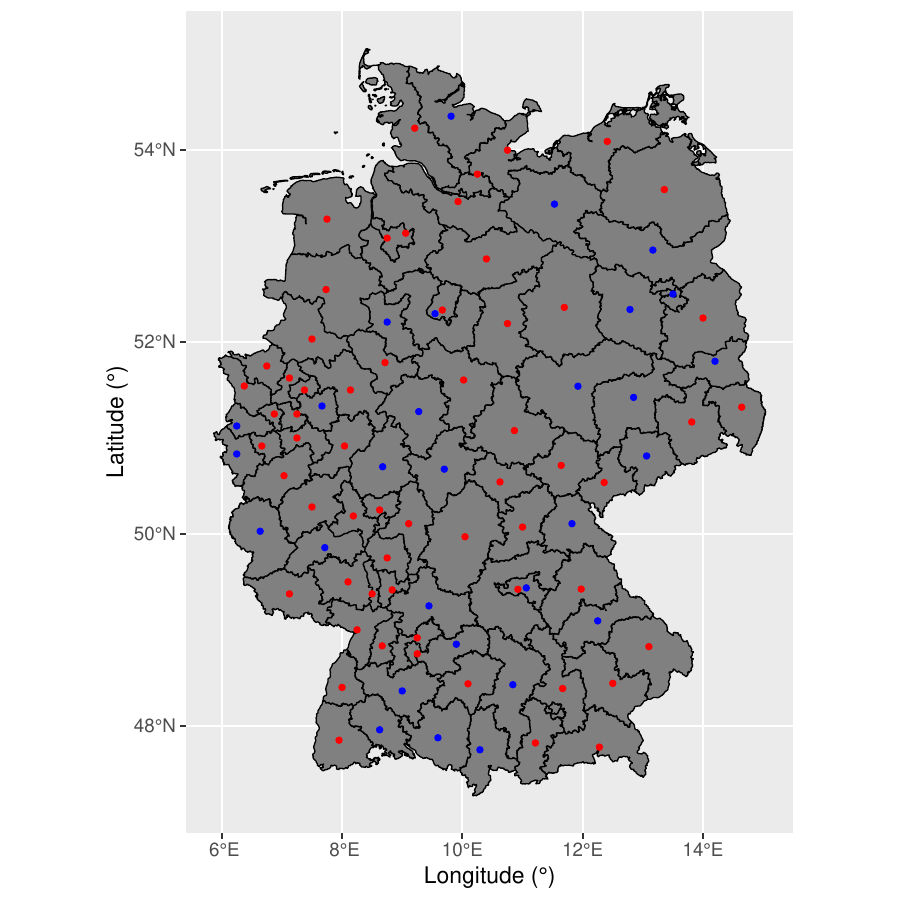}
\caption{Map showing the $89$ CRESTA zone centroids used for our analysis. We consider $60$ centroids for calibration (red dots) and $29$ for validation (blue dots).}\label{fig:mapStations2} 
\end{center}
\end{figure}

As in Section \ref{sec:calibevents_apdx}, we consider a split of the grid points into two groups: we randomly choose 59 grid points for calibration and the remaining 30 are used for validation. 
We display the map of the stations in Figure \ref{fig:mapStations2}, where the red dots correspond to calibration stations and the blue dots to validation stations.
As in Section \ref{sec:calibevents_apdx}, we jointly fit the marginal distributions with the dependence structure arising from various max-stable fields (Smith, powered exponential, Whittle--Matérn, Cauchy and Brown--Resnick, see Section \ref{sec:max-stable}) to the wind speed maxima using the \texttt{fitmaxstab} function from the {\sf R} package~\texttt{SpatialExtremes}.

For the GEV parameters, we choose the following trend surfaces:
\begin{align*}
\eta(\bm{x}) &= \eta_0 + \eta_1  \mathrm{lon}(\bm{x}) + \eta_2 \mathrm{lat}(\bm{x}) + \eta_3 \mathrm{alt}(\bm{x}), \\
\tau(\bm{x}) &= \tau_0 + \tau_1 \mathrm{lon}(\bm{x}) + \tau_2 \mathrm{lat}(\bm{x}),  \\
\xi(\bm{x}) &= \xi_0,
\end{align*}
where $\mathrm{lon}(\bm{x})$, $\mathrm{lat}(\bm{x})$ and $\mathrm{alt}(\bm{x})$ denote the longitude, latitude and altitude of location $\bm{x}$, respectively.
For this alternative calibration, adding the longitude in $\tau$ and not transforming the latitude (as in the calibration on events) will lead to an improvement in the fit.

\begin{table}
\begin{center}
\begin{tabular}{l|c|c}
Model & CLIC & Number of model parameters \\
\hline
Extremal-$t$ Whittle--Matérn & 4,882,594  & 12  \\
Extremal-$t$ Cauchy & 4,882,837  & 12  \\
Extremal-$t$ powered exponential & 4,886,564 & 12   \\
Brown--Resnick  & 4,889,217 & 10  \\
Schlather powered exponential & 4,909,067 & 11   \\
Smith  & 4,909,293  & 11  \\
Schlather Whittle--Matérn & 4,909,718 & 11   \\
Schlather Cauchy & 4,909,925 & 11
\end{tabular}
\end{center}
\caption{CLIC values for the different fitted models (in ascending order).}
\label{Table_CLIC_Values}
\end{table}

\begin{figure}[!ht]
    \centering
    \begin{subfigure}[b]{0.48\textwidth}
        \centering
       \includegraphics[width=\textwidth]{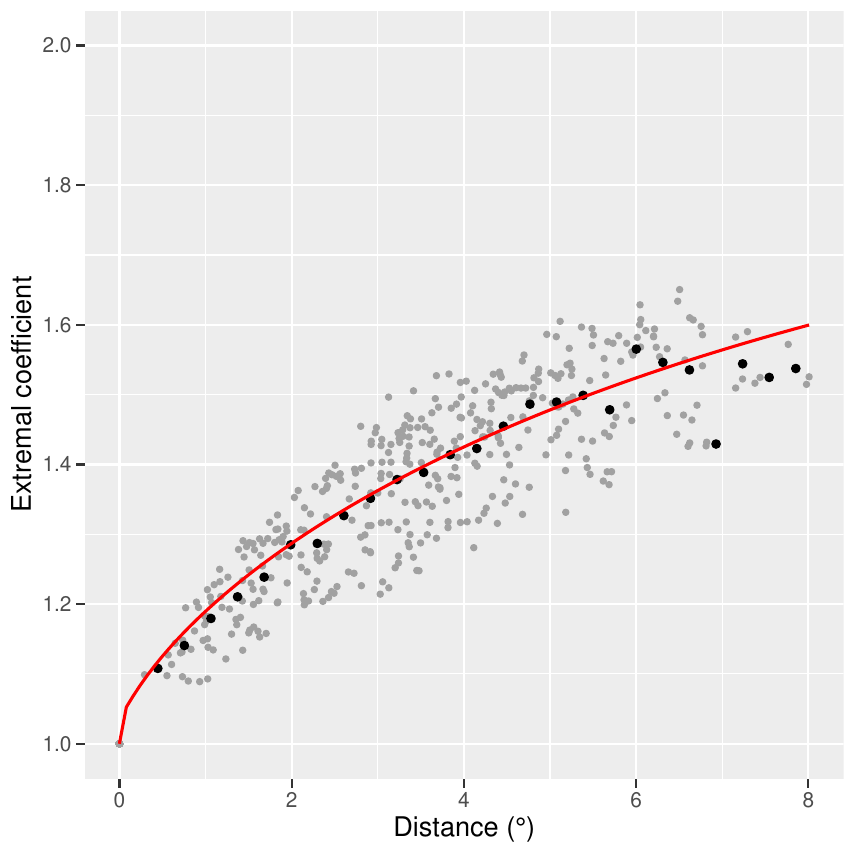}
        \end{subfigure}
        \hfill
        \begin{subfigure}[b]{0.48\textwidth}  
            \centering 
            \includegraphics[width=\textwidth]{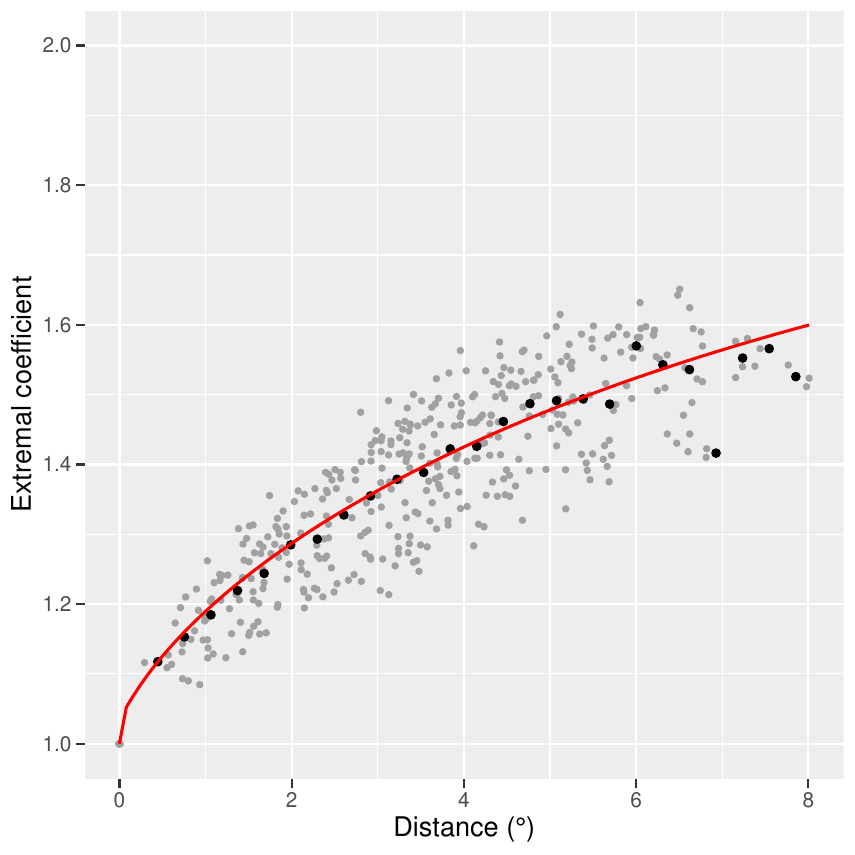}
        \end{subfigure}
        \caption{Model performance on the validation centroids. Theoretical pairwise extremal coefficient function from the extremal-$t$ Whittle--Matérn model (red line), and empirical pairwise extremal coefficients (dots). The gray and black
dots are pairwise and binned estimates, respectively. The empirical extremal coefficients have been computed using the empirical distribution functions (left) and the obtained GEV parameters (right).}
\label{fig:Fig_GoodnessFitExtCoeffExtCauchy}
\end{figure}

\begin{figure*}
    \centering
    \begin{subfigure}[b]{0.32\textwidth}
        \centering
       \includegraphics[width=\textwidth]{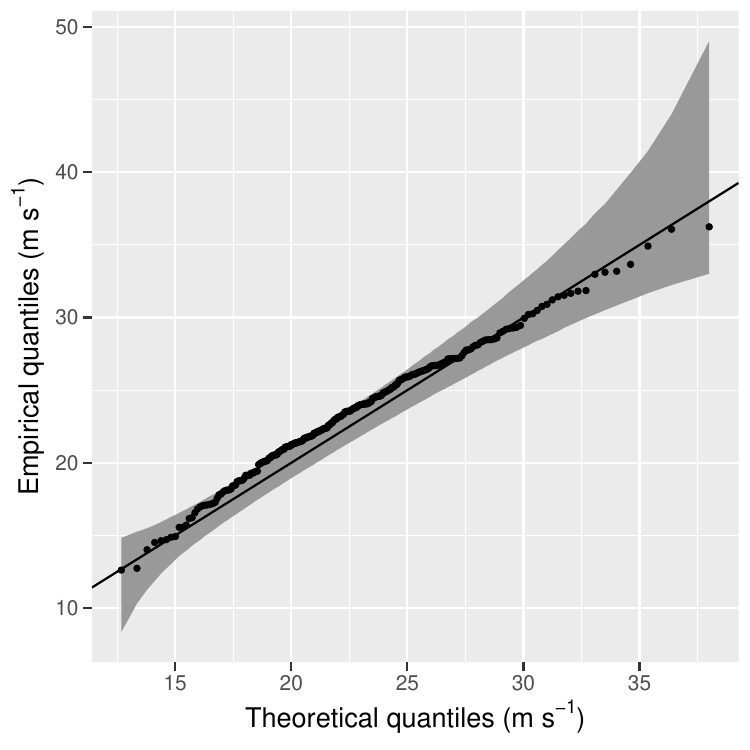}
        \end{subfigure}
        \hfill
        \begin{subfigure}[b]{0.32\textwidth}  
            \centering 
            \includegraphics[width=\textwidth]{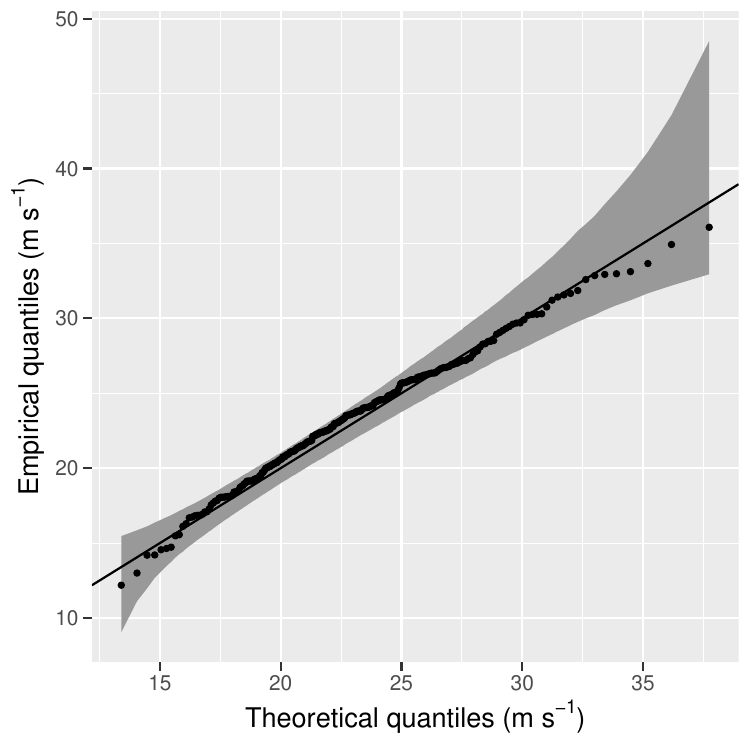}
        \end{subfigure}
        \hfill
        \begin{subfigure}[b]{0.32\textwidth}  
            \centering 
            \includegraphics[width=\textwidth]{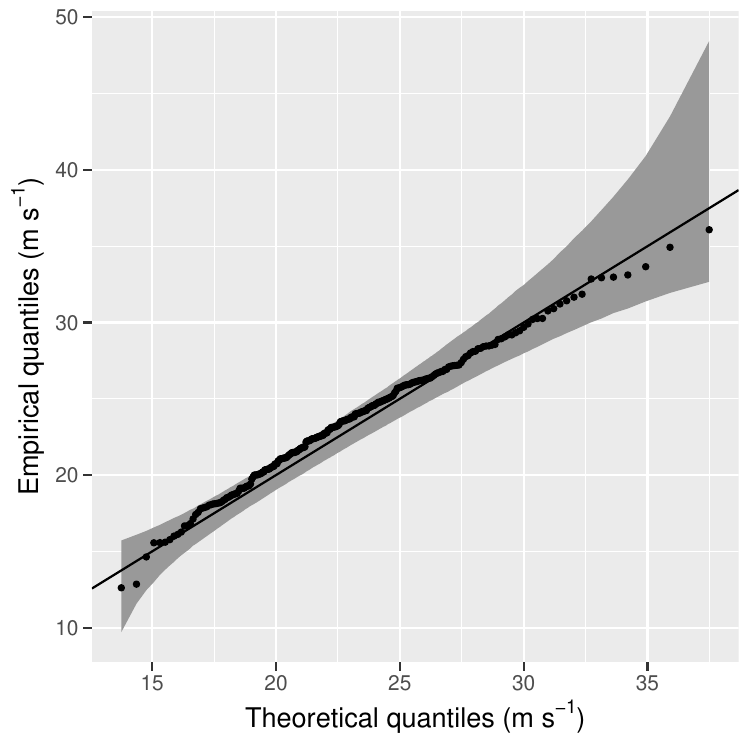}
        \end{subfigure}
        \vskip\baselineskip
        \begin{subfigure}[b]{0.32\textwidth}   
            \centering 
            \includegraphics[width=\textwidth]{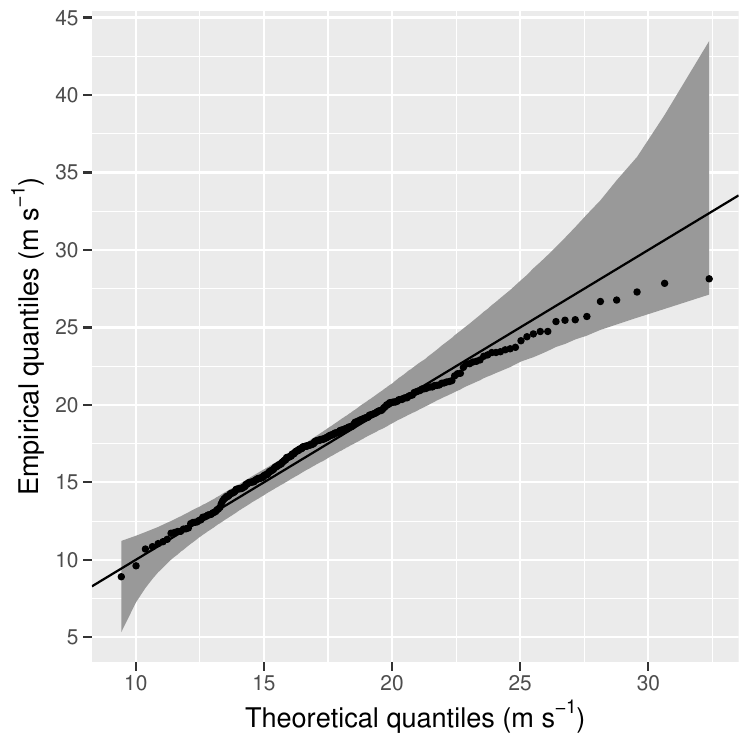}
        \end{subfigure}
        \hfill
        \begin{subfigure}[b]{0.32\textwidth}   
            \centering 
            \includegraphics[width=\textwidth]{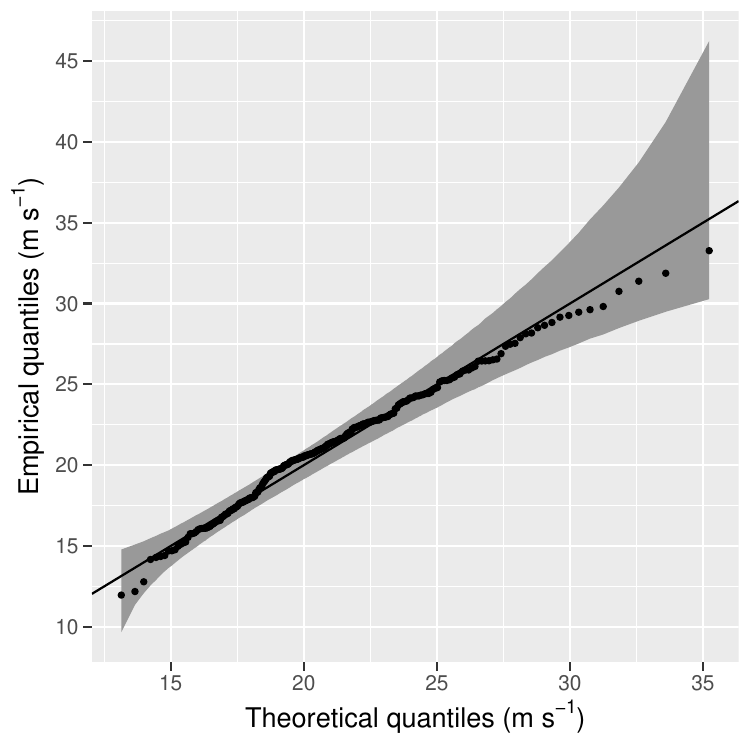}
        \end{subfigure}
        \hfill
        \begin{subfigure}[b]{0.32\textwidth}   
            \centering 
            \includegraphics[width=\textwidth]{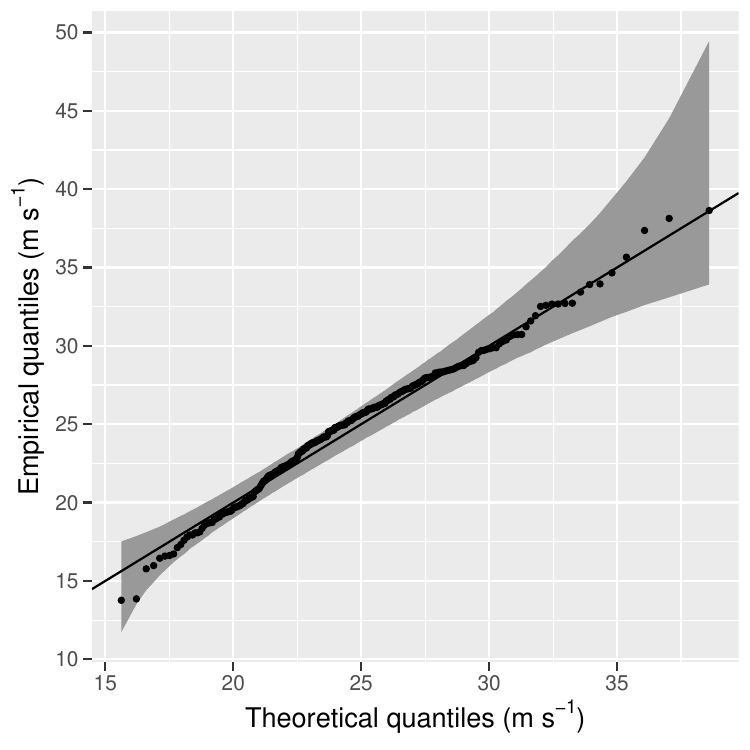}
        \end{subfigure}
        \vskip\baselineskip
        \begin{subfigure}[b]{0.32\textwidth}   
            \centering 
            \includegraphics[width=\textwidth]{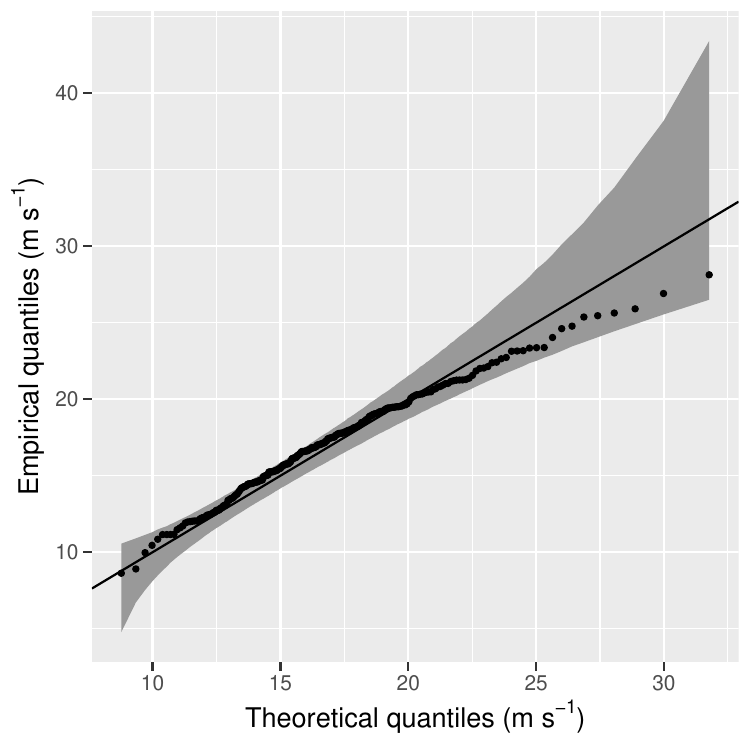}
        \end{subfigure}
        \hfill
        \begin{subfigure}[b]{0.32\textwidth}   
            \centering 
            \includegraphics[width=\textwidth]{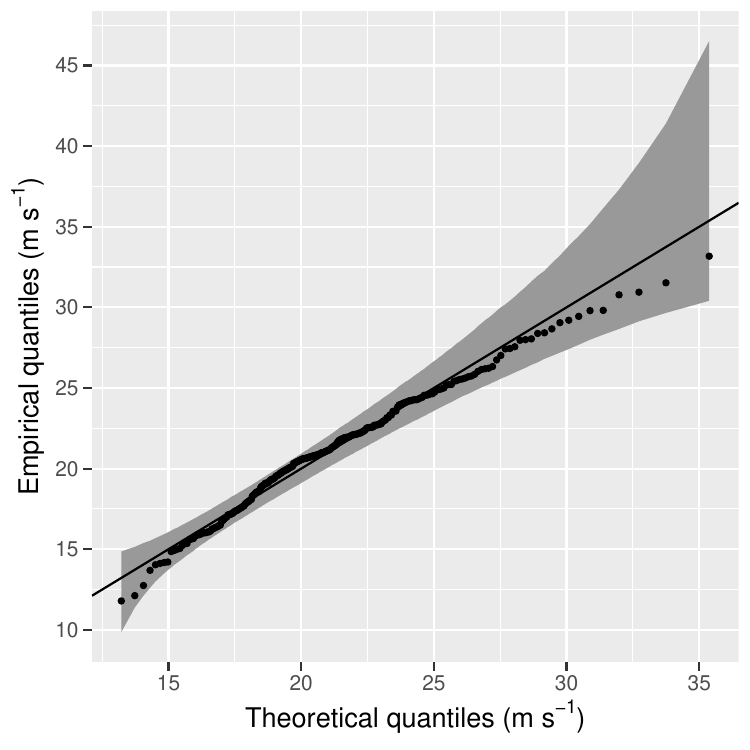}
        \end{subfigure}
        \hfill
        \begin{subfigure}[b]{0.32\textwidth}   
            \centering 
            \includegraphics[width=\textwidth]{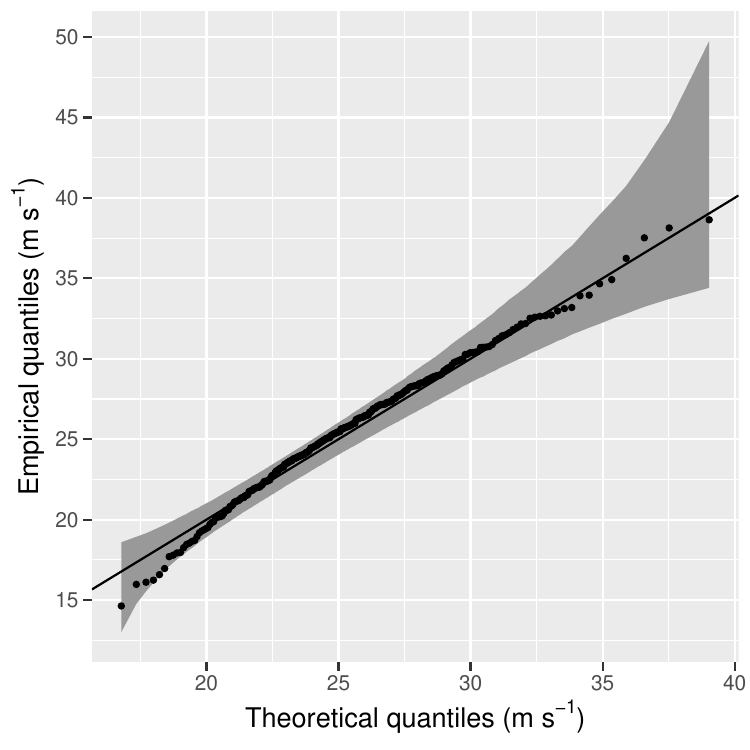}
        \end{subfigure}
        \caption{Performance of the chosen model on the validation centroids. The top row concerns maxima for pairs of validation centroids separated by a low (left), moderate (middle) and a long (right) distance. The middle row focuses on minima (left), mean (middle) and maxima (right) for a group of $10$ validation centroids chosen randomly. The bottom row concerns  minima (left), mean (middle) and maxima (right) for all $29$ validation centroids. Overall envelopes at the 95\% confidence level are depicted in dark gray.}
        \label{fig:Fig_GoodnessFitQQPlotExtCauchy}
    \end{figure*}

We perform the same max-stability tests as in Section \ref{sec:calibevents_apdx}.
As for the calibration on events, we observe from the plots in the first row in Figure \ref{fig:Fig_GoodnessFitQQPlotExtCauchy} that pairwise dependencies seem to be modeled properly, no matter the distance between the two stations under consideration. The last two rows confirm that the higher dimensional properties are also correctly taken into account, using different summary statistics.
The plot for the fitted extremal coefficient on the validation stations is shown in Figure \ref{fig:Fig_GoodnessFitExtCoeffExtCauchy}. The satisfactory fit provides supplementary evidence that the dependence structure is adequately captured. Note that the better fit suggested by Figure \ref{fig:Fig_GoodnessFitExtCoeffExtCauchy} compared to Figure \ref{fig:Fig_GoodnessFitExtCoeffExtCauchyEve} reflects the fact that this second calibration is based on block maxima and not on events for which extreme value theory is a priori less suitable. However, as cat bonds are structured on event basis, the calibrations in Sections \ref{sec:calibevents} and \ref{sec:calibevents_apdx} are more consistent with what is done in the industry.

Refitting the model using all locations produces the following estimates for the model parameters:
\begin{table}[h!]
    \centering
    \begin{tabular}{c|r}
    $a_0$ & 0.002 (0.001) \\ 
    $\kappa$ & 17.984 (4.001) \\ 
    $\psi$ & 0.742 (0.050) \\
    $\nu$ & 5.572 (0.784) \\ 
    $\eta_0$ & -1.893 (2.358) \\ 
    $\eta_1$ & -0.271 (0.034) \\ 
    $\eta_2$ & 0.459 (0.045) \\ 
    $\eta_3$ & 0.002 (0.0001) \\ 
    $\tau_0$ & 11.507 (1.382) \\ 
    $\tau_1$ & -0.036 (0.019) \\ 
    $\tau_2$ & -0.139 (0.026) \\ 
    $\xi_0$ & -0.113 (0.019)
    \end{tabular}
    \caption{Parameter estimates (standard errors inside parentheses) of the best model when using all centroids for the fit.}
    \label{table_Param_Estimates_Final_reanalysis}
\end{table}

\begin{figure}[!ht]
    \centering
       \includegraphics[width=0.6\textwidth]{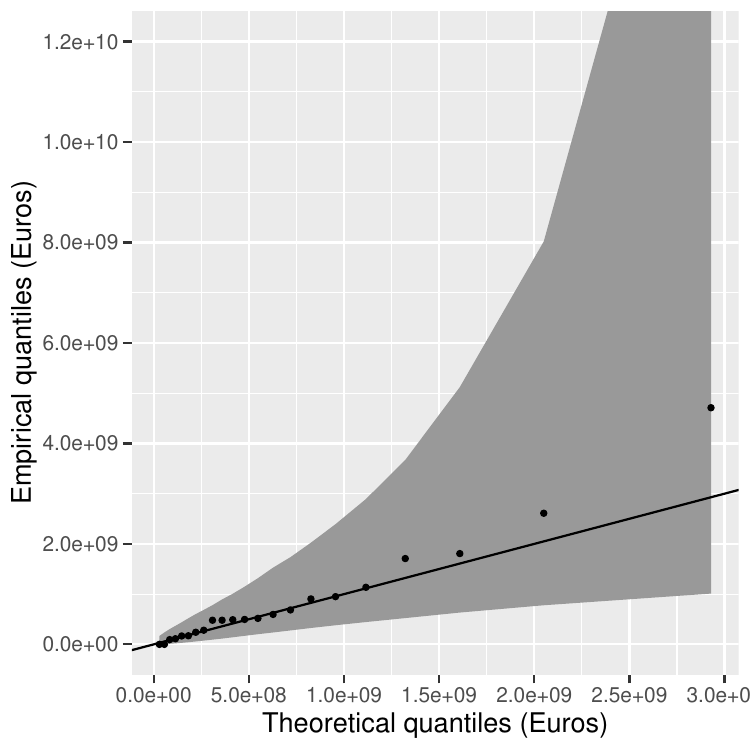}            
        \caption{Q-Q plot of the observed versus simulated yearly trigger values obtained from the fitted extremal-$t$ model with Whittle--Matérn correlation using our trigger. 
        Overall envelopes at the 95\% confidence level are depicted in dark gray.}
\label{fig:QQplotWM}
\end{figure}

As in Section \ref{sec:modelrisk}, we compare the empirical distribution of the yearly losses as given by our simulations $\{S_j, \mbox{ }j=1,\ldots,J\}$ obtained with the extremal-$t$ model (with Whittle--Matérn correlation) with the observed yearly trigger values by means of the Q-Q plot in Figure \ref{fig:QQplotWM}. We observe a stronger underestimation in the tail of the distribution compared to the Q-Q plots obtained with the extremal-$t$ model with Cauchy correlation and the Brown--Resnick models in Figure \ref{fig:QQplotTrig}. 
Considering the same examples as in Section \ref{sec:results}, we obtain for tranche A a price of 102.84\euro{} and for tranche B a price of 103.26\euro{}, suggesting lower losses under the extremal-$t$ model with Whittle--Matérn correlation compared to the extremal-$t$ model with Cauchy correlation, see Section \ref{sec:results}.

\subsection{Discussion of coverage type}\label{sec:covtype}
As mentioned in Section \ref{sec:general}, cat bonds can be structured to provide either per-occurrence or per-aggregate cover. In the context of the calibration on monthly maxima, modeling the physical hazard with a max-stable random field results in a cover for the maximal event during a month. The calibration can also be based on the maximal event during a year, which would more closely satisfy the asymptotics of extreme value theory.  
Even though several extreme windstorms could occur during an extended winter period, a coverage based on the most extreme event in a year (based on wind speeds) is relevant for the protection of remote-layer per-event reinsurance contracts, per-event contracts with no reinstatements and also for natural perils where events are infrequent, such as earthquakes for which the calibration would be analogously based on maxima of observable variables such as spectral acceleration or peak ground acceleration (PGA). 
Moreover, the bond is not necessary liquidated following the first triggering event. Indeed we have to wait until the end of the period as there could potentially be a bigger event taking place within this time frame. However, there is no need to wait until the end of the observation period if one of the known events already fully exhausts the structure, leading to a total loss of principal.

The coverage type also depends on the type of peril: in the case of wind, insured losses are usually due to heavy winds which are caused by extra-tropical cyclones and occur during autumn and winter months. As discussed previously, these winds are not localized but rather live on a global scale. In this sense, high wind speed values are often observed simultaneously across the entire region. The coverage we propose in Section \ref{sec:appendixcalib} is thus closer to occurrence-based covers. Indeed, if we consider the maximal wind speeds at each site, it is likely that they will occur during the same (short) period and that the resulting losses will originate from the same event. One could argue in a similar way for other large-scale perils such as hurricanes. On the other hand, tornadoes are very localized and events would have to be considered separately.

One could also consider a specific event which would correspond to a per-occurrence cover in the standard sense, or the aggregation of losses caused by several events during a year, which is the calibration we propose in Section \ref{sec:modelrisk}. 
In these cases the max-stable strategy is less adapted and the modeling can be more complicated from a statistical viewpoint. We refer to \cite{fondeville} for a study on $r$-Pareto processes which could help to tackle this issue.
One could also question whether using the second and third largest wind speed observations at each site would increase the model fit. However, we believe the dataset is large enough to calibrate a max-stable random field to the annual or monthly maxima at each grid point, which provide 42 and 252 observations for each of the 89 representative grid points, respectively.
Moreover, by analyzing the time stamps of the second and third largest wind speeds at each location for each winter period, we observe some rather strong clustering in the time stamps when these events occur, which is consistent with the fact that winter storms live on large temporal and spatial scales. Taking into account only the largest monthly wind speed observations at each location and for each six-month window between October and March thus allows us to work with temporally uncorrelated variables and to leverage the max-stable machinery. Furthermore, from a practical point of view, only the largest wind speeds cause insured losses, and for cat bonds, which typically exhibit high attachment points and cover remote (low-frequency) layers, in the rare cases where the bond is triggered, it is usually only by the largest event and not also by the second largest event. These practical considerations justify calibrating a max-stable random field to the wind speed maxima.

\end{document}